\documentstyle[12pt]{article}
\catcode`\@=11
\def\newsymbol#1#2#3#4#5{\let\next@\relax%
 \ifnum#2=\@ne\else%
 \ifnum#2=\tw@\let\next@\msyfam@\fi\fi%
 \mathchardef#1="#3\next@#4#5}
\def\mathhexbox@#1#2#3{\relax%
 \ifmmode\mathpalette{}{\m@th\mathchar"#1#2#3}r
 \else\leavevmode\hbox{$\m@th\mathchar"#1#2#3$}\fi}
\def\hexnumber@#1{\ifcase#1 0\or 1\or 2\or 3\or 4\or 5\or 6\or 7\or 8%
\or 9\or A\or B\or C\or D\or E\or F\fi}
\font\tenmsy=msbm10
\font\sevenmsy=msbm7
\font\fivemsy=msbm5
\newfam\msyfam
\textfont\msyfam=\tenmsy
\scriptfont\msyfam=\sevenmsy
\scriptscriptfont\msyfam=\fivemsy
\edef\msyfam@{\hexnumber@\msyfam}
\def\Bbb#1{\fam\msyfam\relax#1}
\catcode`\@=\active
\load{\footnotesize}{\sf}

\newtheorem{theorem}{Theorem}[section]
\newtheorem{proposition}[theorem]{Proposition}
\newtheorem{lemma}[theorem]{Lemma}
\newtheorem{corollary}[theorem]{Corollary}

\newtheorem{remark}[theorem]{Remark}

\newcommand{\proof}{{\noindent \it Proof:\ }}
\newcommand{\qed}{\hfill $\Box$\par\hfill\\}
\newcommand{\BR}{{{\Bbb R}^3}}

\newcommand{\RR}{{{\Bbb R}}}
\newcommand{\RI}{{\Bbb R}_{\mbox{\tiny IR}}^{3}}
\newcommand{\RU}{{\Bbb R}_{\mbox{\tiny UV}}^{3}}

\newtheorem{lem}[theorem]{\bf Lemma}
\newtheorem{cor}[theorem]{\bf Corollary}

\newcommand{\vkl}{V_{\k\la}} 
\renewcommand{\o}{\otimes}

\newcommand{\LR}{{L^2({\Bbb R}^3)}}

\newcommand{\g}[1]{e^{-i  g^{[j,#1]}\cdot x}} 
\newcommand{\pg}[1]{e^{i g^{[j,#1]}\cdot x}}

\newcommand{\mf}[1]{e^{-i f_j^{[#1]} x_j }}
\newcommand{\pf}[1]{e^{i f_j^{[#1]} x_j }}
\renewcommand{\iff}{e^{i f_j x_j}} 
\newcommand{\ig}{e^{ig\cdot x}}
\newcommand{\iif}{e^{-i f_j x_j}}

\newcommand{\fff}{{\ff}}
\newcommand{\fffI}{{\ff}_1}
\newcommand{\fffU}{{\ff}_2}

\newcommand{\ff}{{\cal F}}

\newcommand{\f}{^{-1}}

\newcommand{\eppp}{E_{\mathrm at}}
\newcommand{\teppp}{{E_{\mathrm at}^{\mbox{\tiny $
\left( \tau\right)$}}}}

\newcommand{\zeroeppp}{{E_{\mathrm at}^{\mbox{\tiny $
\left( 0\right)$}}}}

\newcommand{\tepppp}{{E_{1}^{\mbox{\tiny $
\left( \tau\right)$}}}}

\newcommand{\tQ}{{Q^{\mbox{\tiny $\left( \tau\right)$}}}}

\newcommand{\teqqq}{E_{2}^{\mbox{\tiny $
\left( \tau\right)$}}}
\newcommand{\zerogsat}{{\psi_{\mathrm at}^{\mbox{\tiny $
\left( 0\right)$}}}}

\newcommand{\tgsat}{{\psi_{\mathrm at}^{\mbox{\tiny $
\left( \tau\right)$}}}}
\newcommand{\gsat}{{\psi_{\mathrm at}}}
\newcommand{\pap}{H_{\mathrm at}} 
\newcommand{\tpap}{{H_{\mathrm at}^{\mbox{\tiny $
\left( \tau\right)$}}}} 

\newcommand{\zeropap}{{H_{\mathrm at}^{\mbox{\tiny $
\left( 0\right)$}}}}

\newcommand{\euv}{{\mathsl e}_{{}_{\mbox{\tiny UV}}}}
\newcommand{\eir}{{\mathsl e}_{{}_{\mbox{\tiny IR}}}}
\newcommand{\airi}{{{\mathsl e}_{\mbox{\tiny 
IR}}^{\mbox{\tiny $(1)$}}}}
\newcommand{\airii}{{{\mathsl e}_{\mbox{\tiny 
IR}}^{\mbox{\tiny $(2)$}}}}
\newcommand{\rc}{\varrho}

\newcommand{\add}{a^{\ast}}

\newcommand{\tai}{{A(\tau)}}

\newcommand{\akl}{A_{\kappa \Lambda}}
\newcommand{\takl}{{A_{\kappa \Lambda}^{
\mbox{\tiny $\left(\tau\right)$}}}}

\newcommand{\aklj}{A_{\kappa \Lambda, j}}
\newcommand{\akljast}{{A_{\kappa \Lambda, j}^{*}}}
\newcommand{\akljs}{{A_{\kappa \Lambda, j}}}
\newcommand{\aklis}{{A_{\kappa \Lambda, 1}}}
\newcommand{\akliis}{{A_{\kappa \Lambda, 2}}}
\newcommand{\akliiis}{{A_{\kappa \Lambda, 3}}}

\newcommand{\bkl}{B_{\kappa \Lambda}}

\newcommand{\bakl}{B_{\kappa \Lambda}^{\mbox{\tiny $(\tau)$}}}

\newcommand{\bzi}{B}

\newcommand{\han}{{1/2}}

\newcommand{\hhh}{{\cal H}}

\newcommand{\N}{H_{\mathrm N}} 
\newcommand{\M}[2]{\langle#1 \phi\, ,\, #2 \psi\rangle_{\cal H}} 
\newcommand{\G}{H_{\k\la}} 
\newcommand{\tG}{{H_{\k\la}^{
\mbox{\tiny $\left(\tau\right)$}}}} 
 
\newcommand{\oneG}{{H_{\k\la}^{\mbox{\tiny $\left( 1\right)$}}}} 
 
\newcommand{\z}[1]{g^{[j,#1]}} 
\newcommand{\w}[1]{f_j^{[#1]}}

\newcommand{\od}{\omega}

\newcommand{\Gi}{H}

\newcommand{\zz}{{H_0}}
\newcommand{\tzz}{{H_{0}^{
\mbox{\tiny $\left(\tau\right)$}}}}

\newcommand{\El}{E_{\k \la}}
\newcommand{\binEl}{E^{\mathrm bin}_{\kappa\Lambda}} 
\newcommand{\tEl}{{E_{\k \la}^{
\mbox{\tiny $\left(\tau\right)$}}}}
\newcommand{\oneEl}{{E_{\k \la}^{
\mbox{\tiny $\left( 1\right)$}}}}
\newcommand{\zeroEl}{{E_{\k \la}^{
\mbox{\tiny $\left( 0\right)$}}}}

\newcommand{\bk}{\beta} 
\newcommand{\k}{\kappa}

\newcommand{\lk}{\left(}

\newcommand{\rk}{\right)}
\newcommand{\lkk}{\left\{}
\newcommand{\rkk}{\right\}}

\newcommand{\hf}{{H_{\mathrm f}}}
\newcommand{\ho}{{H_{0}}} 

\newcommand{\tho}{{H_{0}^{
\mbox{\tiny $\left(\tau\right)$}}}} 
\newcommand{\zeroho}{{H_{0}^{
\mbox{\tiny $\left( 0\right)$}}}}

\newcommand{\hfi}{{H_{{\mathrm f}1}}}
\newcommand{\hfu}{{H_{{\mathrm f}2}}}
\newcommand{\nf}{{N_{\mathrm f}}}
\newcommand{\nfi}{{N_{{\mathrm f}1}}}
\newcommand{\nfu}{{N_{{\mathrm f}2}}}

\newcommand{\la}{\Lambda}

\newcommand{\vp}{{\widehat{\varphi}}}

\newcommand{\vvp}{\varphi} 
\newcommand{\kx}{e^{ik\cdot x}}
\newcommand{\kxx}{e^{-ik\cdot x}}

\newcommand{\eq}[1]{\begin{equation}
\label{#1}}
\newcommand{\en}{\end{equation}}
\newcommand{\bl}[1]{\begin{lemma}
\label{#1}}
\newcommand{\el}{\end{lemma}}

\newcommand{\bt}[1]{\begin{theorem}
\label{#1}}
\newcommand{\et}{\end{theorem}}

\newcommand{\brem}[1]{\begin{remark}
\label{#1}}
\newcommand{\erem}{\end{remark}}

\newcommand{\bi}{\begin{description}}
\newcommand{\ei}{\end{description}}

\newcommand{\bp}[1]{\begin{proposition}
\label{#1}}
\newcommand{\ep}{\end{proposition}}

\newcommand{\bc}[1]{\begin{corollary}
\label{#1}}
\newcommand{\ec}{\end{corollary}}

\newcommand{\kak}[1]{(\ref{#1})}

\newcommand{\gr}{\psi_{\k\la}}
\newcommand{\tgr}{\psi_{\k\la}^{\mbox{\tiny $\left( 
\tau\right)$}}}
\newcommand{\onegr}{\psi_{\k\la}^{\mbox{\tiny $\left( 
1\right)$}}}

\newcommand{\limk}{\lim_{\k\rightarrow 0}}

\newcommand{\liml}{\lim_{\la\rightarrow \infty}}

\newcommand{\e}{\epsilon}
\newcommand{\charge}{{\mathsl e}}

\newcommand{\hi}{H_{\rm I}} 
\newcommand{\thi}{{H_{\rm I}^{\mbox{\tiny $\left(\tau\right)$}}}}

\newcommand{\zerohi}{{H_{\rm I}^{\left( 0\right)}}}

\makeatletter
\@addtoreset{equation}{section}
\makeatother

\begin{document}
\setlength{\baselineskip}{18pt}
\title
{Ground State for Point Particles 
Interacting Through 
a Massless Scalar Bose Field}
\author{
Masao Hirokawa\footnote{Department of Mathematics, 
Okayama University.}, 
Fumio Hiroshima\footnote{Department of 
Mathematics and Physics, 
Setsunan University, 
572-8508, Osaka, Japan.} 
and Herbert Spohn\footnote{Technische Universit\"at 
M\"unchen, Zentrum Mathematik,  
85747, Garching, Germany.} \\ 
\qquad \\ 
{\it Dedicated to} \\ 
{\it Hiroshi Ezawa, Elliott H. Lieb and Edward J. Nelson} \\ 
{\it on the occasion of their $70^{\it th}$ birthdays.}}
\maketitle
\begin{abstract}
We consider a massless 
scalar Bose field interacting with two particles, 
one of them infinitely heavy. 
Neither an infrared nor an ultraviolet 
cutoff is imposed. 
In case the charge of the particles 
is of the same sign and sufficiently small, 
we prove the existence of a ground state.  
\end{abstract}

\section{Introduction}
\label{sec:intro}

In a famous paper \cite{ne} Nelson studies 
the Hamiltonian of $N$ particles interacting 
through a massive scalar Bose field. 
He proves that the ultraviolet cutoff can be removed 
at the expense of an infinite energy renormalization. 
In the sequel some qualitative aspects of the non-cutoff Nelson 
Hamiltonian were investigated. 
Fr\"{o}hlich \cite{fro} studies the energy-momentum 
relation for $N=1$. 
Ammari \cite{am} also assumes $N=1$ and in addition 
supposes that the particle is confined by an external 
potential growing at least quadratically at infinity. 
He proves existence of a ground state and 
asymptotic completeness for the scattering of bosons 
from the bound particle. 
In our work we consider two particles, 
one of them is infinitely heavy and has charge 
$\charge Z$ with $Z > 0$ the atomic number,  
and the other one has charge $\charge$. 
Following \cite{ne} the ultraviolet cutoff 
for the $N = 2$ Hamiltonian is removed. 
We also remove the infrared cutoff by taking the mass 
of the bosons to zero. 
As a byproduct the effective Coulomb interaction between 
the two particles becomes explicit. 
It is attractive for charges of equal sign. 
In other words, we study here a hydrogen-like atom, where 
the ``electron'' and ``nucleus'' interact through 
a scalar Bose field. 
Ultimately one would like to establish that 
the binding energy of the atom is approximately still given 
by Balmer's formula 
$(\charge^{2}Z)^{2}m_{\mathrm eff}/2(4\pi)^{2}$ 
plus small radiative correction of order $\charge^{6}$, 
where $m_{\mathrm eff}$ is the effective mass 
as determined through the energy-momentum relation, 
see \cite{ll} and the discussion in the last section. 
Here we make a first step by proving the existence 
of a ground state for $\charge$ sufficiently small.

The infinitely heavy nucleus is nailed down at the origin. 
The other particle has bare mass $m$. 
Its position is denoted by $x$, 
momentum by $p = -i\nabla_{x}$. 
We set $\hbar=1, c=1$ throughout. 
The two particles are coupled 
through a massless scalar Bose field. 
In momentum representation  
the creation and annihilation operators of the field satisfy 
the standard CCR, 
$$[a(k),a^{*}(k')]=\delta(k-k'),$$
$$[a(k),a(k')]=0,\ \ \ [a^{*}(k),a^{*}(k')]=0,\ \ \ 
k,k'\in\BR.$$
The field energy is given by 
$$H_{\mathrm f} =\int_{\BR} \omega(k) a^{*}(k) a(k) d^{3}k$$
as an operator acting on the symmetric Fock space, 
$\fff$, over $\LR=L^2(\BR,d^3k)$. 
For zero mass bosons the dispersion relation 
is given by  
$$\omega(k)=|k|.$$
In position space the scalar Bose field is 
then defined through 
$$\phi(x) = \int_{\BR}\frac{1}{\sqrt{2\omega(k)}}
\lk \kx a(k) + \kxx a^{*}(k)\rk d^{3}k.$$ 
When smoothened by a real test function 
$\varphi:\BR\rightarrow \RR$, 
the field is denoted by 
$\phi_{\vvp}(x)=\int\vvp(x-y)\phi(y) d^{3}y.$ 
If $\int|\vp(k)|^2/\omega(k)d^{3}k<\infty$,  
$\vp$ the Fourier transform of $\varphi$, 
then $\phi_\vvp(x)$ is a self-adjoint operator 
on $\fff$ for every $x\in\BR$. 
\hfill\break 
$\Diamond$ 
To distinguish we use $e$ for Napier's number, 
$e = 2.718 \cdots$, and the slanted $\charge$ for 
the charge. 
According to the standard conventions 
the fine structure constant is 
$\alpha = \charge^{2}/4\pi.$ 
$\Diamond$ 
\hfill\break  
With these conventions the cutoff Nelson Hamiltonian 
for our system reads 
\eq{h} 
\N=\frac{1}{2m}p^2\otimes 1+1\otimes \hf + 
\charge Z\phi_\vvp(0)+\charge\phi_\vvp(x), 
\en 
acting on 
$$\hhh=\LR\otimes\fff.$$
In contrast to the more familiar Maxwell field, 
under forces transmitted by a scalar field, 
charges of {\it equal}  sign attract. 
Hence $Z > 0$ is assumed.  
For the form factor $\vvp$ we choose specifically  
\eq{S-1.2a}
\vp(k) =\lkk\begin{array}{ll} 
0,&|k|<\k,\\
(2\pi)^{-3/2},&\k\leq |k|\leq \la,\\
0,&|k|>\la,
\end{array}
\right.
\en 
which means $\varphi (x) = \delta (x)$ in the limits 
$\kappa\to 0$, $\Lambda\to\infty$. 

\bp{self}
$\N$ is self-adjoint on the domain $D(p^2)\cap D(\hf)$ 
and bounded from below for arbitrary values of $\charge Z$. 
\ep 
\proof 
The interaction is infinitesimally small relative 
to the free Hamiltonian 
and the claim follows from the Theorem of 
Kato-Rellich. 
\qed

In spirit one would like to have $\vvp(x)=\delta(x)$ in \kak{h}. 
Unfortunately $\N$ is both infrared and ultraviolet divergent: 
As $\varphi(x)\rightarrow \delta(x)$, 
the anticipated ground state does not 
lie in Fock space and the ground state energy 
tends to $-\infty$. 
These divergences are of a mild nature however and 
can be unraveled through the unitary Gross transformation 
$e^{-T}$ which is generated by 
\eq{9/4-1.2a}
T = - \int_{\BR} 
\frac{\charge\widehat{\varphi}(k)}{\sqrt{2\omega(k)}}
\lkk \left( \beta(k) e^{ik\cdot x} 
+ \frac{Z}{\omega(k)}\right) a(k) 
- \left( \beta(k) e^{-ik\cdot x} 
+ \frac{Z}{\omega(k)}\right) \add(k)\rkk d^{3}k
\en
with  
$\bk(k)= \lk \omega(k)+|k|^2/2m \rk\f$. 
As explained by Gross \cite{Gr}, 
the Gross transformation goes back to 
the intermediate coupling approximation by Tomonaga \cite{tomo}, 
who used it in a variational treatment of nuclear forces. 
It was picked up by Lee, Low, and Pines \cite{llp,LP1,LP2} 
and Gross \cite{Gr1} in the context of the ground state energy 
of the polaron problem, for which $\omega(k) = 1$, 
$\vp(k) = 1/|k|$. 
In addition to $e^{-T}$, it is convenient to scale out 
the mass $m$ through the unitary transformation $U_{[m]}$ 
defined by $U_{[m]}^{*}xU_{[m]} = m^{-1}x$, 
$U_{[m]}^{*}pU_{[m]} = mp$, 
and 
$U_{[m]}^{*}a(k)U_{[m]} 
= m^{-3/2}
a\left(m^{-1}k\right)$. 
A more complete discussion of scale changes will be 
given in Section \ref{sec:CS}.

Here and in the following we mostly omit 
the tensor notation $\otimes$. 
We set $U = e^{-T}U_{[m]}$. 
Note that $\bk\in\LR$. 
Then 
$U$ maps $D(p^2)\cap D(\hf)$ onto itself with 
$$U^{*}pU 
= m(p + \charge\akl + \charge\akl^\ast),$$
$$U^{*}xU = 
m^{-1}x,$$
$$U^{*}a(k)U = 
m^{-3/2}a\left( 
m^{-1}k\right) 
+ \frac{\charge}{\sqrt{2\omega(k)}} 
\left( 
\beta(k)e^{-ik\cdot x /m} 
+ \frac{Z}{\omega(k)}
\right),$$ 
where 
\eq{S-1.2b}
\akl 
= \int_{\BR}\frac{\widehat{\varphi}_{0}(k)}{
\sqrt{2\omega(k)}} 
k\beta_{0}(k) a(k)e^{ik\cdot x}d^{3}k 
\en 
with   
$\widehat{\varphi}_{0}(k) = 
\widehat{\varphi}(mk)$ and 
\eq{beta-0}
\beta_{0}(k) = 
\left( 
|k| + \frac{|k|^{2}}{2}
\right)^{-1}. 
\en 
\hfill\break 
Following \cite{ne} one obtains 
$$
m^{-1}\left\{
U^{*} \N U 
+ \int_{\BR} 
\frac{\charge^{2}|\widehat{\varphi}(k)|^{2}}{2\omega(k)}
\lk \bk(k) 
+ \frac{Z^{2}}{\omega(k)}\rk d^{3}k
\right\} 
= \G + \vkl, 
$$ 
where 
\eq{gr} 
\G = \pap + \hf + \charge (p\cdot \akl +\akl^\ast\cdot p) 
+ \frac{\charge^{2}}{2} \lk 
\akl^2 + 2\akl^\ast \cdot \akl 
+ \akl^{\ast2} \rk, 
\en  
and 
\eq{Hatom} 
\pap=\frac{1}{2}p^2 + V_{\mathrm ex}(x)
\en 
with the external Coulomb potential 
\eq{coulomb} 
V_{\mathrm ex}(x) = -\frac{\charge^{2}Z}{4\pi|x|}.  
\en 
The Gross transformation generates 
in addition the correction potential  
$$\vkl= \frac{\charge^{2}Z}{m}
\int_{\BR}\left( \left(2\pi\right)^{-3} 
- |\widehat{\varphi}\left( k\right)|^{2}\right) 
\omega\left( k\right)^{-2}e^{ik\cdot x/m}d^{3}k.$$

In the form \kak{gr} we remove both cutoffs. 
The simple piece is the correction potential $\vkl$. 
It satisfies $|\vkl| \le const |x|^{-1}$ 
and hence is relatively bounded with respect to 
$\pap$. 
We conclude that 
${\displaystyle \lim_{\kappa\to 0, \Lambda\to\infty} 
\vkl = 0}$ 
in the $L^{2}$-norm. 
The convergence of $\G$ is the content of

\bt{new-main1} 
(removal of IR and UV cutoffs). 
There exist a strictly positive constant $\euv$ 
and self-adjoint operators $H_{\kappa\infty}$ 
and $H$ such that for every $\charge$ with 
$|\charge| < \euv$
\bi 
\item[](i)\,\,  
$\G$ converges to $H_{\kappa\infty}$ as 
$\Lambda\to\infty$ in the norm resolvent sense,   
\item[](i)\,\,  
$H_{\kappa\infty}$ converges to $H$ as 
$\kappa\to 0$ in the norm resolvent sense. 
\ei
\et 
\proof 
The claim follows immediately from Proposition \ref{N}. 
\qed

The limit $H$ can be defined only as quadratic form. 
It has the obvious form Hamiltonian 
\eq{9/4-(1)} 
H = \pap + \hf 
+ \charge\left(p\cdot A + A^{*}\cdot p \right) 
+ \frac{\charge^{2}}{2}\left( A^{2} + 2A^{*}A + A^{2}\right), 
\en
where 
\eq{9/4-(2)} 
A = (2\pi)^{-3/2}\int_{\BR}\frac{1}{\sqrt{2\omega(k)}}
k\beta_{0}(k)a(k)e^{ik\cdot x}d^{3}k.  
\en
\hfill\break 
{\it Remark}. 
Note that at $\kappa = 0$ the transformation 
$e^{-T}$ from \kak{9/4-1.2a} is no longer unitarily implemented 
and the Fock representation in \kak{9/4-(1)} is disjoint from 
the one of \kak{h}. 
This point has been recently studied by Arai \cite{ar} 
and is also reflected in the structure of the corresponding 
functional measures \cite{lms}.  
\hfill\break

As our main result we establish $H$ 
to have a ground state in case $|\charge|Z > 0$ 
and $|\charge|$ sufficiently small. 

\bt{new-main2}
(existence of a ground state). 
There exists a strictly positive 
constant $\eir$ 
such that for charge $\charge$ with 
$0 < |\charge| < \min\left\{ \eir , 
\euv\right\}$  
the Hamiltonian $H$ in \kak{9/4-(1)} has 
a ground state $\psi_{\mathrm g}$. 
Let  
\eq{number-op} 
\nf = \int_{\BR}a^{*}(k)a(k)d^{3}k
\en
be the number operator for the bosons. 
Then $\psi_{\mathrm g} \in D(\nf^{1/2})$. 
\et

\hfill\break 
{\it Remark}. 
The smallness condition $|\charge| < \euv$ is needed to have 
$H$ of Theorem \ref{new-main1} well-defined. 
It is an ultraviolet condition. 
The smallness condition $|\charge| < \eir$ comes from 
estimating the overlap of an approximating sequence 
of ground states with the Fock vacuum, 
which is limited through the infrared behavior. 
Whether the restriction on $\charge$ is artifact of 
the proof is not understood 
at present. 
It is conceivable that the ground state is lost 
at strong coupling. 
Clearly, $H$ has no ground state for $|\charge|Z \ge 0$. 
\hfill\break 

The constant $\euv$ is fairly explicit and given by 
the unique positive solution of 
\eq{restriction1}
C_{\mbox{\tiny UV}}(\euv) = 1,
\en
where $C_{\mbox{\tiny UV}}$ is given by 
\eq{eq:C*} 
C_{\mbox{\tiny UV}}(\charge) =  
\frac{2\charge}{\pi}
\sqrt{1 + \frac{1}{2}\left( 
\frac{\charge^{2}Z}{4\pi}\right)^{2}} 
+ 
\frac{1}{4\pi^{2}}
\left( 14 + \sqrt{6}\,\,\pi\right) 
\charge^{2}.
\en 
Note that $\euv$ depends on $Z$, 
$\euv \to c_{1} > 0$ for $Z\to 0$ and 
$\euv \cong c_{2}Z^{-1/3}$ for $Z\to\infty$ 
with certain positive constants $c_{1}$ and 
$c_{2}$.
$\eir$ is slightly more indirectly defined and 
discussed in Section \ref{sec:Proof-Lemma-2}.

The ground state issue for models of type \kak{gr} 
has been studied extensively in the past years. 
To put our result in perspective, we recall that the relevant 
parameters are $\kappa, \Lambda$, and the behavior of 
the external potential at infinity, 
$V_{\mathrm ex}(x) = {\mathcal O}(|x|^{\gamma})$ as $|x|\to\infty$. 
Here we study the case $\kappa = 0$, $\Lambda = \infty$, 
and $\gamma = - 1$, since, by the definition of the Nelson 
Hamiltonian, $V_{\mathrm ex}$ has to be the Coulomb potential. 
Ammari \cite{am} considers $\kappa > 0$, $\Lambda = \infty$, 
$\gamma \ge 2$. 
Arai \cite{ar} and L\H{o}rinczi et al. \cite{lms} 
allow for $\kappa=0$, $\Lambda < \infty$, $\gamma \ge 2$.
Arai has no restriction on the magnitude of $\charge$, 
since in his case the resolvent of $\pap$ is compact. 
Bach {\it et al.} \cite{bfs2} have 
$\kappa=0$, $\Lambda < \infty$, $\gamma \ge 0$. 
In their work there is a restriction on the size 
of $|\charge|$. 
This condition is removed in the beautiful work of 
Griesemer, Lieb, and Loss \cite{gll} who require 
$\kappa=0$, $\Lambda < \infty$, $\gamma \le 0$. 
They also provide an extensive bibliography on 
earlier work which is mostly concerned with $\kappa > 0$, 
$\Lambda < \infty$. 
 
Our basic strategy to prove Theorem \ref{new-main2} has been 
used before. 
$\G$ has a unique ground state $\gr$. 
If $\gr$ converges to a non-zero limit vector $\psi_{\mathrm g}$ 
as $\kappa\to 0$, $\Lambda\to\infty$, then $\psi_{\mathrm g}$ 
is a ground state of $H$. 
To make sure that $\psi_{\mathrm g} \ne 0$, 
one estimates its overlap with $\gsat\otimes\Omega$, 
$\gsat$ the ground state of $\pap$ and $\Omega$ 
the Fock vacuum, from which one immediate difficulty 
becomes apparent: 
As $\charge\to 0$, $\gsat$ delocalizes and the overlap is not 
so easily controlled. 
In this context, we emphasize that, in contrast to all 
previous work, the external potential $V_{\mathrm ex}$ 
is fixed by the theory and not at our disposal. 
In particular for $\charge\to 0$, both the coupling to the field 
and the strength of $V_{\mathrm ex}$ vanish. 
The overlap is controlled by a bound on the average photon 
number $\langle \gr\, ,\, \nf\gr\rangle_{\mathcal H}$ 
which in essence  relies on the spatial localization in 
the form $\langle \gr\, ,\, f(x)\gr\rangle_{\mathcal H}$ 
with suitable $f : {\Bbb R}^{3}\to {\Bbb R}$. 
For the choice $f(x) = |x|$, the overlap estimate 
becomes poor as $\charge\to 0$. 
To improve we can only allow for a much slower increase of $f$, 
as $f(x) = \log(1 + |x|)$, thereby shifting the problem 
to the soft photon bound, i.e., to a bound on 
$\int_{|k|\le 1}\langle\gr\, ,\, a^{*}(k)a(k)\gr\rangle_{\mathcal}
d^{3}k$, for which we develop a novel and 
rather powerful iteration scheme. 
In comparison, the ultraviolet cutoff causes less difficulties, 
except for the fact that $H$ is defined 
only as quadratic form, which means that the resolvent 
of $\G$ cannot be controlled through norm estimates. 
This difficulty ``propagates'' to the infrared regime, 
where standard estimation techniques can no longer be used. 

To give a brief summary: The removal of cutoffs 
and the existence of $H$ are studied in Section \ref{sec:ELH}. 
Section \ref{sec:GE-BE} deals with the binding energy 
and the existence of the cutoff ground state $\gr$, 
for which spatial localization is established in Section \ref{sec:LPS}. 
The average photon number bound is divided into 
$\left\{ |k| \ge 1\right\}$, hard photons, and 
$\left\{ |k| < 1\right\}$, soft photons (Sections \ref{sec:HPNB} 
and \ref{sec:SPNB}). 
Both estimates together yield a bound on the overlap 
with the decoupled ground state for $0 < |\charge| < \min\left\{ 
\eir, \euv\right\}$, Section \ref{sec:Proof-Lemma-2}, 
from which the main theorem then easily follows. 
In Section \ref{sec:CS} we collect a few properties for 
scale changes which will be useful throughout. 
We conclude with some open problems in 
Section \ref{sec:outlook}.

\section{Change of scale}
\label{sec:CS} 

The Hamiltonian $H$ and its cutoff $\G$ are written 
in relativistic units. 
Since the coupling $\charge$ enters into both the 
strength of the Coulomb potential and the particle-field 
interaction, for small $\charge$ the ground state is spatially 
delocalized and it seems more natural to transform $H$ 
to atomic units, which will be implemented through 
a change of spatial scale by the factor 
\eq{10-2-1}
r(\tau) = \rc^{\tau}, \qquad 
\tau \in {\Bbb R}, \rc > 0. 
\en 
The free parameter $\rc$ will be used to optimize 
our bounds. 
It will be convenient to set 
\eq{10-2-2}
\rc = \alpha Z\lambda_{1}
\en 
with a constant $\lambda_{1}$ and the fine structure constant. 

\hfill\break
{\it Remark}.  
Relativistic units correspond to $\tau = 0$. 
On the other hand, the standard atomic units are 
$r = \alpha$, i.e., $\tau = 1$ and $\lambda_{1} = 1/Z$. 
\hfill\break 

For $\Psi = \left\{ \Psi^{(0)}, \Psi^{(1)}, \Psi^{(2)}, 
\cdots, \Psi^{(n)}, \cdots, \right\} \in {\cal H}$, 
we set 
\eq{eq:units}
(U_{\tau}\Psi)^{(n)}(x , k_{1}, \cdots, k_{n}) 
= \rc^{3\tau/2}\rc^{- 3n\tau}
\Psi^{(n)}
\biggl( \rc^{\tau}x, \rc^{-2\tau}k_{1}, \cdots, 
\rc^{-2\tau}k_{n}\biggr).
\en
Then, $U_{\tau_{1}}U_{\tau_{2}} = U_{\tau_{1}+\tau_{2}}$, 
$U_{0} = I$, 
and $x, p, a(k)$ transform as 
\eq{trans-x-p} 
U_{\tau}^{*}xU_{\tau} 
= \rc^{-\tau}x, \,\,\, U_{\tau}^{*}pU_{\tau} 
= \rc^{\tau}p,
\en
\eq{trans-a}
U_{\tau}^{*}a(k)U_{\tau} 
= \rc^{-3\tau}\,\, 
a(\rc^{-2\tau}k). 
\en
As a consequence $\takl$ transforms to 
\eq{tauA}
\takl 
= \rc^{-2\tau}U_{\tau}^{*}\akl U_{1,\tau}
\en 
with 
\eq{S-1.2b-t}
\takl 
= \int_{\BR}\frac{\widehat{\varphi}_{\tau}(k)}{
\sqrt{2|k|}} 
k\beta_{\tau}(k) a(k)e^{ir(\tau) k\cdot x}d^{3}k 
\en 
and    
\eq{beta-e}
\widehat{\varphi}_{\tau}(k) = 
\widehat{\varphi}\biggl( m\rc^{2\tau}k\biggr),\,\,\, 
\beta_{\tau}(k) = 
\left( 
|k| + \rc^{2\tau}|k|^{2}/2
\right)^{-1}.   
\en 
Note that $U_{\tau}^{*}\nf U_{\tau} = 
\nf$ as it should be.

Let us define the Hamiltonians $\tG$ through 
\eq{tG} 
\tG 
= \rc^{-2\tau}U_{\tau}^{*}\G U_{\tau}.
\en
Then we have 
\eq{tG'} 
\tG = \tpap + \hf + \thi,
\en
where the atomic part is given by  
\eq{tpap}
\tpap 
= \frac{1}{2}p^{2} 
-\,\, \frac{\alpha Zr(-\tau)}{|x|}
= \frac{1}{2}p^{2} 
-\,\, \frac{r(1-\tau)}{\lambda_{1} |x|} 
\en 
with $\lambda_{1}$ of \kak{10-2-2}, 
and the interaction part by 
\eq{thi} 
\thi  =  
\charge\rc^{\tau}
(p\cdot \takl +\takl^\ast\cdot p) 
+ \frac{\charge^{2}\rc^{2\tau}}{2} \lk 
\takl^2 + 2\takl^\ast \cdot \takl 
+ \takl^{\ast2} \rk.
\en
For $\tG$ we set 
\eq{difference-gse'}
\tEl = \inf\sigma\left( 
\tG\right) 
= \rc^{-2\tau}\El. 
\en

$\tpap$ has the ground state energy 
\eq{gse-atom-tau} 
\teppp = \inf\sigma\left(\tpap\right) 
= -\,\, \frac{(\alpha Z)^{2}\rc^{-2\tau}}{2}
= -\,\, \frac{r(2-2\tau)}{2\lambda_{1}^{2}} 
\en 
and the normalized ground state 
\eq{tgsat} 
\tgsat (x) = 
\pi^{-1/2}(\alpha Z)^{3/2}\rc^{-3\tau/2}
e^{- \alpha Zr(-\tau)|x|}. 
\en
We set 
\eq{9/4-(null)}
\tho = 
\tpap - \teppp +\hf 
\en 
and for $\tau = 0$, 
$$\ho = \zeroho,\,\,\, 
\hi = \zerohi,\,\,\, 
\El = \zeroEl;\,\,\, 
\pap = \zeropap,\,\,\, 
\eppp = \zeroeppp,\,\,\, 
\gsat = \zerogsat.$$

\section{Existence of the limit Hamiltonian} 
\label{sec:ELH}

The free Hamiltonian $\tho$ defines the quadratic form 
\eq{9/5-(null)}
B_{0}^{\mbox{\tiny $(\tau)$}}(\phi,\psi) 
= \langle \tho^{1/2}\phi\, ,\, 
\tho^{1/2}\psi\rangle_{\cal H}
\en 
for all $\phi, \psi \in D(\tho^{1/2})$. 

\bl{new-forms} 
The quadratic form 
\begin{eqnarray*} 
\bakl(\phi\, ,\,\psi) 
&=&  
\charge\rc^{\tau}\M{p}{\takl} 
+ \charge\rc^{\tau}\M{\takl}{p} \\ 
&{}& 
+ 
\frac{\charge^{2}\rc^{2\tau}}{2} 
\left(\M{}{\takl^2} 
+ 2\M{\takl}{\takl}
+ \M{\takl^2}{}
\right)
\end{eqnarray*} 
is well defined on 
$D(\tzz^\han)\times D(\tzz^\han)$ 
for all $\k\geq 0$ and  $\la\leq \infty$. 
Moreover the bound  
\eq{rel}
|\bakl(\phi,\phi)| 
\leq 
C_{*}(\charge, \tau)\|(\tzz+1)^\han \phi\|^2
\en 
holds, where $C_{*}(\charge, \tau)$ is defined 
by 
\begin{eqnarray} 
\nonumber  
&{}& 
C_{*}(\charge, \tau) \equiv C_{*}(\alpha, \tau) \\ 
&{}& =  
\sqrt{6}\,\, \alpha \rc^{-\tau} 
+ 4\alpha^{1/2}\sqrt{(1-\teppp)/\pi} 
+ \frac{2}{\pi}\alpha 
\biggl( \rc^{2\tau} + 3\rc^{\tau} + 3\biggr). 
\label{c-*-tau} 
\end{eqnarray}  
\el

\proof 
As easy facts, for every $\psi \in D(\tho^{1/2})$ 
one has  
$$|\langle p\psi , \takl\psi\rangle_{\cal H}| 
\le \|p(\tho + 1)^{-1/2}\|\,\, 
\|\takl(\tho + 1)^{-1/2}\|\,\, 
\| (\tho + 1)^{1/2}\psi\|^{2}.$$ 
We set 
$C^{(\tau)} = 1 - \teppp$.
Then 
\begin{eqnarray*} 
&{}& 
2C^{(\tau)} 
\left( \tpap - \teppp + 1\right) - p^{2} \\ 
&{}& =   
2\left( C^{(\tau)} - 1\right) 
\left[ 
 \left\{\frac{1}{2}p^{2} 
         - 
        \left(\frac{C^{(\tau)}}{
                          C^{(\tau)} - 1}
        \right)
        \frac{r(1-\tau)}{\lambda_{1} |x|}\right\} 
+  
\frac{C^{(\tau)}}{C^{(\tau)} - 1} 
\left( - \teppp + 1\right)
\right] \\ 
&{}& \ge   
2\left( C^{(\tau)} - 1\right) 
\left[ 
  \left( \frac{C^{(\tau)}}{
                       C^{(\tau)} - 1}
    \right)^{2} \teppp
+  
\frac{C^{(\tau)}}{C^{(\tau)} - 1} 
\left( -\teppp  + 1\right)
\right] \\ 
&{}& = 
\frac{2C^{(\tau)}}{C^{(\tau)} - 1}
\left(\teppp + C^{(\tau)} - 1
\right) 
= 0,
\end{eqnarray*}  
which implies that 
\eq{ineq-p} 
\| p (\tho + 1)^{-1/2}\| 
\le \sqrt{2(1 - \teppp)}.
\en 
On the other hand, by sandwiching 
$(\hf +1)^{-1/2}(\hf + 1)^{1/2}$ in 
between $\takl (\tho + 1)^{-1/2}$ 
and applying Lemma \ref{ine} (i) 
of Appendix A, one has 
\begin{eqnarray} 
\nonumber 
&{}& \|\takl (\tho + 1)^{-1/2}\| 
= \|(\tho + 1)^{-1/2}\takl^{*}\| \\ 
\nonumber 
&{}&\le  \|\takl (\hf +1)^{-1/2}\| 
= \|(\hf +1)^{-1/2}\takl^{*}\|  \\ 
&{}&\le 
\frac{1}{\sqrt{2}(2\pi)^{3/2}}
\Biggl\{ \int_{\BR} 
\left( 
|k| + \frac{\rc^{2\tau}}{2}|k|^{2}
\right)^{-2}d^{3}k
\Biggr\}^{1/2} 
\le \frac{1}{\sqrt{2}\pi}\rc^{-\tau}. 
\label{maru0}
\end{eqnarray} 
Thus  
\eq{maru1} 
|\langle p\psi , \takl\psi\rangle_{\cal H}| 
\le 
\frac{\rc^{-\tau}}{\pi}
\sqrt{1 - \teppp}   
\| (\tho + 1)^{1/2}\psi\|_{\cal H}^{2}.
\en 
It is easy to show that 
\eq{maru2}
|\langle \takl\psi , p\psi\rangle_{\cal H}| 
= 
|\langle p\psi , \takl\psi\rangle_{\cal H}|.
\en

By putting $(\tho + 1)^{-1/2}(\tho + 1)^{1/2}$ 
in front of both $\psi$ and using $\hf \le \tho$, 
it follows that  
$$
|\langle \psi , \takl^{2}\psi\rangle_{\cal H}| 
\le \|(\hf + 1)^{-1/2}
\takl^{2}(\hf + 1)^{-1/2}\|\,\, 
\| (\tho + 1)^{1/2}\psi\|^{2}. 
$$ 
On the other hand, by Lemma \ref{ine} (ii) 
of Appendix A, one has 
\eq{maru4}
\|(\hf + 1)^{-1/2}
\takl^{2}(\hf + 1)^{-1/2}\| 
\le 
\Xi(f^{\tai},f^{\tai}), 
\en 
where 
\eq{fA} 
f^{\tai}(k) 
= 
\frac{\widehat{\varphi}_{\tau}(k)}{\sqrt{2|k|}} 
k\beta_{\tau}(k)e^{ir(2\tau)k\cdot x}. 
\en
$f^{\tai}$ can be estimated as 
\eq{maru5} 
\| f^{\tai}_{\mbox{\tiny IR}}\|_{L^{2}} 
\le  
\frac{1}{\sqrt{2}(2\pi)^{3/2}}
\Biggl\{ 
\int_{|k| < 1} 
|k|\left( |k| + \frac{\rc^{2\tau}}{2}|k|^{2}\right)^{-2}
d^{3}k 
\Biggr\}^{1/2} 
< \frac{1}{2\pi}. 
\en
In the same way,  
\eq{maru6} 
\| \omega^{-1/2}f^{\tai}_{\mbox{\tiny IR}}\|_{L^{2}} 
\le  
\frac{1}{\sqrt{2}(2\pi)^{3/2}}
\Biggl\{ 
\int_{|k| < 1} 
\left( |k| + \frac{\rc^{2\tau}}{2}|k|^{2}\right)^{-2}
d^{3}k 
\Biggr\}^{1/2} 
< \frac{1}{2\pi}. 
\en 
Also,  
\begin{eqnarray} 
\| \omega^{-1/2}f^{\tai}_{\mbox{\tiny UV}}\|_{L^{2}} 
\le 
\frac{1}{\sqrt{2}(2\pi)^{3/2}}
\Biggl\{ 
\int_{|k| \ge 1} 
\left( |k| + \frac{\rc^{2\tau}}{2}|k|^{2}\right)^{-2}
d^{3}k 
\Biggr\}^{1/2} 
\le \frac{\rc^{-\tau}}{\sqrt{2}\pi}. 
\label{maru6'}
\end{eqnarray}   
Moreover, one obtains 
\begin{eqnarray}
\nonumber 
\| \omega^{-1/4}f^{\tai}_{\mbox{\tiny UV}}\|_{L^{2}} 
&\le&  
\frac{1}{\sqrt{2}(2\pi)^{3/2}}
\Biggl\{ 
\int_{|k| \ge 1} 
|k|^{1/2}
\left( |k| + \frac{\rc^{2\tau}}{2}|k|^{2}\right)^{-2}
d^{3}k 
\Biggr\}^{1/2} \\ 
&\le&  
\rc^{-3\tau/2} 
\left( 
\frac{1}{2\sqrt{2}\pi} 
+ \frac{\rc^{\tau}}{2\pi^{2}}
\right)^{1/2}.
\label{maru7} 
\end{eqnarray}

From \kak{maru4}-\kak{maru7} one concludes that  
\eq{3-12-1} 
\| (\tho + 1)^{-1/2}\takl^{2}(\tho + 1)^{-1/2}\| 
\le 
\frac{1}{2\pi^{2}} + \frac{\sqrt{2}}{\pi^{2}}
\rc^{-\tau} 
+ \frac{\sqrt{3}}{2\pi^{2}}\rc^{-2\tau} 
+ \frac{\sqrt{3}}{2\sqrt{2}\pi}\rc^{-3\tau},  
\en  
which implies  
\eq{maru9}
|\langle \psi , \takl^{2}\psi\rangle_{\cal H}| 
\le 
\Biggl( 
\frac{1}{2\pi^{2}} + \frac{\sqrt{2}}{\pi^{2}}
\rc^{-\tau} 
+ \frac{\sqrt{3}}{2\pi^{2}}\rc^{-2\tau} 
+ \frac{\sqrt{3}}{2\sqrt{2}\pi}\rc^{-3\tau}  
\Biggr)
\| (\tho + 1)^{1/2}\psi\|_{\cal H}^{2}.
\en

By \kak{maru0}, one has 
\eq{maru11}
|\langle \takl\psi , \takl\psi\rangle_{\cal H}| 
\le 
\| \takl (\tho + 1)^{-1/2}\|^{2} 
\|(\tho + 1)^{1/2}\psi\|_{\cal H}^{2}
\le 
\frac{\rc^{-2\tau}}{2\pi^{2}}
\|(\tho + 1)^{1/2}\psi\|_{\cal H}^{2}. 
\en 
Lemma \ref{new-forms} follows from 
\kak{maru1}, \kak{maru2}, \kak{maru9}-\kak{maru11}.    
\qed

\bl{hirohiro3}
Let $\charge$ and $\tau$ be such that 
$C_{*}(\charge, \tau) < 1$.
Then, 
for arbitrary $\kappa, \Lambda, \e$ with 
$0 < \e$ and $0 < \kappa < \Lambda < \infty$   
\eq{i1} 
\|(\tho + 1)^\han (\tG - \tEl + \e)^{-\han}\| 
\leq 
\frac{\displaystyle 
C_{1}(\charge, \tau)}{\displaystyle  
\min\left\{\sqrt{\e},\,\,\, 1\right\}}, 
\en 
where $C_{1}(\charge, \tau)$ is defined by 
\eq{C1}
C_{1}(\charge, \tau) = 
\biggl(1 - C_{*}(\charge, \tau)\biggr)^{-1/2}.    
\en 
\el 
\proof 
Using \kak{difference-gse'} and \kak{rel}, 
we have 
$\ho \le C_{1}\left(\charge, \tau\right)^{2}$ 
$\times 
\left(\tG - \tEl 
+ C_{*}\left(\charge, \tau\right)\right)$.   
Since we assume 
$C_{*}\left(\charge, \tau\right) < 1$, 
we have  
$\tho + 1 \le 
C_{1}\left(\charge, \tau\right)^{2}
(\tG - \tEl + C_{*}(\charge, \tau))$ 
and    
$$
\|(\tG - \tEl + 1)^{\han}(\tG - \tEl + \e)^{-\han}\| 
\le 
\left\{ 
  \begin{array}{@{\,}ll}
\e^{-1/2}
& \mbox{if $\e < 1$,} \\  
\qquad & \mbox{\qquad} \\ 
1
& \mbox{if $\e \ge 1$} 
\end{array}
\right.   
$$
which is the assertion \kak{i1}. 
\qed

In the case $\tau = 0$ let  
\eq{S-1.7}
\bkl(\phi,\psi) = B_{0}^{\mbox{\tiny $(0)$}}(\phi,\psi) 
+ B_{\kappa\Lambda}^{\mbox{\tiny $(0)$}}(\phi,\psi) 
= \langle\phi , \G \psi\rangle_{\cal H}.
\en 
Since $C_{\mbox{\tiny UV}}(\charge) = 
C_{*}(\charge, 0)$,
by \kak{rel} and the KLMN theorem 
(cf. \cite[Theorem X. 17]{rs2}), 
the second equality in \kak{S-1.7} 
holds on the form domain 
$Q(\G)=  D(\zz^\han)$, 
provided that $|\charge| < \euv$.

Nelson \cite{ne} proves the following    
\bp{N}
Suppose $|\charge| < \euv$. 
Then  
\eq{nelson}
\liml \bkl(\psi,\psi)=B_{\k\infty}(\psi,\psi),\ \ \ \k>0,
\en 
and 
\eq{nelson2}
\limk B_{\k\infty}(\psi,\psi)=\bzi(\psi,\psi)
\en 
uniformly on any set of $\psi$ in $D(\zz^\han)$ for which 
$\|\zz^\han \psi\|+\|\psi\|$ is bounded. 
\ep 
In fact Nelson proves only \kak{nelson}. 
However, the claim \kak{nelson2} follows directly 
from Lemma \ref{new-forms}. 

An important consequence of 
\kak{nelson} and \kak{nelson2} 
is the validity of Theorem \ref{new-main1}. 
It follows from 
the KLMN theorem and \cite[VIII.25]{rs1}, 
see also \cite[Theorem A.1]{am}. 

\hfill\break 
{\it Remark}. 
Clearly, by unitary equivalence, 
$\tG$ is well defined for all $\tau$. 
However, if, for example, we had 
applied Lemma \ref{new-forms} directly to $\oneG$ 
with $\lambda_{1} = 1/Z$, i.e., for standard atomic units, 
we would obtain a lower bound as $C_{*}(\charge, 1) 
\ge \sqrt{6} > 1$, 
and thus too large for the KLMN theorem. 
\hfill\break

\section{Ground state energy and binding energy}
\label{sec:GE-BE}

We set 
\eq{gse}
\El = \inf\sigma(\G). 
\en 
Then 
\bl{s-hiro}
For every $\charge$ with $|\charge| < \euv$ and 
arbitrary $\kappa, \Lambda$ with 
$0 < \kappa < \Lambda < \infty$  
\eq{s-hi} 
\eppp - 1 < \eppp - C_{\mbox{\tiny UV}}(\charge) \le 
 \El\leq \eppp.
\en 
\el
\proof 
The upper bound is variational with trial 
function $\psi_{\mathrm at}(x)\otimes \Omega$, 
where $\psi_{\mathrm at}(x)$ is the ground state 
for $\pap$ and $\Omega$ is the Fock vacuum. 
For the lower bound, since $- H_{I} \le 
C_{\mbox{\tiny UV}}(\charge)H_{0} 
+ C_{\mbox{\tiny UV}}(\charge)$ for 
$|\charge| < \euv$ by \kak{rel},  
\eq{ss2}
\ho
\leq \frac{1}{1-C_{\mbox{\tiny UV}}(\charge)}
(\G - \eppp) 
+ \frac{C_{\mbox{\tiny UV}}(\charge)}{1-C_{\mbox{\tiny UV}}(\charge)}. 
\en 
Since $H_{0} \ge 0$, it follows that 
\eq{ss3}
\El \ge 
\eppp - C_{\mbox{\tiny UV}}(\charge)
\ge \eppp - C_{\mbox{\tiny UV}}(\euv)  
> \eppp - 1  
\en 
using $C_{\mbox{\tiny UV}}(\charge) < 1$ 
for $|\charge| < \euv$. 
\qed

The bounds \kak{s-hi}, together with 
Theorem \ref{new-main1} yield  

\bl{convgrs} We set $E_{\kappa\infty} := 
\inf\sigma\left( H_{\kappa\infty}\right)$ 
and $E_{\mathrm g} := \inf\sigma\left( \Gi\right)$. 
Then, for every $\charge$ with $|\charge| < \euv$  
\begin{eqnarray} 
\lim_{\Lambda\to\infty}E_{\kappa\Lambda} 
= E_{\kappa\infty},\,\,\, 
\lim_{\kappa\to 0}E_{\kappa\infty} 
= E_{\mathrm g}.  
\label{convgse1} 
\end{eqnarray}
\el

\bp{masa3} 
For every $\charge$ with $0 < |\charge| < \euv$ and 
arbitrary $\kappa, \Lambda$ with 
$0 < \kappa < \Lambda < \infty$, 
$\G$ has a unique ground state 
$\gr \in  D(x^{2}) \cap D(\hf)$. 
\ep 

\proof 
The results in \cite{bfs2,sp} are applicable to $\G$ 
and establish existence. 
In fact, the results of \cite{bfs2,sp} 
should be applied to $\oneG$ defined in \kak{tG} 
which equals $\G$ after a change of units such that 
$\charge^{2}$ is removed from the external potential. 
By unitary equivalence, this implies 
the existence of the ground state 
for $\G$.  
Using the technique of \cite{gll}, presumably, 
the result can be extended to all $\charge$, $\charge \ne 0$. 
In function space the semigroup $e^{-t\N}$, $t > 0$, 
is positivity improving \cite{bfs1}, 
which implies uniqueness. 
Finally, it follows from Proposition \ref{1stmomentum} (ii) 
that $\gr \in D(x^{2})$. 
\qed

Let $\El^{V=0} = \inf\sigma 
\left( \G^{V=0}\right)$ where the superscript means 
that in \kak{gr} the external potential $V_{\mathrm ex}$ 
is omitted. 
The (positive) binding energy is defined by 
\eq{def-binding-energy} 
\binEl 
= \El^{V=0} - \El, 
\en 
which is the difference in energy for 
the electron at infinity and in its 
ground state. 
Following the proof of Theorem 3.1 in \cite{gll}, 
we show that $\binEl > 0$. 
However, a slight modification is needed, 
since $\G$ is normal ordered. 
\bp{binding-energy} 
(strict positivity of binding energy). 
For every $\charge$ with $0 < |\charge| < \euv$, 
$$\binEl \ge - \eppp.$$ 
\ep

\proof 
Let $\gsat$ be the ground state of $\pap$,  
$\gsat > 0$ and $\|\gsat\|_{L^{2}} = 1$. 
For every $\e > 0$ there exists a vector 
$F \in {\cal H}$ such that 
$\langle F\, ,\, \G^{V=0}F\rangle_{\cal H} 
< \El^{V=0} + \e$. 
Then  
\begin{eqnarray} 
\nonumber 
&{}& 
\langle \gsat F \, ,\, 
\left\{ \G - \left( \El^{V=0} + \e + \eppp 
\right)\right\}\gsat F\rangle_{\cal H} \\ 
\nonumber 
&{}&= 
\langle  \gsat F\, ,\, 
\gsat\left\{ \G^{V=0} - \left( \El^{V=0} + \e\right)
\right\}F\rangle_{\cal H} 
+ 
\langle  \gsat F\, ,\, 
\left( \pap\gsat - \eppp\gsat \right)F
\rangle_{\cal H}  \\ 
&{}&\quad   
+ 
\langle  \gsat F\, ,\, 
(p\gsat)\cdot \left\{ \left( 
p + \charge\akl + \charge\akl^{*}
\right)F\right\}
\rangle_{\cal H}.
\label{binding-1}
\end{eqnarray} 
The term 
\begin{eqnarray*} 
&{}&  
\langle\gsat F\, ,\, (p\gsat)\cdot 
\left\{\left( 
\charge\akl + \charge\akl^{*}\right)F\right\}
\rangle_{\cal H} \\ 
&{}& 
= - i\charge \int_{\BR} 
\gsat(x)(\nabla\gsat)(x) 
\langle F\, ,\, \left(\akl + \akl^{*}\right)F
\rangle_{\cal F}(x) 
d^{3}x = 0, 
\end{eqnarray*}
since it is purely imaginary 
and all other terms of \kak{binding-1} are real. 
Hence 
\begin{eqnarray} 
\nonumber 
&{}& 
\langle \gsat F\, ,\, 
\left\{ \G - \left( \El^{V=0} + \e + \eppp 
\right)\right\}\gsat F\rangle_{\cal H} \\ 
\nonumber 
&{}&=   
\int_{\BR} 
\left\{ 
\langle F\, ,\, \G^{V=0}F\rangle_{{\cal F}}(x) 
- \left(\El^{V=0} + \e\right)
\| F\|^{2}_{{\cal F}}(x)
\right\}\gsat(x)^{2} d^{3}x \\ 
&{}&\quad 
+ 
\int_{\BR} 
\langle F\, ,\, pF\rangle_{{\cal F}}(x)
\gsat(x)(p\gsat)(x)d^{3}x.
\label{binding-2}
\end{eqnarray} 
By the translation invariance of 
$\G^{V=0}$, for arbitrary $y \in  \BR$ 
there exists a translated vector $F_{y} 
\in {\cal H}$ so that 
$\langle F , \G^{V=0}F\rangle_{\cal F}(x) 
\to 
\langle F_{y} , \G^{V=0}F_{y}\rangle_{\cal F}(x) 
= \langle F , \G^{V=0}F\rangle_{\cal F}(x+y)$ 
and $\| F\|^{2}_{\cal F}(x) 
\to 
\| F_{y}\|^{2}_{\cal F}(x) 
= \| F\|^{2}_{\cal F}(x+y)$. 
Set 
\begin{eqnarray*} 
\Omega_{y} 
&=& 
\langle\gsat F_{y}\, ,\, 
\left\{ \G - \left(\El^{V=0} + \e + \eppp\right)
\right\}\gsat F_{y}\rangle_{\cal H} \\  
&=&  
\int_{\BR} 
\left\{ 
\langle F\, ,\, \G^{V=0}F\rangle_{\cal F}(x) 
- \left(\El^{V=0} + \e\right)
\| F\|^{2}_{\cal F}(x)
\right\}\gsat(x-y)^{2} d^{3}x \\ 
&{}& \quad 
+ 
\int_{\BR} 
\langle F\, ,\, pF\rangle_{\cal F}(x)
\gsat(x-y)(p\gsat)(x-y)d^{3}x,
\end{eqnarray*}  
where we used \kak{binding-2}. 
Then we have 
$$\int_{\BR}\Omega_{y}d^{3}y 
= 
\int_{\BR}\int_{\BR} 
\langle F\, ,\, 
\left\{ 
\G^{V=0} - \left(\El^{V=0} + \e\right)
\right\}
F\rangle_{\cal F}(x) 
\gsat(x-y)^{2} d^{3}x d^{3}y, 
$$ 
since 
${\displaystyle \int_{\BR}\gsat(x-y)(p\gsat)(x-y) 
d^{3}y = 0}$,  
and  
$$\int_{\BR}\Omega_{y}d^{3}y 
= 
\langle F\, ,\, \left\{ 
\G^{V=0} - 
\left( \El^{V=0} +\e\right)
\right\}F\rangle_{\cal H} < 0,$$ 
which implies that there exists $y_{0} \in \BR$ 
such that $\Omega_{y_{{}_{0}}} < 0$. 
We conclude that 
$\El^{V=0} - \El \ge - \eppp - \e$. 
Since $\e$ is arbitrary, taking $\e\searrow 0$, 
we obtain $\El^{V=0} - \El \ge - \eppp$ 
for all $0 < \kappa < \Lambda < \infty$. 
\qed

We set  
\begin{eqnarray}
\tEl^{V=0} = \inf\sigma\left( 
\tG^{V=0}\right) 
= \rc^{-2\tau}\El^{V=0}, 
\label{difference-gse} 
\end{eqnarray} 
where 
$$\tG^{V=0} 
= \rc^{-2\tau}U_{\tau}^{*}\G^{V=0}U_{\tau}.$$ 
Then, it follows from Lemmas \ref{s-hiro} and 
\ref{binding-energy}  
that 
\eq{s-hiro'} 
\teppp - \rc^{-2\tau} < \tEl \le \teppp 
\en
and 
\eq{binding-energy-t} 
\inf_{0<\kappa < \Lambda < \infty}
(\tEl^{V=0} - \tEl) \ge - \teppp 
\en
for $|\charge| < \euv$.

\section{Localization in position space}
\label{sec:LPS}

The coupling to the Bose field makes the electron 
effectively heavier. 
Thus, the spatial localization should become better 
with increasing coupling. 
In this section we prove exponential localization 
uniformly in $\kappa, \Lambda$ for $|\charge| < \euv$. 
In fact the localization length is off only by a factor 
$2$ from the one of the uncoupled hydrogen atom.

As already used in \cite{bfs2,gll}, 
the localization serves as an input to the soft photon bound. 
Unfortunately, in our estimate the constant in front of 
the exponential depends so badly on $\charge, Z$ that eventually 
one would have to impose a lower bound on $Z$, which should be 
avoided since also $\eir$, $\euv$ depend on $Z$. 
The constant can be improved by estimating only lower order moments. 
We will explain the bound for the average of $|x|$ and $|x|^{2}$. 
In our proof, in fact, we can allow only for 
the average of $\log(1 + |x|)$.

Following the idea in the first part of 
the proof of \cite[Lemma 6.2]{gll}, 
we have the following lemma. 

\bl{SL1}  
Let $G$ be in $C^{\infty}(\BR)$, 
non-negative function with 
$$
\sup_{x}|\nabla G(x)|,\,\,\,  
\sup_{x}|x|^{-1}G(x)^{2} < \infty.
$$  
Then, for $|\charge| < \euv$, $\onegr \in D(G)$ and 
\eq{eq:1stmomentum} 
\|G\onegr\| _{\cal H}^{2} 
\le \lambda_{1}^{2}\biggl\{ 
\sup_{x}|\nabla G(x)|^{2} 
+ 2 
\sup_{x}\frac{G(x)^{2}}{\lambda_{1}|x|} 
\biggr\} 
\|\onegr\|^{2}_{\cal H}.  
\en 
\el

\proof 
First we prove \kak{eq:1stmomentum} for 
in case of $G \in  C_{0}^{\infty}(\BR)$. 
Then, from direct conputations, 
we have $\onegr \in  D(G)$ and 
$G\onegr, G^{2}\onegr \in  D(\oneG)$. 
Since 
$$\left[\left[\oneG - \oneEl\, ,\, G\right]\, ,\, 
G\right] 
= (\oneG - \oneEl)G^{2} 
- 2 G(\oneG - \oneEl)G 
+ G^{2}(\oneG - \oneEl),$$
we have 
\eq{moment-2}
\langle G\onegr\, ,\, (\oneG - \oneEl) 
G\onegr
\rangle_{\cal H} 
= \langle \onegr \, ,\, 
\frac{1}{2}|\nabla G|^{2} \onegr
\rangle_{\cal H}.
\en  
On the other hand, we have  
\begin{eqnarray} 
\nonumber 
&{}& 
\langle G\onegr\, ,\, 
\left(\oneG - \oneEl\right)G\onegr\rangle_{\cal H} 
=  
\Biggl< G\onegr\, ,\, 
\left(\oneG^{V=0} - \frac{1}{\lambda_{1}|x|} - \oneEl\right)
G\onegr\Biggr>_{\cal H} \\ 
&{}&\ge  
(\oneEl^{V=0} - \oneEl)
\| G\onegr\|_{\cal H}^{2} 
- 
\frac{1}{\lambda_{1}}\Biggl<\onegr\, ,\, 
\frac{G^{2}}{|x|}\onegr
\Biggr>_{\cal H}.
\label{moment-3}
\end{eqnarray} 
We have by \kak{moment-2} and \kak{moment-3} 
\begin{eqnarray}  
(\oneEl^{V=0} - \oneEl)\| G\onegr\|_{\cal H}^{2} 
&\le&  
\langle \onegr \, ,\, 
\frac{1}{2}|\nabla G|^{2} \onegr
\rangle_{\cal H}
+ 
\frac{1}{\lambda_{1}}
\Biggl<\onegr , \frac{G^{2}}{|x|}\onegr
\Biggr>_{\cal H}.  
\label{koredayon} 
\end{eqnarray} 
The assertion \kak{eq:1stmomentum} for $G \in  
C_{0}^{\infty}(\BR)$ follows from 
\kak{binding-energy-t} and \kak{koredayon}.
For $G$ satisfying assumptions in Lemma \ref{SL1}, 
we take a sequence of $G_{n} \in  C_{0}^{\infty}(\BR)$ 
such that $G_{n}(x) \le  G_{n+1}(x) \le  G(x)$ 
for each $n \in {\Bbb N}$, $G_{n}(x) \to G(x)$ 
for almost every $x \in  \BR$, and  
$|\nabla G_{n}(x)| \le n^{-1}G_{n}(x) + |\nabla G(x)|$. 
Then, by Lebesgue's monotone convergence theorem and 
\kak{eq:1stmomentum} for $G_{n} \in C_{0}^{\infty}(\BR)$, 
we have 
\begin{eqnarray*}
\int_{\BR}\lim_{n\to\infty}|G_{n}(x)|^{2}
\|\onegr\|_{\mathcal F}^{2}(x) &=&  
\lim_{n\to\infty}\int_{\BR}|G_{n}(x)|^{2}
\|\onegr\|_{\mathcal F}^{2}(x) \\ 
&\le&  
\lambda_{1}^{2}\biggl\{ 
\sup_{x}|\nabla G(x)|^{2} 
+ 2 
\sup_{x}\frac{G(x)^{2}}{\lambda_{1}|x|} 
\biggr\} 
\|\onegr\|^{2}_{\cal H}.
\end{eqnarray*}
Thus, $\onegr \in  D(G)$ and 
\kak{eq:1stmomentum} holds for $G$. 
\qed 

To have a more concrete estimate we set 
\eq{GR} 
G_{{}_{R}}(x) = \chi_{{}_{R}}(|x|) g(x).
\en
Here $\chi_{{}_{R}}(r) = 0$ for $r < R/2$ and 
$\chi_{{}_{R}}(r) = 1$ for $r > R$ with linear 
interpolation.  
The parameter $R > 0$ serves as a variation 
which will be optimized at the end. 
$g(x)$ is a twice differentiable satisfying 
\eq{condition-g}
\sup_{R/2 < |x|}|\nabla g(x)|^{2} < \infty.
\en  
Then, 
we have 
\eq{5/27-1}
\sup_{x} |\nabla G_{{}_{R}}|^{2}  
\le  
4R^{-2}
\sup_{R/2 < |x| < R}|g(x)|^{2} 
+ 4R^{-1}\sup_{R/2 < |x| < R}|g(x)||\nabla g(x)| 
+ \sup_{R/2 < |x|}|\nabla g(x)|^{2}. 
\en

For example, if $g(x) = \sqrt{\log (3 + c|x|)}$ for a 
strictly positive constant $c$, then 
\eq{log}  
\sup_{x} |\nabla G_{{}_{R}}|^{2}  
\le 4R^{-2}\log (3+cR) + 5R^{-2}. 
\en
Similarly if $g(x) = |x|^{1/2}$, then 
\eq{|x|1/2}  
\sup_{x} |\nabla G_{{}_{R}}|^{2}  
\le 
7R^{-1},
\en
and if $g(x) = |x|$, then 
\eq{|x|}  
\sup_{x} |\nabla G_{{}_{R}}|^{2}  
\le 
9.
\en 
By Lemma \ref{SL1} 
we have the following. 

\bl{1stmomentum'}
(spatial localization). 
Let $g$ be differentiable and non-negative, satisfy 
\kak{condition-g}, and $G_{{}_{R}}(x)$ be defined 
in \kak{GR}. 
\bi 
\item[] (i) If $\sup_{R < |x|}|x|^{-1}|g(x)|^{2} < \infty$, 
then $\onegr \in D(g)$ and 
\eq{eq:1stmomentum'} 
\|g\onegr\|_{\cal H}^{2} 
\le \biggl\{ 
\sup_{|x| \le R}|g(x)|^{2} + 
\lambda_{1}^{2}\sup_{x}|\nabla G_{{}_{R}}(x)|^{2} 
+ 2\lambda_{1} 
\sup_{x}\frac{G_{{}_{R}}(x)^{2}}{|x|} 
\biggr\} 
\|\onegr\|^{2}_{\cal H}  
\en 
for $|\charge| < \euv$. 
\item[] (ii) 
If $4\lambda_{1} < R$, then $\onegr \in  D(g)$
\eq{eq:1stmomentum''} 
\|g\onegr\|_{\cal H}^{2} 
\le \biggl\{ 
\sup_{|x| \le R}|g(x)|^{2} + 
\frac{\lambda_{1}}{2}\left( 
\frac{1}{2\lambda_{1}} - \frac{2}{R}\right)^{-1}
 \sup_{x}|\nabla G_{{}_{R}}(x)|^{2}  
\biggr\} 
\|\onegr\|^{2}_{\cal H}  
\en 
for $|\charge| < \euv$. 
\ei 
\el

\proof 
We note that $(1 - \chi_{{}_{R}})g$ is 
a bounded operator. 
So, $D(g) = D(G_{{}_{R}})$ and 
the assertion (i) follows from 
Lemma \ref{SL1} and     
\begin{eqnarray}
\nonumber 
\|g\onegr\|_{\cal H}^{2} 
&=&  
\|\left(  
1 - \chi_{{}_{R}}
+ \chi_{{}_{R}} 
\right)
g\onegr\|_{\cal H}^{2} \\ 
&\le&  
\sup_{|x| \le R}|g(x)|^{2}
\|\onegr\|_{\cal H}^{2} 
+ 
\| G_{{}_{R}}\onegr\|_{\cal H}^{2}.
\label{moment-1}
\end{eqnarray} 
On the other hand, 
by \kak{koredayon} and the same limiting argument 
as in the proof of Lemma \ref{SL1}, 
we have 
\eq{6/20-1}
\| G_{{}_{R}}\onegr\|_{\cal H}^{2} 
\le 
\frac{1}{2}
\left\{ 
\left(\oneEl^{V=0} - \oneEl\right) 
- \frac{2}{\lambda_{1} R}\right\}^{-1} 
\langle \onegr \, ,\, 
|\nabla G_{{}_{R}}|^{2} \onegr
\rangle_{\cal H}. 
\en 
Thus, the assertion (ii) follows from 
\kak{binding-energy-t}, 
\kak{moment-1} and \kak{6/20-1}. 
\qed

\bp{1stmomentum}
(ground state expectation and Bohr radius). 
Let $0 < |\charge| < \euv$. Then, 
\bi
\item[(i)] for every $R > 0$, $\gr \in  D(\log (3 + |x|))$ 
and 
\begin{eqnarray*} 
\nonumber 
&{}& 
\langle\gr\, ,\,  
\log (3 + |x|)\gr\rangle_{\cal H} \\ 
&{}& 
\le  
\Biggl[
\Biggl\{ 
\log \left( 3 + \frac{4\pi}{\charge^{2}Z}R\right)
\Biggr\}^{2} 
+ 4\left( R^{-2} + R^{-1}\right)
\log \left( 3 + \frac{4\pi}{\charge^{2}Z}R\right) 
+ 5R^{-2}
\Biggr]
\|\gr\|_{\cal H}^{2}.
\end{eqnarray*}
\item[(ii)] $\gr \in  D(|x|)$ and  
${\displaystyle 
\langle\gr\, ,\,  |x|\gr\rangle_{\cal H} 
\le 
\frac{40\pi}{\charge^{2}Z}
\|\gr\|^{2}_{\cal H}}$  
\item[(iii)] For every $R$ with $4 < R$, 
$\gr \in  D(|x|^{2})$ and  
$$ 
\langle\gr\, ,\,  |x|^{2}\gr\rangle_{\cal H} 
\le 
\left(\frac{4\pi}{\charge^{2}Z}\right)^{2} 
\left\{ R^{2} 
+ 5\left(\frac{\displaystyle 1}{\displaystyle 2} - 
\frac{\displaystyle 2}{\displaystyle R}\right)^{-1}
\right\}\|\gr\|_{\cal H}^{2}. 
$$ 
\ei
\ep

\proof 
It follows from Lemma \ref{1stmomentum'}(i) that 
$\gr$ is in $D(\log(3+|x|))$. 
We have 
\begin{eqnarray*} 
\nonumber 
&{}& 
\langle\gr\, ,\, 
\log(3 + |x|)\gr\rangle_{\cal H} 
=   
\langle U_{1}^{*}\gr\, ,\, 
U_{1}^{*}\log(3 + |x|)\gr\rangle_{\cal H}   \\ 
\nonumber 
&{}&=   
\langle \onegr\, ,\, 
U_{1}^{*}\log(3 +|x|)U_{1}
\onegr\rangle_{\cal H}  
=    
\langle \onegr\, ,\, 
\log(3 + \frac{4\pi|x|}{\charge^{2}Z})
\onegr\rangle_{\cal H}  \\ 
\nonumber 
&{}&=   
\langle \onegr\, ,\, 
\log(3 + \frac{4\pi|x|}{\charge^{2}Z})
\onegr\rangle_{\cal H} 
=  
\langle \onegr\, ,\, 
\log( 3 + \frac{4\pi|x|}{\charge^{2}Z})
\onegr\rangle_{\cal H}. 
\end{eqnarray*}  
Thus, by Lemma \ref{1stmomentum'}(i) with \kak{log} 
and taking $\lambda_{1} = 1$ 
so that $r(1) = \alpha Z$,  
we have the assertion (i). 
On the other hand, 
it follows from Lemma \ref{1stmomentum'}(ii) that 
$\gr$ is in $D(|x|)$ and $D(|x|^{2})$. 
By Lemma \ref{1stmomentum'}(i) with \kak{|x|1/2} 
and $\lambda_{1} = 1$, we have 
$$
\langle\gr\, ,\, |x|\gr\rangle_{\cal H} 
\le \frac{4\pi}{\charge^{2}Z} 
\left( R + 7R^{-1} + 2\right) 
\|\gr\|_{\cal H}^{2} 
$$ 
for every $R > 0$, which implies the assertion (ii). 
In the same way, the assertion (iii) follows from 
Lemma \ref{1stmomentum'}(ii) with \kak{|x|}. 
\qed

\bp{exp-decay} 
(exponential decay). 
For every $R$ with $4 < R$ and $\beta$ with 
$$\frac{\displaystyle 1}{\displaystyle 2} - 
\frac{\displaystyle 2}{\displaystyle R} 
- \frac{\displaystyle \beta^{2}}{\displaystyle 4} 
\left( \frac{\displaystyle 4\pi}{\charge^{2}Z}\right)^2 > 0,
$$  
$\gr \in  D(\exp[\beta |x|])$ and 
\begin{eqnarray*}
&{}& \langle\gr\, ,\,  \exp[\beta |x|]\gr\rangle_{\cal H} \\ 
&{}& \le  
\Biggl[ 1 +  
\left( \frac{4}{R^{2}} 
+ \frac{8\beta\pi}{R\charge^{2}Z}\right)
\left(\frac{1}{2} - \frac{2}{R^{2}} 
- \frac{\beta^{2}}{4}
\left( \frac{4\pi}{\charge^{2}Z}\right)^{2}\right)^{-1}
\Biggr] 
e^{4\pi\beta R/(\charge^{2}Z)}\|\gr\|_{\cal H}^{2}. 
\end{eqnarray*} 
\ep

\proof 
Let $g(x) = e^{\widetilde{\beta}|x|/2}$, where 
$\widetilde{\beta} = 4\pi\beta/(\charge^{2}Z)$. 
Then, in the same way as \kak{5/27-1}, 
we have 
\eq{9/9-1}
|\nabla G_{{}_{R}}|^{2} 
\le 4R^{-2}e^{\widetilde{\beta}R} 
+ 2\widetilde{\beta}R^{-1}e^{\widetilde{\beta}R} 
+ \frac{\widetilde{\beta}^{2}}{4}e^{\widetilde{\beta}|x|}. 
\en  
By \kak{9/9-1} and the same limiting argument 
as in the proof of Lemma \ref{SL1} 
with $\lambda_{1} = 1$ ($r(1) = \alpha Z$),  
we get 
\eq{9/9-2} 
\langle\onegr\, ,\, G_{{}_{R}}^{2}\onegr
\rangle_{\cal H} 
\le 
\left( 4R^{-2} + 2\widetilde{\beta}R^{-1}\right)
\left( \frac{1}{2} - \frac{2}{R} 
- \frac{\widetilde{\beta}}{4}\right)^{-1}
e^{\widetilde{\beta}R}
\|\onegr\|_{\cal H}^{2}.
\en 
By \kak{moment-1} and \kak{9/9-2}, 
we have 
\eq{9/9-3} 
\langle\onegr\, ,\, e^{\widetilde{\beta}|x|}\onegr
\rangle_{\cal H} 
\le 
\Biggl[ 1 +
\left( 4R^{-2} + 2\widetilde{\beta}R^{-1}\right)
\left( \frac{1}{2} - \frac{2}{R} 
- \frac{\widetilde{\beta}}{4}\right)^{-1}
\Biggr]
e^{\widetilde{\beta}R}
\|\onegr\|_{\cal H}^{2}.
\en 
By \kak{9/9-3} with noting $\langle \gr\, ,\, e^{\beta|x|}
\gr\rangle_{\cal H} = 
\langle \onegr\, ,\, e^{\widetilde{\beta}|x|}
\onegr\rangle_{\cal H}$, 
we obtain our assertion. 
\qed

\section{Hard photon number bound} 
\label{sec:HPNB} 

Let $$h_{x}(k)= 2^{-1/2}k\beta_{0}(k) 
e^{-i k\cdot x}
 \omega(k)^{-1/2}\widehat{\varphi}_{0}(k).$$ 
From the pull-through formula for $a(k)$ 
one concludes 
\begin{eqnarray} 
\nonumber 
a(k)\G\gr 
&=& 
\sqrt{4\pi\alpha}\,\, h_{x}(k)\cdot 
\left( p + 2\sqrt{\pi}\alpha^{1/2}\akl 
+ 2\sqrt{\pi}\alpha^{1/2}\akl^\ast\right)\gr \\ 
&{}& \quad  
+\od(k)a(k)\gr + \G a(k)\gr, 
\label{pull-through}
\end{eqnarray}
from which it follows that 
$$(\G - \El + \od(k)) a(k) \gr =
- \sqrt{4\pi\alpha}\,\, h_{x}(k)\cdot 
\left( p + 2\sqrt{\pi}\alpha^{1/2}\akl 
+ 2\sqrt{\pi}\alpha^{1/2}\akl^\ast\right) \gr.$$ 
Thus one has 
\eq{soft0}
\langle\gr\, ,\, \nf \gr\rangle_{\hhh} 
= \int_{\BR}\|a(k)\gr\|_{\cal H}^2d^{3}k
\en 
and 
\eq{soft}
\|a(k)\gr\|_{\cal H}^2 
=
4\pi\alpha 
\| (\G - \El + \od(k))\f h_{x}(k)\cdot  
\left( p + 2\sqrt{\pi}\alpha^{1/2}\akl 
+ 2\sqrt{\pi}\alpha^{1/2}\akl^\ast\right)\gr\|_{\cal H}^2.   
\en

In order to establish a uniform photon 
number bound, we divide the momentum space 
into the low energy infrared (IR) region, 
$|k| < 1$, and the high energy ultraviolet (UV) region, 
$|k| \ge 1$. 
Accordingly $\langle\gr\, ,\, \nf \gr\rangle_{\cal H}$ 
is separated into an IR part and a UV part, 
$$\langle\gr\, ,\,\nf\gr\rangle_{\cal H}
=\int_{|k|<1}\|a(k)\gr\|_{\cal H}^2d^{3}k
+\int_{|k|\geq 1}\|a(k)\gr\|_{\cal H}^2d^{3}k.$$

Here we consider the UV region ($|k| \ge 1$) only 
and we do not need to worry about infrared singularities. 
In accordance we set $\tau = 0$ throughout. 

If one pays attention to the increase of the order of 
ultraviolet divergence resulting from 
the commutator 
\eq{hoshi}
[p\, ,\, e^{ik\cdot x}] = ke^{ik\cdot x},
\en 
the following inequality is useful.

\begin{lem}
\label{lem:UV} 
For every $\alpha$ with $\alpha < \euv^{2} /4\pi$  
and arbitrary $\kappa, \Lambda$ with 
$0 < \kappa < \Lambda < \infty$ 
\begin{eqnarray} 
\nonumber  
&{}& 
\| (\G - \El + \omega(k))\f  {h}_{x}({k})\cdot 
\left( p + \sqrt{4\pi\alpha}\akl
+ \sqrt{4\pi\alpha}\akl^{*} 
\right)\gr\|_{\cal H}  \\ 
&{}&\le  
C_{D}(\alpha)\,\,  
\omega({k})^{-1/2}|{h}_{0}({k})| \,\, 
\|\gr \|_{\cal H}
\label{eq:UV1}
\end{eqnarray} 
in the UV region, where 
\eq{CD} 
C_{D}(\alpha) =  
C_{1}(\alpha)
\left(
\sqrt{2 + \left(\alpha Z\right)^{2}}
+ 2\sqrt{2\alpha/\pi}\right) 
\en 
with 
\eq{C1-alpha}
C_{1}(\alpha) = 
\biggl( 1 - C_{\mbox{\tiny UV}}(\charge)\biggr)^{-1/2}.    
\en 
\end{lem} 
\proof 
Let us set  
\eq{B} 
B = (\G - \El +\omega(k))\f.
\en 
Since 
$$ 
B{h}_{x}({k})\cdot 
\left( p + \sqrt{4\pi\alpha}\akl
\right)\gr
= {h}_{0}({k})\cdot B e^{-i{k}\cdot{x}} 
\left( p + \sqrt{4\pi\alpha}\akl\right) 
B^{1/2}
\omega\left({k}\right)^{1/2}\gr,   
$$
by using $\omega(k)^{1/2}B^{1/2}\gr = \gr$ 
and applying \kak{i1} for $\e > 1$, 
it follows from \kak{ineq-p} and 
\kak{maru0} for $\tau = 0$ that 
\begin{eqnarray*} 
&{}& 
\| B{h}_{x}({k})\cdot 
\left( p + \sqrt{4\pi\alpha}\akl
\right)\gr \|_{\cal H} \\ 
&{}&\le  
|{h}_{0}({k})| \,\, \| B\| \omega({k})^{1/2} 
\left( \| pB^{1/2}\| 
+ \sqrt{4\pi\alpha}\|\akl B^{1/2} \|\right)\,\, 
\|\gr \|_{\cal H} \\ 
&{}&\le   
C_{1}(\alpha)\omega({k})^{-1/2}|{h}_{0}({k})| \,\, 
\left( \| p(\ho + 1)^{-1/2}\| 
+ \sqrt{4\pi\alpha}
\|\akl (\ho + 1)^{-1/2}\|\right)\,\, 
\|\gr \|_{\cal H} \\ 
&{}&\le  
C_{1}(\alpha)\omega(k)^{-1/2}|{h}_{0}({k})| \,\, 
\left( \sqrt{2 + (\alpha  Z)^{2}} 
+ \sqrt{2\alpha/\pi}\right)\,\, 
\|\gr \|_{\cal H}. 
\end{eqnarray*} 
Moreover, by using \kak{maru0} for $\tau = 0$ 
and again \kak{i1} for $\e > 1$ 
and sandwiching $(\ho + 1)^{1/2}(\ho + 1)^{-1/2}
(\hf + 1)^{1/2}(\hf + 1)^{-1/2}$ in 
between $B^{1/2} {h}_{x}({k})$,  
\begin{eqnarray*} 
&{}& 
2\sqrt{\pi}
\| B{h}_{x}({k})\cdot  
\alpha^{1/2}\akl^{*}\gr \|_{\cal H}  
\le 
\sqrt{4\pi\alpha}\|B^{1/2}\|\,\, 
\|B^{1/2} 
{h}_{x}({k})\cdot  
\akl^{*}\gr \|_{\cal H} \\ 
&{}&\le   
C_{1}(\alpha)\sqrt{4\pi\alpha} 
\omega({k})^{-1/2} 
\| (\ho + 1)^{-1/2}(\hf + 1)^{1/2}\| \\ 
&{}& \qquad\qquad  
|{h}_{0}({k})| \,\, 
\| (\hf + 1)^{-1/2} 
\akl^{*}\| \,\, 
\|\gr \|_{\cal H} \\ 
&{}&\le  
C_{1}(\alpha)\omega({k})^{-1/2} |{h}_{0}({k})| 
\sqrt{2\alpha/\pi}\|\gr \|_{\cal H}. 
\end{eqnarray*} 
Thus we infer \kak{eq:UV1} and the lemma follows.
\qed

As an immediate consequence we state 
\begin{cor} 
\label{cor:1} 
For every $\alpha$ with $\alpha < \euv^{2}/4\pi$    
and arbitrary $\kappa, \Lambda$ with 
$0 < \kappa < \Lambda < \infty$ 
it holds 
\begin{eqnarray}
\int_{|{k}| \ge 1}
\| a({k})\gr \|_{\cal H}^{2} d^{3}k 
\le 
\frac{4\alpha C_{D}(\alpha)^{2}}{3\pi} 
\|\gr \|_{\cal H}^{2}. 
\label{eq:UV*}
\end{eqnarray}
\end{cor} 
\proof 
\begin{eqnarray*}  
&{}& \int_{|k|\ge 1} \left( 
\omega\left( k\right)^{-1/2}|h_{0}(k)|\right)^{2} 
\le 
\frac{1}{2(2\pi)^{3}} 
\int_{|k|\ge 1} 
|k|^{-2}
\left( 
1 + \frac{1}{2}|k|^{2}
\right)^{-2} d^{3}k  
= \frac{1}{3\pi^{2}}.
\end{eqnarray*}  
Our corollary follows now from \kak{soft}, 
Lemma \ref{lem:UV}, and the above inequality. 
\qed

\section{Soft photon number bound}
\label{sec:SPNB}

To show the soft photon bound, 
one relies on the localization of 
the ground state $\gr$. 
More precisely, following \cite{bfs2}, 
it can be proved that 
\eq{bfs} 
\int_{|k|<1}\| a(k)\gr\|_{\cal H}^{2}d^{3}k 
\le const\,\, \alpha\| |x|\gr\|_{\cal H}^{2}
\en
by using $i\left[ \G\, ,\, x\right] 
= p + \sqrt{4\pi\alpha}\akl 
+ \sqrt{4\pi\alpha}\akl^{*}$. 

By Proposition \ref{1stmomentum} a uniform bound on 
$\| |x|\gr\|^{2}$ is available. 
However the prefactor in the bound in 
Proposition \ref{1stmomentum} (iii) contains 
the factor $(\charge^{2}Z)^{-2}$ which becomes 
large for $\charge\to 0$. 
Since $\eir$ and $\euv$ also depend on $\charge^{2}Z$ 
conflicting requirements result. 
As can be seen from Proposition \ref{1stmomentum}, 
the only resolution is to improve the prefactor by 
lowering the estimated moment. 
The main task of this section is to device 
an iterative scheme \`{a} la renormalization 
introducing an energy scale parameter,  
which systematically reduces the order of the moment bound.   
 
Throughout this section, we consider the infrared region only.

\subsection{Basic estimate}

We denote the components of 
$\akl^{\sharp}$ by 
$\akljs$\, ($j = 1, 2, 3$), 
i.e., $\akl^{\sharp} = 
(\aklis, \akliis, \akliiis)$. 
The following estimate is too rough 
for controlling the infrared singularity, 
nevertheless it is useful. 

\begin{lemma} 
\label{lem:(3)}
Let $F, G$ be $\BR$-valued functions of 
$k \in \BR$, independent of $x \in \BR$. Then 
for every $\alpha$ with $\sqrt{\alpha} < \euv$ and 
arbitrary $\kappa, \Lambda$ with 
$0 < \kappa < \Lambda < \infty$ 
\begin{eqnarray} 
\nonumber 
&{}& 
\| Be^{iF\cdot{x}} 
\left(p_{j} + \sqrt{4\pi\alpha}\aklj
+ \sqrt{4\pi\alpha}\akljast 
\right) 
e^{iG\cdot{x}}
\gr \|_{\cal H} \\ 
&{}&\le  
\biggl(  C_{D}(\alpha) + {\rm min}\{|F_j|,|G_j|\} 
\biggr) 
\omega({k})^{-1}
\|\gr \|_{\cal H}
\label{(3)}
\end{eqnarray}  
in the IR region. 
In particular in the case either $F_j=0$ or $G_j=0$, 
$$
\| Be^{iF\cdot{x}} 
\left(p_{j} + 2\sqrt{\pi}\alpha^{1/2}\aklj
+ 2\sqrt{\pi}\alpha^{1/2}\akljast 
\right) 
e^{iG\cdot{x}}
\gr \|_{\cal H}  
\le 
C_D(\alpha)\,\, \omega({k})^{-1} 
\|\gr \|_{\cal H}.
$$
\end{lemma} 
\proof 
By using \kak{maru0} and \kak{i1} with $\e < 1$, we have 
\begin{eqnarray*}
\| Be^{iF\cdot{x}} 
\akljast 
e^{iG\cdot{x}}
\gr \|_{\cal H} 
&\le& 
\omega({k})^{-1/2} 
\| B^{1/2}(\ho + 1)^{1/2}\| \,\, 
\| (\ho + 1)^{-1/2}\akljast\| \,\, 
\| \gr \|_{\cal H}  \\ 
&\le& 
C_{1}(\alpha)
\left( \sqrt{2}\pi\omega(k)\right)^{-1} 
\| \gr \|_{\cal H}, 
\end{eqnarray*}  
and, by using 
$\gr = \omega(k)^{1/2}B^{1/2}\gr$,  
\begin{eqnarray*} 
&{}& 
\| Be^{iF\cdot{x}} 
\aklj
e^{iG\cdot{x}}
\gr \|_{\cal H}   
\le 
\omega({k})^{-1/2}
\| B^{1/2}e^{iF\cdot{x}} 
\aklj 
e^{iG\cdot{x}}
\gr \|_{\cal H}  \\ 
&{}&\le  
\| B^{1/2}e^{iF\cdot{x}}\|\,\, 
\| \aklj
e^{iG\cdot{x}} 
(\hf + 1)^{-1/2}(\hf + 1)^{1/2}
(\ho + 1)^{-1/2}(\ho + 1)^{1/2}
B^{1/2}
\gr \|_{\cal H} \\ 
&{}&\le   
\| B^{1/2}\|\,\, 
\| \aklj
(\hf + 1)^{-1/2} 
e^{iG\cdot{x}} 
(\hf + 1)^{1/2}
(\ho + 1)^{-1/2}(\ho + 1)^{1/2}
B^{1/2}
\gr \|_{\cal H} \\ 
&{}&\le  
\omega({k})^{-1/2} 
\| \aklj
(\hf + 1)^{-1/2}\| \,\, 
\| (\hf + 1)^{1/2}
(\ho + 1)^{-1/2}\| \,\, 
\| (\ho + 1)^{1/2}B^{1/2}\| \,\, 
\|\gr \|_{\cal H} \\ 
&{}&\le  
C_{1}(\alpha)
\left(\sqrt{2}\pi\omega(k)\right)^{-1} 
\| \gr \|_{\cal H}. 
\end{eqnarray*} 
Next we estimate 
$\|Be^{iJ\cdot x} p_j e^{i\widetilde{J}\cdot x} 
\gr\|_{\cal H}$. 
Using 
$ 
e^{iF\cdot{x}}p_{j} = 
p_{j}e^{iF\cdot{x}} - 
F_{j}e^{iF\cdot{x}}$,
we obtain 
by \kak{ineq-p} and \kak{i1} with $\e < 1$   
\begin{eqnarray*} 
&{}& 
\| Be^{iF\cdot{x}} 
p_{j} 
e^{iG\cdot{x}}
\gr \|_{\cal H}  
\le 
\| Bp_{j}e^{iF\cdot{x}} 
e^{iG\cdot{x}}
\gr \|_{\cal H}   
+ 
|F_{j}|\,\, 
\| Be^{iF\cdot{x}} 
e^{iG\cdot{x}}
\gr \|_{\cal H} \\ 
&{}&\le  
\omega({k})^{-1/2} 
\| B^{1/2}(\ho + 1)^{1/2}
(\ho + 1)^{-1/2}p_{j}e^{iF\cdot{x}} 
e^{iG\cdot{x}}
\gr \|_{\cal H}   
+ 
|F_{j}|\,\, 
\| B\|\,\, 
\| \gr \|_{\cal H} \\ 
&{}&\le  
\omega({k})^{-1} 
\left( C_{1}\left(\alpha\right)
\sqrt{2 + (\alpha Z)^{2}} 
+ |F_{j}|\right) 
\| \gr \|_{\cal H}. 
\end{eqnarray*} 
Alternatively using 
$p_{j}e^{iG\cdot{x}} 
= e^{iG\cdot{x}}p_{j} 
+ G_{j}e^{iG\cdot{x}}$
we have 
\begin{eqnarray*} 
&{}& 
\| Be^{iF\cdot{x}} 
p_{j} 
e^{iG\cdot{x}}
\gr \|_{\cal H}  
\le 
\| Be^{iF\cdot{x}} 
e^{iG\cdot{x}}
p_{j}\gr \|_{\cal H}   
+ 
|G_{j}|\,\, 
\| Be^{iF\cdot{x}} 
e^{iG\cdot{x}}
\gr \|_{\cal H} \\ 
&{}&\le  
\omega({k})^{-1} 
\| p_{j}\gr \|_{\cal H}   
+ 
|G_{j}|\,\, 
\| B\|\,\, 
\| \gr \|_{\cal H} 
\le 
\omega({k})^{-1/2} 
\| p_{j}B^{1/2}\gr \|_{\cal H} 
+ 
\omega({k})^{-1}
|G_{j}| \,\, 
\| \gr \|_{\cal H} \\ 
&{}&\le   
\omega({k})^{-1}
\left( C_{1}\left(\alpha\right)\sqrt{2 + (\alpha Z)^{2}}  
+ |G_{j}|\right) 
\| \gr \|_{\cal H},  
\end{eqnarray*} 
where we used \kak{i1} with $\e < 1$ 
in the last inequality. 
Hence \kak{(3)} follows. 
\qed

\subsection{Energy scale renormalization}

Let $\gamma(k) = \left(2\omega(k)\right)^{-1/2}\beta_{0}(k)
\widehat{\varphi}_{0}(k)$. 
From the pull-through formula \kak{pull-through}, 
one has 
\eq{a+I0}
a(k)\gr = 
\sqrt{4\pi\alpha}\,\,\gamma(k)BI_{0}\gr
\en
with $I_{0} = 
e^{-i k\cdot{x}}
{k}\cdot {D}$ 
and 
${D} 
= p + \sqrt{4\pi\alpha}\,\,\akl 
+ \sqrt{4\pi\alpha}\,\,\akl^{*}$. 
Therefore, 
\begin{eqnarray}
\nonumber  
\int_{|k|<1} \|a(k)\gr \|_{\cal H}^2 d^{3}k 
&\leq&  
{\rm const.}\times \int_{|k|<1} 
\omega(k)^{-2}\gamma(k)^2|\widehat{\varphi}_{m}(k)|^{2} d^{3}k \\ 
&\cong& - \log \kappa 
\quad \mbox{(for small $\kappa$).}
\label{eq:IR-div}
\end{eqnarray} 
    
As mentioned already, 
to avoid such infrared divergence, 
one considers the commutator  
\begin{eqnarray} 
i\left[ \G\, ,\, x\right] = 
p + \sqrt{4\pi\alpha}\,\,\akl 
+ \sqrt{4\pi\alpha}\,\,\akl^{*},
\label{velocity}
\end{eqnarray}  
which yields then an infrared convergent bound proportional 
to $\langle\gr\, ,\, |x|^{2}\gr\rangle_{\cal H}$. 
We have to reduce to moment bound from $|x|^{2}$ 
to $\log(1+ |x|)$ and first explain the iterative construction. 
In our specific application, in fact, two iteration steps suffice.

Let us start with the identity 
\begin{eqnarray} 
[ \G\, ,\, \iff ] 
= \frac{1}{2}f_{j}^{2} \iff
+ f_{j} \iff D_{j},
\label{eq:IR0} 
\end{eqnarray}
where 
$f_{j}$ depends only on $k \in \BR$, and 
$D_{j} = 
p_{j} + \sqrt{4\pi\alpha}\aklj 
+ \sqrt{2\pi\alpha}\akljast.$  
Then 
\begin{eqnarray} 
D_{j} =  
\frac{1}{f_{j}} \iif 
[\G\,  ,\,  \iff ] 
- \frac{1}{2}f_{j}. 
\label{eq:IR1} 
\end{eqnarray} 
Moreover 
$$ 
[\G\,  ,\, \ig ] 
= 
\frac{1}{2}|{g}|^{2} \ig  
+ \ig g \cdot{D}. 
$$ 
Let $\{f^{[n]}\}_{n=1}^\infty$ be a given scale 
of $\BR$-valued functions on $\BR$. 
Their proper adjustment will be part of the proof.   
We define 
$$
\z n_\ell  
=
k_\ell +\delta_{j\ell}  \sum_{m = 1}^{n}f_{j}^{[m]},\ \ \ 
j=1,2,3,\ \ \ \ell=1,2,3,$$
and set $\z n =e^{2}(\z n_1,\z n _2,\z n_3)$ 
and $\z 0 =(k_1,k_2,k_3)$. 
Note that 
\eq{eq:IR3}
 \g n \mf{n+1} =\g{n+1}.
\en 
Since 
$$ 
[ \G\, ,\, \g {n-1}] = 
[\G\,  ,\, \g n \pf n ] 
=
[\G\,  ,\, \g n ] \pf n +\g n 
[\G\,  ,\, \pf n ],$$ 
one obtains that 
\eq{eq:IR4}
[\G\,  ,\, \pf n ] 
= 
\pg n [\G\,  ,\, \g {n-1}  
] 
- 
\pg n 
[\G\,  ,\, \g n ] 
\pf n .
\en 
From (\ref{eq:IR1}) and (\ref{eq:IR3}) one infers 
\eq{eq:S1-1}  
I_{0}\onegr  
= 
\sum_{j=1}^{3}
\g 0 
k_{j}
\lk 
\frac{1}{\w 1} 
\mf 1 
[\G\,  ,\, \pf 1 ] 
- \frac{1}{2}\w 1
\rk 
\gr  
=
I_{11}\gr  
+ \Psi_{10},  
\en 
where 
\begin{eqnarray*} 
I_{11} 
&=& 
\sum_{j=1}^{3} 
\frac{k_{j}}{\w 1} 
\g 1 
[\G\,  ,\, \pf 1 ], \\ 
\Psi_{10} 
&= &
- \frac{1}{2}
\sum_{j=1}^{3} 
k_{j}\w 1 
\g 0 
\gr. 
\end{eqnarray*}

By (\ref{eq:IR3})
and (\ref{eq:IR4}) one gets 
$$ 
I_{11}\gr   
=
\sum_{j=1}^{3}\frac{k_{j}}{\w 1} 
[\G\,  ,\, \g 0 ]\gr  
- \sum_{j=1}^{3}\frac{k_{j}}{\w 1} 
[\G\,  ,\, \g 1 ] \pf 1 
\gr  
$$
\eq{eq:S1-2} 
= 
\Psi_{11} + \Psi_{12} + \widetilde{\Psi}_{13},
\en 
where 
\begin{eqnarray*} 
\Psi_{11} 
&=& 
\sum_{j=1}^{3}\frac{k_{j}}{\w 1} 
\left(\G - \El\right)
\g 0 
\gr,  \\
\Psi_{12} 
&=& 
- \frac{1}{2}
\sum_{j=1}^{3}\frac{k_{j}}{\w 1} 
|\z 1 |^{2}
\g 0 
\gr,  \\ 
\widetilde{\Psi}_{13} 
&=& 
\sum_{j=1}^{3}\frac{k_{j}}{\w 1} 
\g 1 
\lk 
\z 1 \cdot {D}
\rk 
\pf 1 
\gr. 
\end{eqnarray*} 
Moreover we decompose $\widetilde{\Psi}_{13}$ 
as 
$$
\widetilde{\Psi}_{13}  
=
\sum_{j=1}^{3}\frac{k_{j}}{\w 1} 
\g 1 
\z 1 _j  D_j 
\pf 1 
\gr  
+ 
\sum_{j=1}^{3}\mathop{\sum_{\ell=1}^{3}}_{\ell\ne j}
\frac{k_{j}}{\w 1} 
\g 1 
\z 1 _\ell D_{\ell}
\pf 1 
\gr  
$$
$$=
\widetilde{I}_{1}\gr  + \Psi_{13}, 
$$ 
where 
\begin{eqnarray*} 
\widetilde{I}_{1} 
&=& 
\sum_{j=1}^{3}\widetilde{I}_{1}^{j}, \qquad  
\widetilde{I}_{1}^{j} 
:= 
\frac{k_{j}
\left( k_{j} + \w 1\right)}{\w 1} 
 \g 1 D_{j} \pf 1 , \\ 
\Psi_{13} 
&=& 
\sum_{j=1}^{3}\mathop{\sum_{\ell=1}^{3}}_{\ell\ne j}
\frac{k_{j}k_{\ell}}{\w 1} 
 \g 1 
D_{\ell}\pf 1 
\gr, 
\end{eqnarray*} 
and we used the definition of $\z 1$.

Set 
\begin{eqnarray*} 
I_{1} 
= 
\sum_{j=1}^{3} 
\widetilde{I}_{1}^{j} 
\mf 1 
= \sum_{j=1}^{3}\frac{k_{j}
\left( k_{j} + \w 1\right)}{\w 1} 
 \g 1 D_{j}. 
\end{eqnarray*}
Then the error term $\Psi_{1}^{\mathrm er}$ 
is defined as 
$$
\Psi_{1}^{\mathrm er} 
:= 
\left(\widetilde{I}_{1} - I_{1}\right)\gr  
= 
\sum_{j=1}^{3}\widetilde{I}_{1}^{j}
\left( 1 -  \mf 1 \right)\gr. 
$$
Therefore we obtain that 
\eq{R1} 
I_{0}\gr  
= 
\sum_{\nu=0}^{3}\Psi_{1\nu} + 
I_{1}\gr + \Psi_{1}^{\mathrm er}.
\en

The scale function $f^{[1]}$ is chosen such that 
$|\gamma(k)|^{2}\|B\sum_{\nu=0}^{3}
\Psi_{1\nu}\|_{\cal H}^{2}$ 
is integrable at small $k$. 
Therefore, the first term of \kak{R1} can be retained. 
The map from $I_{0}\gr$ to $I_{1}\gr$ is called 
the first iteration ${\cal R}_{1}$. 
We repeat the same procedure as applied to $I_{1}\gr$.
Then, by (\ref{eq:IR1}) and (\ref{eq:IR3}), we get 
$$
I_{1}\gr 
= 
\sum_{j=1}^{3}
\frac{k_{j}
\left( k_{j} + \w 1 \right)}{\w 1} 
\g 1
\lkk 
\frac{1}{\w 2 }
\mf 2 
[\G\,  ,\, \pf 2 ] 
- \frac{1}{2}\w 2 
\rkk 
\gr$$ 
\eq{eq:S2-1}  
=
I_{21}\gr  
+ \Psi_{20},  
\en 
where 
\begin{eqnarray*} 
I_{21} 
&=& 
\sum_{j=1}^{3}
\frac{k_{j}\left( k_{j} 
+ \w 1
\right)}{\w 1 
\w 2}
 \g 2 
[\G\, ,\, \pf 2 ], \\ 
\Psi_{20} 
&=& 
- \frac{1}{2}
\sum_{j=1}^{3} 
\frac{k_{j}\left( k_{j} 
+ \w 1
\right) \w 2}{
\w 1}
 \g 1 \gr. 
\end{eqnarray*} 
By (\ref{eq:IR3})
and (\ref{eq:IR4}) one obtains  
$$
I_{21}\gr  
=
\sum_{j=1}^{3}\frac{k_{j}\left( k_{j} 
+ \w 1 \right)}{\w 1 \w 2 } 
\lk
[\G\,  ,\, \g 1 ] 
- 
[\G\,  ,\, \g 2 ] 
\pf 2 \rk
\gr  $$
$$
=
\sum_{j=1}^{3}\frac{k_{j}\left( k_{j} 
+ \w 1 \right)}{\w 1\w 2 } 
\left(\G  - \El\right) 
\g 1 
\gr  
- \sum_{j=1}^{3}
\frac{k_{j}\left( k_{j} 
+ \w 1 \right)}{
\w 1\w 2 } 
[\G\,  ,\, \g 2 ] 
\pf 2 
\gr$$
$$
=
\Psi_{21} + \Psi_{22} + \widetilde\Psi_{23}, 
$$ 
where 
\begin{eqnarray*} 
\Psi_{21} 
&=& 
\sum_{j=1}^{3}
\frac{k_{j}\left( k_{j} 
+ \w 1 \rk}{\w 1
\w 2 } 
\left(\G - \El\right)
\g 1 
\gr ,  \\
\Psi_{22} 
&=& 
- \frac{1}{2}
\sum_{j=1}^{3}\frac{k_{j}\left( k_{j} 
+ \w 1 \rk }{
\w 1\w 2 } 
|\z 2|^{2}
\g 1 
\gr ,  \\ 
\widetilde{\Psi}_{23} 
&=& 
\sum_{j=1}^{3}\frac{k_{j}\left( k_{j} 
+ \w 1 \rk}{
\w 1\w 2 } 
 \g 2 
\lk 
\z 2\cdot {D}
\rk 
\pf 2 
\gr . 
\end{eqnarray*} 
Moreover 
$\widetilde{\Psi}_{23}$ is decomposed as 
$ 
\widetilde{\Psi}_{23}  
=
\widetilde{I}_{2}\gr  
+ \Psi_{23}, 
$ 
where 
\begin{eqnarray*} 
\widetilde{I}_{2} 
&=& 
\sum_{j=1}^{3}\frac{k_{j} 
\left( k_{j} + \w 1\rk  
\left( k_{j} + \w 1
+ \w 2\right)}{\w 1\w 2 } 
\g 2 
D_{j}\pf 2, \\ 
\Psi_{23} 
&=& 
\sum_{j=1}^{3}\mathop{\sum_{\ell=1}^{3}}_{\ell\ne j}
\frac{k_{j} 
\left( k_{j} + \w 1 \rk 
k_{\ell}}{\w 1\w 2 } 
\g 2 
D_{\ell} \pf 2 
\gr. 
\end{eqnarray*} 
Here we used the definition of $\z 2_\ell $. 
Therefore, we obtain that 
\eq{R2}
I_{1}\gr  
= 
\sum_{\nu=0}^{3}\Psi_{2\nu} + 
\widetilde{I}_{2}\gr.
\en 
Again $f^{[2]}$ is chosen such that 
$|\gamma(k)|^{2}\| B\sum_{\nu=0}^{3}
\Psi_{2\nu}\|_{\cal H}^{2}$ 
is integrable at small $k$. 
Therefore, the first term of \kak{R2} may be retained. 
Then, in the same way as for $I_{1}$ we can get $I_{2}$ 
by using the dipole approximation only for 
the last factor $\pf 2$ in each component of 
$\widetilde{I}_{2}$. 
The map from $I_{1}\gr$ to $I_{2}\gr$ 
is called the second iteration ${\cal R}_{2}$. 
In particular the iterations may be 
repeated ad infinitum. 
However, we stop our iteration here 
and set 
$$I_{2} = \widetilde{I}_{2}.$$
So, there is no error term from the second step.  

\begin{lem}
\label{lem:IR} 
There exists $\delta$ 
with $0 < \delta < 1/2$ 
such that for every $\alpha$ with $\alpha < 
\euv^{2} /4\pi$ 
and arbitrary $\kappa, \Lambda$ with 
$0 < \kappa < \Lambda < \infty$ 
\begin{eqnarray*}
&{}& 
\biggl\| (\G - \El + \omega(k))\f {h}_{x}({k})\cdot 
\left( p + 2\sqrt{\pi}\alpha^{1/2}\akl 
+ 2\sqrt{\pi}\alpha^{1/2}\akl^{*} 
\right)\gr - 
\gamma(k)B\Psi_{1}^{\mathrm er}
\biggr\|_{\cal H} \\   
&{}&\le  
3
\left( 
8C_{D}\left(\alpha\right) + \frac{21}{2}
\right)
|\gamma({k})|\,\, 
|{k}|^{\delta}\,\, 
\|\gr \|_{\cal H}
\end{eqnarray*} 
holds in the IR region.
\end{lem}

\begin{lem}
\label{lem:IR'} 
For all $\alpha$ with 
$\alpha < \euv^{2}/4\pi$ 
and $\alpha Z < 1$, and 
every $\varepsilon$ with $1/2 < \varepsilon < 1$, 
it holds  
$$
\alpha\int_{|k| < 1}|\gamma(k)|^{2} 
\|B\Psi_{1}^{\mathrm er}\|_{\cal H}^{2}d^{3}k 
\le 
M\alpha 
\left\{ 
9 + 2L(\alpha,Z)^{2} + 
9\times 10^{-2}L(\alpha, Z)
\right\}\|\gr\|_{H}^{2},
$$
where 
\begin{eqnarray*} 
&{}& 
L(\alpha, Z) = 
\log \left( 3 + \frac{10^{2}}{\alpha Z}\right), \\ 
&{}& 
M 
= \frac{18C_{1}(\alpha)^{2}
\left( C_{D}\left(\alpha\right) + 2\right)^{2}}{
\varepsilon\pi^{2}}.
\end{eqnarray*}
\end{lem}

In the same way as for the proof of 
\ref{cor:1}, from Lemmas 
\ref{lem:IR} and \ref{lem:IR'} 
it follows 
\begin{cor}
\label{cor:2} 
We have for every $\alpha$ with $\alpha 
< \euv^{2}/4\pi$ and 
arbitrary $\kappa, \Lambda$ with 
$0 < \kappa < \Lambda < \infty$ 
\begin{eqnarray}
\nonumber 
&{}&  
\int_{|{k}| < 1} 
\,\,\, 
\| a({k})\gr \|_{\cal H}^{2} d{k} \\ 
\nonumber 
&{}&\le 
\Biggl[ 
\frac{9\alpha}{\pi\delta}
\left( 8C_{D}(\alpha) + \frac{21}{2}
\right)^{2} 
+ 
2M\alpha\left(
9 + 2L(\alpha,Z)^{2} + 
9\times 10^{-2}L(\alpha, Z)
\right)
\Biggr]
\|\gr\|_{\cal H}^{2}. \\ 
&{}& \qquad 
\label{eq:IR*}
\end{eqnarray}
\end{cor}

\subsection{Proof of Lemma \ref{lem:IR}}  
 
We choose our energy scale parameters 
as follows: for fixed $\varepsilon$ with 
$1/2 < \varepsilon < 1$, we set 
\eq{sd}
\w 1(k) = |{k}|^{\varepsilon}, \ \ \ 
\w 2 (k) = - \w 1(k), \ \ \ 
j = 1, 2, 3. 
\en 
Then, the following inequality 
\eq{koremo}
|k_{j} + \w 1| 
\leq 
|k_{j}| + |\w 1| 
\le 
\left( |k| + |k|^{\varepsilon}\right)
\leq 
2|{k}|^{\varepsilon} 
\en 
holds, since $|k|<1$. 

For 
\eq{2times} 
BI_{0}\gr  - \Psi_{1}^{\mathrm er}
= \sum_{n = 1}^{2}\sum_{\nu = 0}^{3} 
B\Psi_{n\nu} + BI_{2}\gr,
\en 
we estimate each term in \kak{2times}. 
Using \kak{sd} $B\Psi_{n\nu}$ is estimated from above.
Let us first consider $B\Psi_{1\nu}$. 
One has  
\begin{eqnarray} 
 \| B\Psi_{10}\|_{\cal H} 
&\leq&  
\frac{1}{2}\sum_{j=1}^{3}|k_{j}|\,\,  
|\w 1|\,\, 
\| B\|\,\, \|\gr \|_{\cal H} 
\leq 
|k|^\varepsilon \|\gr\|_{\cal H}, 
\label{01} \\   
\nonumber  
 \| B\Psi_{11}\|_{\cal H} 
&\leq&  
\sum_{j=1}^{3}\frac{|k_{j}|}{|\w 1|} 
\| B(\G - \El)\|\,\, 
\| \g 0 
\gr \|_{\cal H}  \\ 
&\leq&   
3|k|^{1-\varepsilon}
\|\gr\|_{\cal H} 
\label{02}
\end{eqnarray}  
and 
$$ 
 \| B\Psi_{12}\|_{\cal H} 
\leq 
\frac{1}{2}
\sum_{j=1}^{3}\frac{|k_{j}|}{|\w 1|}\| B\|\,\, 
|\z 1|^{2}
\| \gr \|_{\cal H} 
\leq 
\frac{1}{2|k|}
\sum_{j=1}^{3}\frac{|k_{j}|}{|\w 1|}
|\z 1|^{2}
\| \gr \|_{\cal H}. 
$$ 
Note that 
$$ 
\sum_{j=1}^{3}\frac{|k_{j}|}{|\w 1|}
|\z 1|^{2}  
\le
|{k}|^{1-\varepsilon}
\sum_{j=1}^{3}
|\z 1|^{2}. 
$$ 
Let ${\cal P}_{3}^{+}$ be the set of 
even permutations of $(1, 2, 3)$. 
Then by \kak{koremo}
\begin{eqnarray} 
\nonumber 
\sum_{j=1}^{3}
|\z 1|^{2}  
&=& 
\sum_{\sigma\in {\cal P}_{3}^{+}}
\left\{ 
|k_{\sigma(1)} 
+ f_{\sigma(1)}^{[1]}|^{2} 
+ |k_{\sigma(2)}|^{2} 
+ |k_{\sigma(3)}|^{2}
\right\} \\ 
\nonumber 
&\le& 
\sum_{\sigma\in {\cal P}_{3}^{+}}
\left\{ 
4|{k}|^{2\varepsilon} 
+ 2|{k}|^{2\varepsilon}
\right\}  
=
18|{k}|^{2\varepsilon}. 
\label{eq:theta2}  
\end{eqnarray} 
Hence it follows that 
\eq{03} 
\| B\Psi_{12}\|_{\cal H} 
\leq 
9|{k}|^{\varepsilon}
\|\gr \|_{\cal H}. 
\en 
Moreover it follows from Lemma \ref{lem:(3)} that 
\begin{eqnarray} 
\nonumber  
 \| B\Psi_{13}\|_{\cal H} 
&\leq&  
\sum_{j=1}^{3}\mathop{\sum_{\ell=1}^{3}}_{\ell\ne j}
\frac{|k_{j}|\,\, |k_{\ell}|}{|\w 1|}
\| B \g 1 
D_{\ell}\pf 1 
\gr \|_{\cal H} \\ 
&\leq& 
6 C_D(\alpha)|k|^{1-\varepsilon}\|\gr\|_{\cal H}.
\label{04} 
\end{eqnarray}  

Next let us estimate $B\Psi_{2\nu}$.
By \kak{koremo} one has 
\begin{eqnarray} 
\| B\Psi_{20}\|_{\cal H} 
&\leq&  
\frac{1}{2}\| B\| \sum_{j=1}^{3}
\frac{|k_{j}|\,\, |k_{j} + \w 1|\,\,  
|\w 2 |}{|\w 1|}  \|\gr \|_{\cal H} 
\leq 
3|k|^\varepsilon \|\gr\|_{\cal H}, 
\label{05} \\ 
\nonumber 
 \| B\Psi_{21}\|_{\cal H}
&\leq&  
\sum_{j=1}^{3}\frac{|k_{j}|\,\, 
|k_{j} + \w 1|}{|\w 1|\,\, |\w 2 |} 
\| B(\G - \El)\|\,\, 
\| \g 1 \gr \|_{\cal H} \\ 
&\leq&  
6|k|^{1-\varepsilon}\|\gr\|_{\cal H} 
\label{06}
\end{eqnarray}  
and 
$$
 \| B\Psi_{22}\|_{\cal H} 
\leq 
\frac{1}{2}\| B\|
\sum_{j=1}^{3}\frac{|k_{j}|\,\, 
|k_{j} + \w 1|}{|\w 1|\,\, 
|\w 2 |}
|\z 2|^{2} \,\, 
\| \g 1 \gr \|_{\cal H} $$
$$ 
\leq 
\frac{1}{2|k|} 
\sum_{j=1}^{3}\frac{|k_{j}|\,\, 
|k_{j} + \w 1|}{|\w 1|\,\, 
|\w 2 |}
|\z 2 |^{2}
\|\gr \|_{\cal H}. 
$$ 
Note that $\w 1=-\w2$. 
Therefore  by \kak{koremo}
$$ 
\sum_{j=1}^{3}\frac{|k_{j}|\,\, 
|k_{j} + \w 1|}{|\w 1|\,\, 
|\w 2 |}
|\z 2 |^{2} 
\leq 
\sum_{j=1}^{3}\frac{|{k}|\,\, 
|k_{j} + \w 1|}{|\w 1|^{2}}
|\z 2 |^{2} 
\leq 
2|{k}|^{1-\varepsilon}
\sum_{j=1}^{3}|\z 2 |^{2}  
$$ 
and in addition 
\begin{eqnarray} 
\nonumber 
\sum_{j=1}^{3}|\z 2 |^{2} 
&=& 
\sum_{\sigma\in {\cal P}_{3}^{+}}
\left\{ 
|k_{\sigma(1)} + f_{\sigma(1)}^{[1]} 
+ f_{\sigma(1)}^{[2]}|^{2} 
+ |k_{\sigma(2)}|^{2} 
+ |k_{\sigma(3)}|^{2}
\right\} \\ 
\nonumber 
&=& 
\sum_{\sigma\in {\cal P}_{3}^{+}}
\left\{ 
|k_{\sigma(1)}|^{2} 
+ |k_{\sigma(2)}|^{2} 
+ |k_{\sigma(3)}|^{2}
\right\} 
=3|k|^{2}. 
\label{eq:theta5}
\end{eqnarray}
Since $|k| < 1$, one has  
\eq{07} 
\| B\Psi_{22}\|_{\cal H} 
\leq 
3|{k}|^{2-\varepsilon}
\|\gr \|_{\cal H} 
< 3|{k}|^{1-\varepsilon}
\|\gr \|_{\cal H}. 
\en 
Moreover 
it follows from Lemma \ref{lem:(3)} and \kak{koremo} 
that 
\begin{eqnarray} 
\nonumber 
 \| B\Psi_{23}\|_{\cal H} 
&\leq&  
\sum_{j=1}^{3}\mathop{\sum_{\ell=1}^{3}}_{\ell\ne j}
\frac{|k_{j}|\,\, 
|k_{j} + \w 1|\,\, |k_{\ell}|}{|\w 1|\,\, 
|\w 2 |}
\| B\g 2 
D_{\ell}\pf 2 \gr\|_{\cal H} \\ 
&\le&  
12 C_D(\alpha)|k|^{1-\varepsilon}\|\gr\|_{\cal H}.
\label{08}
\end{eqnarray}  

Finally let us consider $I_2\gr$. 
Note that $\w 1=-\w2$. 
By Lemma \ref{lem:(3)} and \kak{koremo} we have 
\begin{eqnarray} 
\nonumber 
\| BI_{2}\gr \|_{\cal H} 
&\le&  
\sum_{j=1}^{3} 
\frac{|k_{j}|\,\, 
|k_{j} + \w 1|\,\, 
|k_{j} + \w 1 + \w 2 |}{|\w 1|\,\, 
|\w 2 |} 
\| B\g2 
D_{j} \pf 2 
\gr \|_{\cal H} \\ 
\nonumber 
&\le&  
2|{k}|^{1-\varepsilon} 
\sum_{j=1}^{3} 
\left( 
C_{D}(\alpha) + |{k}|^{\varepsilon}\right)
\|\gr \|_{\cal H}  \\ 
&\le&  
6(C_{D}(\alpha) + 1)|{k}|^{1-\varepsilon}
\|\gr \|_{\cal H},  
\label{09}
\end{eqnarray}  
where we used $|k|^{\varepsilon} < 1$ 
in the last inequality.
\hfill\break

Combining \kak{2times}, \kak{01}--\kak{08} 
and \kak{09},  
we complete the proof 
of Lemma \ref{lem:IR}.   
Set $\delta = 1 - \varepsilon$. 
Then $0 < \delta < 1/2$. 
By \kak{01}--\kak{09}
one has  
\begin{eqnarray*} 
\| BI_{0}\gr - B\Psi_{1}^{\mathrm er} \|_{\cal H} 
&\leq&  
\sum_{n=1}^{2}\sum_{\nu=0}^{3}\|B\Psi_{n\nu}\|_{\cal H}   
+ \| BI_{2}\gr \|_{\cal H} \\ 
&\le&  
3|k|^{1-\varepsilon}
\left\{ 
\frac{9}{2} |k|^{2\varepsilon -1} 
+ 2\left( 3 + 4C_{D}\left(\alpha\right)\right)
\right\}
\|\gr \|_{\cal H}.  
\end{eqnarray*}  
Noting $|k|^{2\varepsilon - 1} < 1$, 
the above inequality implies Lemma \ref{lem:IR}.

\subsection{Proof of Lemma \ref{lem:IR'}}  
\label{subsec:IR'}

The estimate of the error term $\Psi_{1}^{\mathrm er}$ 
shows how our iteration improves the moment bound. 
 
We use the same $\varepsilon$ as 
in the proof of Lemma \ref{lem:IR}, i.e., 
$$1/2 < \varepsilon < 1.$$
First, by \kak{koremo} and using 
$\gr = \omega(k)^{1/2}B^{1/2}\gr$ and 
$(\hf + 1)^{-1/2}(\hf + 1)^{1/2} = I$, 
one has 
\begin{eqnarray} 
\nonumber 
&{}& 
\| B\widetilde{I}_{1}^{j}
\left( 1 - \mf 1\right)\gr\|_{\cal H} \\ 
\nonumber 
&{}&\le  
2|k|\omega(k)^{-1/2}
\| B^{1/2}\g 1 D_{j}\left( \pf 1 - 1\right)\gr\|_{\cal H} \\ 
\nonumber 
&{}&= 
2|k|\,\, 
\| B^{1/2}\g 1 D_{j}(\hf + 1)^{-1/2}
\left( \pf 1 - 1\right)
(\hf + 1)^{1/2}
B^{1/2}
\gr\|_{\cal H} \\ 
\nonumber 
&{}&\le  
2|k|\,\, 
\| B^{1/2}\g 1 D_{j}(\hf + 1)^{-1/2}\|\,\,
\|\left( \pf 1 - 1\right)
(\hf + 1)^{1/2}
B^{1/2}
\gr\|_{\cal H}. \\ 
&{}& \qquad 
\label{add1}
\end{eqnarray}  

By \kak{hoshi} one gets 
$\left[ D_{j}\, ,\, \g 1\right] 
= \left[ p_{j}\, ,\, \g 1\right] 
= - g_{j}^{[j,1]} \g 1$. 
Thus by \kak{koremo} one has 
\begin{eqnarray} 
\nonumber 
&{}& 
\| B^{1/2}\g 1 D_{j}(\hf + 1)^{-1/2}\| \\ 
\nonumber
&{}&\le 
\| B^{1/2}D_{j}(\hf + 1)^{-1/2}\| 
+ 
|g_{j}^{[j,1]}|\,\, \|B^{1/2}\|\,\, \|(\hf + 1)^{-1/2} \g 1\| \\ 
&{}&\le  
\| B^{1/2}D_{j}(\hf + 1)^{-1/2}\| 
+ 
2|k|^{\varepsilon - 1/2}. 
\label{add2}
\end{eqnarray}  

On the other hand, by \kak{ineq-p}, \kak{maru0}, 
and \kak{i1} in the case $\e < 1$,  
one gets  
\begin{eqnarray} 
\nonumber 
&{}& 
\| B^{1/2}D_{j}(\hf + 1)^{-1/2}\| \\ 
\nonumber 
&{}&\le  
\| B (\ho + 1)^{1/2}\| 
\left( 
\| (\ho + 1)^{-1/2}p_j\|  
+ 2\sqrt{\pi}\alpha^{1/2} 
\| (\ho + 1)^{-1/2}\aklj^{*}\|
\right) \\ 
\nonumber 
&{}& \qquad 
+ 2\sqrt{\pi}\alpha^{1/2} |k|^{-1/2}
\| \aklj(\hf + 1)^{-1/2}\| \\ 
&{}&\le  
C_{D}(\alpha)|k|^{-1/2}.
\label{add3}
\end{eqnarray} 
By \kak{add1}--\kak{add3}, one obtains 
\eq{add4}
\| B\widetilde{I}_{1}^{j}\left( 
1 - \mf 1\right)\gr\|_{\cal H} 
\le 
2|k|^{1/2}\left( 
C_{D}(\alpha) + 2\right)
\| \left( 
e^{i|k|^{\varepsilon}x_{j}} - 1\right)
(\hf + 1)^{1/2}B^{1/2}\gr\|_{\cal H},  
\en 
where we used $|k|^{\varepsilon} < 1 
< C_{1}(\alpha)$.
Then, by \kak{add4} and 
using the change of variable, 
we have 
\begin{eqnarray} 
\nonumber 
&{}& 
\int_{|k|< 1}
|\gamma(k)|^{2}\| B\Psi_{1}^{\mathrm er}\|_{\cal H}^{2} 
d^{3}k \\ 
\nonumber 
&\le&  
\frac{2\pi}{2(2\pi)^{3}} 
\int_{0}^{\pi}\sin\theta d\theta 
\int_{0}^{1}\frac{1}{r}\times 
3\sum_{j=1}^{3}
\| 
\left( \G - \El + r\right)^{-1}
\widetilde{I}_{1}^{j}\left( 
1 - \mf 1 \right)\gr\|_{\cal H}^{2}dr \\ 
\nonumber 
&\le&  
\frac{3}{\pi^{2}}
\left( C_{D}\left(\alpha\right) + 2\right)^{2} 
\int_{0}^{1} 
\sum_{j=1}^{3} 
\| \left( 
e^{ir^{\varepsilon}x_{j}} - 1
\right)
\left( \hf + 1\right)^{1/2}
\left( \G - \El + r\right)^{-1/2}\gr\|_{\cal H}^{2}dr \\ 
&\le&   
\frac{3}{\pi^{2}}
\left( C_{D}\left(\alpha\right) + 2\right)^{2} 
\int_{0}^{1}S(r)dr, 
\label{3/20-5}
\end{eqnarray}   
where $S : \left[ 0 , 1\right] 
\to \left[\left. 0 , \infty\right)\right.$ 
is defined by 
\begin{eqnarray}
S(r) 
= 
\lkk \begin{array} {lc}
{\displaystyle \sum_{j=1}^{3}} 
\Biggl\|r^{-1/2} 
\left(e^{ir^{\varepsilon}x_{j}} - 1\right)
\left( \hf + 1\right)^{1/2}
\gr\Biggr\|_{\cal H}^{2}
&  \mbox{for $0 < r \le 1$}, \\ 
\qquad  & \\
0  
& \mbox{for $r = 0$.}
\end{array} \right.
\label{megane1}
\end{eqnarray}
We note the following points: 
(i) $S(r)$ can be written as an integral over $x$ 
according to 
$$
S(r) 
= 
{\displaystyle \sum_{j=1}^{3}} 
\int_{\BR}
\Biggl\| r^{-1/2}\left( 
e^{ir^{\varepsilon}x_{j}} - 1\right)
\left( \hf + 1\right)^{1/2}
\gr\Biggr\|_{\cal F}^{2}(x)d^{3}x.
$$
(ii) $\gr \in D(x^{2}) \cap D(\hf)$, and 
$x$ and $(\hf + 1)^{1/2}$ are (strongly) 
commutable in the sense of the definition 
on p.271 in \cite{rs1}. 
Thereby it can be proved that $\gr \in  D\left( 
x\left( \hf + 1\right)^{1/2}\right)$, 
see Appendix B. Thus, one has 
$$S(r) = 
r^{2(\varepsilon - 1/2)} 
\sum_{j=1}^{3} 
\Biggl\| r^{-\varepsilon}
\left(e^{i r^{\varepsilon}x_{j}} - 1\right)
\left( \hf + 1\right)^{1/2}
\gr\Biggr\|_{\cal H}^{2}
\to 
0 
$$ 
as $r\to 0$.  
On the other hand, it is easy to see 
that $S(r)$ is continuous on 
$\left(\left. 0, 1\right]\right.$.  
Thus, $S(r)$ is continuous on 
$\left[ 0 , 1\right]$, and 
\begin{eqnarray}
\nonumber 
\int_{0}^{1}
S(r)dr 
&=&  
\int_{\BR} 
\|(\hf + 1)^{1/2}\gr\|_{\cal F}^{2}(x)  
\sum_{j=1}^{3}
\int_{0}^{1}\Biggl|
\frac{e^{ir^{\varepsilon}x_{j}} - 1}{\sqrt{r}}
\Biggr|^{2}drd^{3}x \\
&{}&= 
\frac{2}{\varepsilon}
\int_{\BR} 
\|(\hf + 1)^{1/2}\gr\|_{\cal F}^{2}(x)  
\sum_{j=1}^{3}
\int_{0}^{|x_{j}|} 
\frac{1 - \cos s}{s}ds
d^{3}x
\label{3/20-6} 
\end{eqnarray}
by Fubini's theorem. 
Note 
\eq{6/24-1}  
\int_{0}^{10^{2}} 
\frac{1-\cos s}{s} ds 
\le 
3 
+ 
\sum_{n=1}^{15} 
\frac{1}{2\pi n}\int_{2\pi n}^{2\pi(n+1)} 
(1-\cos s)ds  
\le  
\gamma_{{}_{E}} + \log 15 + \frac{91}{30}, 
\en
where $\gamma_{{}_{E}}$ is the Euler number, 
$\gamma_{{}_{E}} = 0.57721\cdots$. 
Thus, altogether  
$$
\int_{0}^{|x_{j}|} 
\frac{1 - \cos s}{s}ds 
\le 
\gamma_{{}_{E}} + \log 15 + \frac{91}{30}
$$ 
for $|x_{j}| \le 10^{2}$. 
On the other hand, 
$$
\int_{0}^{|x_{j}|} 
\frac{1 - \cos s}{s}ds 
\le   
\int_{0}^{10^{2}} 
\frac{1 - \cos s}{s}ds 
+ 2\int_{10^{2}}^{|x_{j}|}\frac{1}{s}ds  
\le   
\gamma_{{}_{E}} + \log 15 + \frac{91}{30} 
+ 2\log|x| 
$$ 
for $10^{2} < |x_{j}|$. 
Thus, we have 
\eq{6/21-1}
\int_{0}^{|x_{j}|} 
\frac{1 - \cos s}{s}ds 
\le 
\max\left\{ \gamma_{{}_{E}} + \log 15 + \frac{91}{30} 
\, ,\, 
\gamma_{{}_{E}} + \log 15 + \frac{91}{30} 
+ 2\log|x|\right\}. 
\en
Set $R = 10^{2}$. Then we have 
\eq{3/20-10'} 
\int_{0}^{|x_{j}|} 
\frac{1 - \cos s}{s}ds 
= 
\left(1 - 
\chi_{{}_{R}}\left( x\right)^{2}\right) 
\int_{0}^{|x_{j}|} 
\frac{1 - \cos s}{s}ds  
+ \chi_{{}_{R}}\left( x\right)^{2}
\int_{0}^{|x_{j}|} 
\frac{1 - \cos s}{s}ds.  
\en 
Since $|x_{j}| \le |x| 
\le R$, so the integral in the first term 
of \kak{3/20-10'} 
is less than $\gamma_{{}_{E}} + \log 15 
+ \frac{91}{30}$, and we get by \kak{6/21-1} 
\begin{eqnarray} 
\nonumber  
\int_{0}^{|x_{j}|} 
\frac{1 - \cos s}{s}ds 
&\le&   
\left(\gamma_{{}_{E}} + \log 15 + \frac{91}{30}\right)
\left(1 - 
\chi_{{}_{R}}\left( x\right)^{2}\right) \\ 
\nonumber 
&{}& 
+ \chi_{{}_{R}}\left( x\right)^{2}
\max\left\{ 
\gamma_{{}_{E}} + \log 15 + \frac{91}{30} 
\, ,\, 
\gamma_{{}_{E}} + \log 15 + \frac{91}{30} 
+ 2\log |x|\right\} \\ 
\nonumber 
&\le& 
\left(\gamma_{{}_{E}} + \log 15 + \frac{91}{30}\right) 
\left(1 - 
\chi_{{}_{R}}\left( x\right)^{2}\right) \\ 
\nonumber 
&{}& 
+ 
\left(\gamma_{{}_{E}} + \log 15 + \frac{91}{30}\right)
\chi_{{}_{R}}\left( x\right)^{2} 
+ 
2\chi_{{}_{R}}\left( x\right)^{2}
\log|x| \\ 
\nonumber 
&=& 
\gamma_{{}_{E}} + \log 15 + \frac{91}{30} 
+ 2\chi_{{}_{R}}\left( x\right)^{2}
\log|x| \\ 
&\le&  
\gamma_{{}_{E}} + \log 15 + \frac{91}{30} 
+ 2\chi_{{}_{R}}\left( x\right)^{2}
\log(3 + |x|). 
\label{3/20-10}  
\end{eqnarray}

By \kak{3/20-6} and \kak{3/20-10}, 
we have 
\eq{3/20-11}    
\int_{0}^{1} S(r)dr 
\le 
\frac{6C_{1}(\alpha)^{2}}{\varepsilon}
\left(\gamma_{{}_{E}} + \log 15 + \frac{91}{30}\right) 
\|\gr\|_{\cal H}^{2} 
+ 
\frac{12}{\varepsilon} 
\|\left(\hf + 1\right)^{1/2}G_{{}_{R}}\gr\|_{\cal H}^{2}, 
\en
where $G_{{}_{R}}(x) = 
\chi_{{}_{R}}(x)\biggl(\log \left( 3 +|x|\right)\biggr)^{1/2}$.
We note 
$$G_{{}_{R}}(\G - \El)G_{{}_{R}} 
= \frac{1}{2}\left( 
|\nabla G_{{}_{R}}|^{2} + (\G -\El)G_{{}_{R}}^{2} 
+ G_{{}_{R}}^{2}(\G - \El)\right)$$ 
with \kak{log}, 
$G_{{}_{R}}^{2}\gr \in D(\G)$ from direct computations, 
and $(\G - \El)\gr = 0$. 
Then, in the same way as \kak{moment-2}, we have 
\eq{3/20-12} 
\| (\G - \El)^{1/2}G_{{}_{R}}\gr\|_{\cal H}^{2} 
= \frac{1}{2} 
\langle \gr\, ,\, |\nabla G_{{}_{R}}|^{2}\gr\rangle_{\cal H}.
\en 
Thus, inserting $(\ho + 1)^{-1/2}
(\ho + 1)^{1/2}(\G - \El + 1)^{-1/2}(\G - \El + 1)^{1/2}$ 
between $(\hf + 1)^{1/2} G_{{}_{R}}$ and 
using Lemma \ref{appendix3}, 
we have 
\begin{eqnarray} 
\nonumber 
&{}& 
\|(\hf + 1)^{1/2} G_{{}_{R}}\gr\|_{\cal H}^{2} \\ 
\nonumber 
&{}&\le   
C_{1}(\alpha)^{2}
\|(\G - \El + 1)^{1/2} G_{{}_{R}}\gr\|_{\cal H}^{2} \\ 
\nonumber 
&{}&=  
C_{1}(\alpha)^{2}
\langle G_{{}_{R}}\gr\, ,\, 
(\G - \El) G_{{}_{R}}\gr\rangle_{\cal H} 
+ 
C_{1}(\alpha)^{2}
\|G_{{}_{R}}\gr\|_{\cal H}^{2} \\ 
&{}&=  
C_{1}(\alpha)^{2}
\| (\G - \El)^{1/2} G_{{}_{R}}\gr\|_{\cal H}^{2} 
+ 
C_{1}(\alpha)^{2}
\|G_{{}_{R}}\gr\|_{\cal H}^{2}. 
\label{3/20-14}
\end{eqnarray}  
By \kak{log}, we have for $R = 10^{2}$ 
\begin{eqnarray} 
\nonumber   
\sup_{x}|\nabla G_{{}_{R}}|^{2} 
&<&  
4\times 10^{-4}\log(3 + 10^{2}) + 5\times 10^{-4} \\ 
&<& 17\times 10^{-4}.
\label{3/20-Cer}
\end{eqnarray}   
By \kak{3/20-12}, \kak{3/20-14},  
\kak{3/20-Cer}, Proposition \ref{1stmomentum} (i), 
and Lemma \ref{appendix3},  
we obtain 
\begin{eqnarray} 
\nonumber  
&{}& 
\|\left(\hf + 1\right)^{1/2}\chi_{{}_{R}}(x) 
\sqrt{\log (3 + |x|)}\gr\|_{\cal H}^{2}  \\ 
\nonumber 
&{}&\le 
C_{1}\left(\alpha\right)^{2}  
\left\{ 
L(\alpha , Z)^{2} 
+ 4\left(10^{-4} + 10^{-2}\right) 
L(\alpha, Z) + 27\times 10^{-4}/2
\right\}
\|\gr\|_{\cal H}^{2}  \\ 
&{}& \qquad 
\label{3/20-15}. 
\end{eqnarray}   
Therefore, \kak{3/20-5}, \kak{3/20-11} and \kak{3/20-15} 
imply the lemma. 
 
\section{Overlap with atomic ground state} 
\label{sec:Proof-Lemma-2}

Let $P_{\mathrm at}$ and $P_\Omega$ be 
the orthogonal projections 
onto the space spanned by $\gsat$ and $\Omega$, 
respectively. 
Set $P = P_{\mathrm at}
\o P_\Omega$ and $Q = (1- P_{\mathrm at})
\o P_\Omega$. 
We define  
\eq{new-C-tau}
C_{\tau}(\charge) = |\charge|^{2-2\tau} + 
|\charge|\sqrt{1+Z^{2}} + \charge^{2}, \quad 
3/4 < \tau \le 1. 
\en
Fix $\tau$ in $\left(\left. 3/4\, ,\, 1\right]\right.$. 
Then, there exists a positive constant 
$\airi$ such that 
\eq{AIRI} 
C_{\tau}(\airi) = 1/2. 
\en 

We prove the following lemma 
in this section.  
\bl{5} 
Fix $\tau$ in $\left(\left. 3/4\, ,\, 1\right]\right.$. 
For every charge $\charge$ satisfying $|\charge| < \min\left\{ 
\euv\, ,\, \airi\, ,\, 1\right\}$, 
and for arbitrary $\kappa, \Lambda$ with 
$0 < \kappa < \Lambda < \infty$,  
\eq{q} 
\langle\gr\, ,\, Q \gr\rangle_{\cal H}  
\le 
8\left(\frac{4\pi}{Z}\right)^{2}
F_{\mbox{\tiny IR}}(\charge) 
\|\gr\|_{\cal H}^{2},  
\en
where 
\eq{new-FIR} 
F_{\mbox{\tiny IR}}(\charge) 
=  
\sqrt{1+Z^{2}} 
\biggl( |\charge|^{4\tau - 3} + 3|\charge|^{1/2}\biggr) 
+ \charge^{2}. 
\en
\el

To find good charge dependence of the cloud 
which electron dresses, 
we develop a way to analyze the cloud 
by using Lemma \ref{key-estimate} in Appendix A. 
After this device, the following conditions work well: 
For $\rc$ of \kak{10-2-2} 
we take 
\eq{10-4-1} 
\rc = \charge^{2} 
\en 
in this section. 
So, $\lambda_{1} = 4\pi/Z$ now, 
and we assume 
\eq{10-2-assumption} 
|\charge| \le 1. 
\en
Namely, 
\eq{10-2-assumption'} 
\rc \le 1. 
\en

\hfill\break 
{\it Remark}.  
The reason why we introduce $\lambda_{1}$ is 
to avoid the trouble mentioned in the remark 
at the end of Section \ref{sec:ELH}. 
By Lemma \ref{new-forms}, we have 
$C_{*}(\alpha , \tau) \sim \sqrt{6}/(\lambda_{1}Z) 
+ o(\alpha^{1/2})$ for sufficiently small $\alpha$. 
So, we can make $C_{*}(\alpha , \tau)$ less than $1$ 
for sufficiently small $\alpha$ in case of \kak{10-4-1} 
though we still cannot control the factor 
$\sqrt{6}/(\lambda_{1}Z) = \sqrt{6}/4\pi$ by $\alpha$.    
\hfill\break

In this section we restrict $\tau$ to lie in the range   
\eq{tau-lambda-1'} 
3/4 <  \tau \le 1.
\en  

For simplicity of notation, we use the fine structure constant 
$\alpha$ for a while, rather than the charge $\charge$. 
Since the external potential $- \alpha Zr(-\tau)/|x|$ 
in $\tpap$ is of long range, as is well known, 
all negative eigenvalues of $\tpap$ have finite multiplicities, 
and they only accumulate at $0$. 
Thus we take a positive ${\mathcal E}$ such that 
\eq{epsilon} 
\tepppp = 
\teppp < - {\mathcal E} < \teqqq = 
- \,\, \frac{(\alpha Z)^{2}r(-2\tau)}{8}.
\en
Let $P_{\mathrm at}^{\mbox{\tiny $\left( \tau\right)$}}$ 
be the orthogonal projection on to the space 
spanned by $\tgsat$.  
Set $\tQ 
= (1 - P_{\mathrm at}^{\mbox{\tiny $\left( \tau\right)$}})
\o P_\Omega$. 
Then, since $U_{\tau}^{*}\G U_{\tau}$ and $\tG$ 
differ only by the multiplicative factor $r(2\tau)$, 
one has $$U_{\tau}^{*} Q U_{\tau} = \tQ.$$
Thus, 
\eq{1-overlap}
\langle \gr\, ,\, Q\gr\rangle_{\cal H} 
= \langle U_{\tau}^{*}\gr\, ,\, U_{\tau}^{*}Q 
U_{\tau}U_{\tau}^{*}\gr\rangle_{\cal H} 
= \langle \tgr\, ,\, Q^{\mbox{\tiny $\left( \tau\right)$}}
\tgr\rangle_{\cal H}.
\en 
Since $\tEl \le \teppp < - {\mathcal E} < \teqqq < 0$ 
by \kak{s-hi} and $\left[ \tho\, ,\, \tQ\right] 
= 0$, one has 
\begin{eqnarray*} 
0 &<& (- {\mathcal E} -\teppp)
\|\tQ\tgr\|_{\cal H}^2
\le
(\teqqq - \tEl)
\langle\tgr\, ,\, \tQ\tgr\rangle_{\cal H} \\ 
&\le& 
\langle \tQ\tgr\, ,\, 
(\tpap + \hf - \tEl)\tgr\rangle_{\cal H} 
= 
- \langle \tQ\tgr\, ,\, 
\thi\tgr\rangle_{\cal H},
\end{eqnarray*} 
which implies that 
$$ 
\|\tQ\tgr\|_{\cal H}^2
\le 
-\,\, \frac{\langle \tQ\tgr\, ,\, 
\thi\tgr\rangle_{\cal H}}{- {\mathcal E} + |\teppp|},
$$ 
since $- {\mathcal E} -\teppp > 0$. 
Taking ${\mathcal E} \to - \teqqq$, one gets 
\eq{q1} 
\|\tQ\tgr\|_{\cal H}^2
\le 
\frac{8\lambda_{1}^{2\tau}}{3(\alpha Z)^{2(1-\tau)}}
\langle \tQ\tgr\, ,\, 
\thi\tgr\rangle_{\cal H}. 
\en 
We will estimate the right-hand side of \kak{q1}. 
Let 
\eq{q2} 
- \langle \tQ\tgr\, ,\, 
\thi\tgr\rangle_{\cal H} 
= (i)+(ii),
\en 
where 
\begin{eqnarray} 
&{}& 
(i) = - \sqrt{4\pi\alpha}r(\tau)
\langle \tQ\tgr\, ,\, (p\cdot \takl 
+ \takl^\ast\cdot p)\tgr
\rangle_{\cal H},
\label{q2-1} \\ 
&{}&  
(ii) = - 2\pi\alpha r(2\tau) 
\langle \tQ\tgr\, ,\,  
(\takl^2 + 2\takl^\ast\cdot \takl + \takl^{\ast 2})\tgr
\rangle_{\cal H}.
\label{q2-2}
\end{eqnarray}   
Noting $\takl \Omega = 0$, in the same way as the proof 
of \cite[Lemma 4.7]{ah} we estimate $(i)$ and $(ii)$.  

Concerning $(i)$, one has by \kak{ineq-p} and \kak{maru0}  
\begin{eqnarray*}
&{}& 
|\langle \tQ\tgr\, ,\, p\cdot \takl\tgr\rangle_{\cal H}| \\ 
&{}&\leq  
\| (\tho + 1)^{1/2} \tQ\tgr\|_{\cal H}
\| (\tho + 1)^{-1/2} p\|\,\, 
\|\takl(\tho + 1)^{-1/2}\|\,\, 
\|(\tho + 1)^{1/2}\tgr\|_{\cal H} \\ 
&{}&\leq  
\frac{1}{\pi}r(-\tau)
\sqrt{2(1 - \teppp)}\,\, 
\|(\tho + 1)^{1/2} \tgr\|_{\cal H}^2,
\end{eqnarray*} 
where we used $[(\tho + 1)^{1/2}, \tQ]=0$.
Since $\takl \Omega = 0$,  
we have 
$$|\langle \tQ\tgr\, ,\,   
\takl^\ast\cdot p\tgr\rangle_{\cal H}|
= 0.$$
It follows from \kak{i1} that  
\eq{masa}
\| (\tho + 1)^{1/2}\tgr\|_{\cal H} 
= \|(\tho + 1)^{1/2} 
(\tG - \tEl + 1)^{-1/2}\tgr\|_{\cal H} 
\le 
C_{1}(\alpha , \tau)\|\tgr\|_{\cal H},  
\en
where $C_{1}(\alpha , \tau)$ is defined by \kak{C1}, 
i.e., $C_{1}(\alpha , \tau) = C_{1}(\charge , \tau)$ 
with the charge $\charge$. 
Then 
\eq{i}
|(i)| 
\leq 
\frac{2C_{1}(\alpha , \tau)^{2}}{\sqrt{\pi}} 
\Theta_{1}(\alpha)\|\onegr\|_{\cal H}^2 
\en 
follows, where 
\eq{i'}
\Theta_{1}(\alpha) 
= \alpha^{1/2}\,\,  
\sqrt{2\left( 1 - \teppp\right)}. 
\en 

Before estimating $(ii)$ we make the following. 

\hfill\break 
{\it Remark}. 
The immediate inclination is to work with revised 
atomic units, i.e., $\tau = 1$. 
Then the prefactor in \kak{q1} is $8\lambda_{1}^{2}/3Z^{2}$. 
Unfortunately, if we had invoked the standard way using Lemma 
\ref{ine}(ii) (e.g., \cite[(4.19)]{ah}), 
the matrix element would have a constant term, 
$\sqrt{6}/(\lambda_{1}Z)$, independent of $\alpha$ 
because of the same reason written in the remark after 
Lemma \ref{5}. 
As mentioned in the remark, $\lambda_{1}$ should satisfy 
$\sqrt{6}/Z < \lambda_{1}$ to make $C_{*}(\alpha,Z)$ 
less than $1$. 
Therefor, we could not make \kak{q1} small enough.  
To resolve such difficulty, we leave the position scale 
parameter $r(\tau)$ open and optimize it at the end of 
the estimate. 
\hfill\break

\bl{extra}
\eq{ii}
|(ii)|  
\le  
\frac{2C_{1}(\alpha, \tau)}{\pi}
\Theta_{2}(\alpha)\|\tgr\|_{\cal H}^{2}
\en 
follows, where 
\begin{eqnarray} 
\nonumber   
&{}& \Theta_{2}(\alpha) \\  
\nonumber 
&{}&=  
\frac{1}{2}\alpha r(2\tau) 
+ \sqrt{2}\alpha r(\tau) \\ 
\nonumber 
&{}& 
+ 
\frac{\sqrt{6}}{\e(1-\e)} 
\Biggl\{ 
\left(\sqrt{2(1 - \teppp)} 
+ \sqrt{\frac{2}{\pi}}
\alpha^{1/2}r(\tau)\right)^{2} 
\frac{\alpha^{3}}{\e(1-2\e)} 
r(-2\tau) \\ 
&{}& \qquad\qquad\qquad\qquad\qquad 
+ 
\frac{\alpha^{4}}{(1-16\e^{2})\pi} 
r(-2\tau)
\Biggr\}^{1/2}. 
\label{ii'} 
\end{eqnarray} 
\el 

\proof 
Let $\RI=\{k\in\BR||k|<1\}$ and $\RU=\BR\setminus \RI$. 
For $f \in L^{2}(\BR)$ we split $f = 
f_{\mbox{\tiny IR}} + f_{\mbox{\tiny UV}}$ 
defined by 
${\displaystyle 
f_{\mbox{\tiny {\rm IR}}}
= f\chi_{{}_{|k|<1}}}$ and  
${\displaystyle 
f_{\mbox{\tiny {\rm UV}}}
= f\chi_{{}_{|k|\ge 1}}}$, 
where 
$\chi_{{}_{|k| < 1}}$ 
is the characteristic (indicator) function 
of $\left\{ k \in \BR \, |\, |k| <1\right\}$ 
and \quad  
$\chi_{{}_{|k|\ge 1}} 
= 1-\chi_{{}_{|k|<1}}$. 

Define $\fffI=\fff(L^2(\RI))$ and $\fffU= \fff(L^2(\RU))$. 
Then there exists the unitary operator $U :\fff\longrightarrow 
\fffI\otimes \fffU$.  
We identify $\fffI\otimes \fffU$ with $\fff$ 
through this $U$, so that
$Ua^\sharp(f) U\f=a^\sharp(f_{\mbox{\tiny IR}})\o 1 
+ 1\o a^\sharp(f_{\mbox{\tiny UV}}) 
= a^\sharp(f_{\mbox{\tiny IR}}) 
+ a^\sharp(f_{\mbox{\tiny UV}})$. 
Set 
\eq{hfiu}
\hfi=\int_{|k| < 1}\omega(k) 
a^{*}(k) a(k) d^{3}k,\ \ \ 
\hfu=\int_{|k|\ge 1}\omega(k)
a^{*}(k) a(k) dk,
\en 
and similarly,  
\eq{nfiu}
\nfi=\int_{|k|<1}\add(k) a(k) d^{3}k,\ \ \ 
\nfu=\int_{|k|\ge 1}\add(k) a(k) d^{3}k.
\en
We set  
\eq{Nrho'} 
N_{\rho} := 
\int_{|k|\ge 1}\omega(k)^{\rho} 
a^{*} (k)a(k)d^{3}k, 
\quad \rho \ge 0.
\en 
So, $N_{0} = \nfu$ and 
$N_{1} = \hfu$.

Since $\takl = a(f^{\tai}) = 
a(f^{\tai}_{\mbox{\tiny IR}}) + 
a(f^{\tai}_{\mbox{\tiny UV}})$, 
we have 
\begin{eqnarray*} 
&{}& 
|\langle \tQ\tgr\, ,\, 
\takl^2\tgr\rangle_{\cal H}| \\ 
&{}&\le  
|\langle \tQ\tgr\, ,\, 
a(f^{\tai}_{\mbox{\tiny IR}})^2\tgr\rangle_{\cal H}|
+ 
2|\langle \tQ\tgr\, ,\, 
a(f^{\tai}_{\mbox{\tiny IR}})a(f^{\tai}_{\mbox{\tiny UV}})
\tgr\rangle_{\cal H}| \\ 
&{}& \qquad
+ 
|\langle \tQ\tgr\, ,\, 
a(f^{\tai}_{\mbox{\tiny UV}})^2\tgr\rangle_{\cal H}| \\ 
&{}&=  
|\langle a(f^{\tai}_{\mbox{\tiny IR}})^{*}\tQ\tgr\, ,\, 
a(f^{\tai}_{\mbox{\tiny IR}})\tgr\rangle_{\cal H}|
+ 
2|\langle a(f^{\tai}_{\mbox{\tiny IR}})^{*}\tQ\tgr\, ,\, 
a(f^{\tai}_{\mbox{\tiny UV}})
\tgr\rangle_{\cal H}| \\ 
&{}& \qquad 
+ 
|\langle \tQ\tgr\, ,\, 
a(f^{\tai}_{\mbox{\tiny UV}})^2\tgr\rangle_{\cal H}|.  
\end{eqnarray*} 
We note that by Lemma \ref{hirohiro3}  
\begin{eqnarray*} 
&{}& 
|\langle a(f^{\tai}_{\mbox{\tiny IR}})^{*}\tQ\tgr\, ,\, 
a(f^{\tai}_{\mbox{\tiny IR}})\tgr\rangle_{\cal H}| \\ 
&{}&\le  
\|a(f^{\tai}_{\mbox{\tiny IR}})^{*}(\hf + 1)^{-1/2}\|\,\, 
\| (\hf + 1)^{1/2}\tQ\tgr\|_{\cal H} 
\|a(f^{\tai}_{\mbox{\tiny IR}})(\hf + 1)^{-1/2}\|\,\,  \\ 
&{}& \qquad \times 
\|(\hf + 1)^{1/2}(\tho + 1)^{-1/2}\|
\|(\tho + 1)^{1/2}(\tG - \tEl + 1)^{-1/2}\| \\ 
&{}& \qquad\qquad\qquad \times  
\| (\tG - \tEl + 1)^{1/2}\tgr\|_{\cal H} \\ 
&{}&\le  
C_{1}(\alpha, \tau)
\left(\|f^{\tai}_{\mbox{\tiny IR}}/\sqrt{\omega}\|_{L^{2}} 
+ \|f^{\tai}_{\mbox{\tiny IR}}\|_{L^{2}}
\right)\|f^{\tai}_{\mbox{\tiny IR}}/\sqrt{\omega}\|_{L^{2}}
\|\tgr\|_{\cal H}^{2} \\ 
&{}&\le  
\frac{C_{1}(\alpha, \tau)}{2\pi^{2}}
\|\tgr\|_{\cal H}^{2},
\end{eqnarray*} 
and 
\begin{eqnarray*} 
&{}& 
|\langle a(f^{\tai}_{\mbox{\tiny IR}})^{*}\tQ\tgr\, ,\, 
a(f^{\tai}_{\mbox{\tiny UV}})\tgr\rangle_{\cal H}| \\ 
&{}&\le  
\|a(f^{\tai}_{\mbox{\tiny IR}})^{*}(\hf + 1)^{-1/2}\|\,\, 
\| (\hf + 1)^{1/2}\tQ\tgr\|_{\cal H} 
\|a(f^{\tai}_{\mbox{\tiny UV}})(\hf + 1)^{-1/2}\|\,\,  \\ 
&{}& \qquad \times 
\|(\hf + 1)^{1/2}(\tho + 1)^{-1/2}\|\,\, 
\|(\tho + 1)^{1/2}(\tG - \tEl + 1)^{-1/2}\| \\ 
&{}& \qquad\qquad\qquad \times  
\| (\tG - \tEl + 1)^{1/2}\tgr\|_{\cal H} \\ 
&{}&\le  
C_{1}(\alpha, \tau)
\left(\|f^{\tai}_{\mbox{\tiny IR}}/\sqrt{\omega}\|_{L^{2}} 
+ \|f^{\tai}_{\mbox{\tiny IR}}\|_{L^{2}}
\right)\|f^{\tai}_{\mbox{\tiny UV}}/\sqrt{\omega}\|_{L^{2}}
\|\tgr\|_{\cal H}^{2} \\ 
&{}&\le  
\frac{r(-\tau)C_{1}(\alpha, \tau)}{\sqrt{2}\pi^{2}}
\|\tgr\|_{\cal H}^{2},
\end{eqnarray*} 
where we used \kak{maru5}, \kak{maru6}, \kak{maru6'}, 
\kak{i1} and the fact that 
$(\hf + 1)^{1/2}\tQ\tgr = \tQ\tgr$. 
So, one has
\begin{eqnarray} 
\nonumber  
&{}& 
|\langle \tQ\tgr\, ,\, 
\takl^2\tgr\rangle_{\cal H}|  \\ 
&{}&\le 
\frac{C_{1}(\alpha,\tau)}{\pi^{2}}
\left(\frac{1}{2} + \sqrt{2}r(-\tau)\right)
\|\tgr\|_{\cal H}^{2} 
+ 
|\langle \tQ\tgr\, ,\, 
a(f^{\tai}_{\mbox{\tiny UV}})^2\tgr\rangle_{\cal H}| 
\label{6/3-1}
\end{eqnarray} 

Before estimating $|\langle \tQ\tgr\, ,\, 
a(f^{\tai}_{\mbox{\tiny UV}})^2\tgr\rangle_{\cal H}|$, 
we introduce some notations to profit from 
Lemma \ref{lem:UV-t} below improving Lemma \ref{lem:UV} 
in the case $0 \le \tau \le 1$. 

Let $$h_{\tau,x}(k)= 2^{-1/2}k\beta_{\tau}(k) 
e^{-ir(2\tau) k\cdot x}
 \omega(k)^{-1/2}\widehat{\varphi}_{\tau}(k).$$ 
From the pull-through formula one concludes 
\eq{pull-through-t} 
a(k) \tgr = 
- \sqrt{4\pi\alpha}\,\, r(\tau)
B_{\tau} h_{\tau,x}(k)\cdot 
\left( p + \alpha^{1/2}r(\tau)\takl 
+ \alpha^{1/2}r(\tau)\takl^\ast\right) \tgr, 
\en 
where 
$$B_{\tau} = 
(\tG - \tEl + \od(k))^{-1}.$$

By Lemma \ref{key-estimate} (i) and (ii) and 
using $(\nf + 1)^{1/2}\tQ\tgr = \tQ\tgr$, 
one has 
\begin{eqnarray*} 
&{}& 
|\langle \tQ\tgr\, ,\, 
a(f^{\tai}_{\mbox{\tiny UV}})^2\tgr\rangle_{\cal H}| \\ 
&{}&=  
|\langle (\nfu + 1)^{1/2}\tQ\tgr\, ,\, 
(\nfu + 1)^{-1/2} a(f^{\tai}_{\mbox{\tiny UV}})^2
N_{2\e}^{-1} 
(\nfu + 1)^{1/2}(\nfu + 1)^{-1/2}N_{2\e}
\tgr\rangle_{\cal H}| \\ 
&{}&\le  
\|(\nfu + 1)^{-1/2} 
a(f^{\tai}_{\mbox{\tiny UV}})^2
N_{2\e}^{-1} 
(\nfu + 1)^{1/2} \|\,\, 
\| (\nfu + 1)^{1/2} \tQ\tgr\|_{\cal H} \\ 
&{}& \qquad \times
\| (\nfu + 1)^{-1/2}N_{2\e}(N_{4\e} + \varepsilon)^{-1/2}\|\,\, 
\|(N_{4\e} + \varepsilon)^{1/2}\tgr\|_{\cal H}  \\ 
&{}&\le 
\sqrt{3}\|f^{\tai}_{\mbox{\tiny UV}}/\omega^{\e}\|_{L^{2}}^{2} 
\|\tgr\|_{\cal H} 
\left( 
\|N_{4\e}^{1/2}\tgr\|_{\cal H}^{2} 
+ \varepsilon \|\tgr\|_{\cal H}^{2} 
\right)^{1/2}
\end{eqnarray*}  
for arbitrary $\varepsilon > 0$. 
So, taking $\varepsilon \searrow 0$, 
one gets 
\eq{6/4-1} 
|\langle \tQ\tgr\, ,\, 
a(f^{\tai}_{\mbox{\tiny UV}})^2
\tgr\rangle_{\cal H}| 
\le 
\sqrt{3}\|f^{\tai}_{\mbox{\tiny UV}}/\omega^{\e}\|_{L^{2}}^{2} 
\|\tgr\|_{\cal H} 
\|N_{4\e}^{1/2}\tgr\|_{\cal H}. 
\en 
On the other hand, one has 
\eq{6/4-2}
\|N_{4\e}^{1/2}\tgr\|_{\cal H}^{2} 
= \langle \tgr\, ,\, 
N_{4\e}\tgr\rangle_{\cal H} 
= 
\int_{|k| \ge 1}|k|^{4\e}\|a(k)\tgr\|_{\cal H}^{2}d^{3}k. 
\en

By \kak{pull-through-t} and Lemma \ref{lem:UV-t} below, 
one has 
\begin{eqnarray} 
\nonumber 
&{}& 
\int_{|k| \ge 1}|k|^{4\e}\|a(k)\tgr\|_{\cal H}^{2}   
d^{3}k \\ 
\nonumber 
&{}&\le  
8\pi\alpha r(2\tau)C_{1}(\alpha, \tau)^{2} 
\Biggl[ 
\left(\sqrt{2(1 - \teppp)} + 
\sqrt{\frac{2}{\pi}}\alpha^{1/2}r(\tau)\right)^{2} 
\int_{|k| \ge 1}|k|^{4\e} 
\frac{|h_{\tau,0}(k)|^{2}}{\omega(k)^{2}}
d^{3}k \\ 
&{}& \qquad 
+ \frac{2\alpha r(2\tau)}{\pi}    
\int_{|k| \ge 1}|k|^{4\e} 
\frac{|h_{\tau,0}(k)|^{2}}{\omega(k)}
d^{3}k\Biggr].
\label{6/4-3}
\end{eqnarray}  
On the other hand, 
\begin{eqnarray} 
\nonumber 
&{}& 
\int_{|k| \ge 1}|k|^{4\e} 
\frac{|h_{\tau,0}(k)|^{2}}{\omega(k)^{2}}
d^{3}k
\le 
\frac{4\pi}{2(2\pi)^{3}} 
\int_{1}^{\infty} 
s^{1+4\e}\left( s + \frac{r(2\tau)}{2}s^{2}\right)^{-2} ds \\ 
\nonumber 
&{}&=  
\frac{1}{4\pi^{2}} 
\int_{1}^{\infty} 
s^{4\e -1}\left( 1 + \frac{r(2\tau)}{2}s\right)^{-2} ds 
= 
\frac{2^{4\e -2}}{\pi^{2}} 
r(- 8\e\tau) 
\int_{r(2\tau)/2}^{\infty} 
s^{4\e}(1+s)^{-2}ds \\ 
\nonumber 
&{}&\le  
\frac{2^{4\e -2}}{\pi^{2}} 
r(- 8\e\tau) 
\left( 
\int_{0}^{1} s^{4\e -1}(1+s)^{-2}ds 
+ 
\int_{1}^{\infty} s^{4\e -1}(1+s)^{-2}ds
\right) \\ 
&{}&\le  
\frac{1}{2^{4(1-\e)}\e(1-2\e))\pi^{2}}
r(- 8\e\tau),
\label{6/4-4}
\end{eqnarray} 
where we note that we assumed 
$\e < 1/4$. 
In the same way, 
\begin{eqnarray} 
\int_{|k| \ge 1}|k|^{4\e} 
\frac{|h_{\tau,0}(k)|^{2}}{\omega(k)}
d^{3}k
\le 
\frac{2^{4\e -1}}{(1-16\e^{2})\pi^{2}}
r(- 2\tau - 8\e\tau),
\label{6/4-5}
\end{eqnarray}  
where we note that we assumed $\e < 1/4$ again. 
Thus, by \kak{6/4-2}-\kak{6/4-5}, 
one gets 
\begin{eqnarray} 
\nonumber 
&{}&  
\|N_{4\e}^{1/2}\tgr\|_{\cal H} \\ 
\nonumber 
&{}&\le  
\sqrt{2}C_{1}(\alpha, \tau) 
\Bigg\{ 
\left( \sqrt{2(1-\teppp)} + 
\sqrt{\frac{2}{\pi}}\alpha^{1/2}r(\tau)\right)^{2} 
\frac{2^{4\e-2}}{\e(1-2\e)\pi}
\alpha r(2\tau - 8\e\tau) \\ 
&{}& \qquad 
+ 
\frac{2^{4\e +3}}{(1-16\e^{2})\pi^{2}}
\alpha^{2}r(2\tau - 8\e\tau)
\Biggr\}^{1/2} 
\|\tgr\|_{\cal H}.  
\label{6/4-6}
\end{eqnarray} 
Moreover we have 
\eq{6/4-7} 
\| f^{\tai}_{\mbox{\tiny UV}}/\omega^{\e}\|_{L^{2}}^{2} \\ 
\le  
\frac{1}{4\pi^{2}}
\int_{1}^{\infty} r^{1-2\e}
\left( 1 + \frac{r(2\tau)}{2}r\right)^{-2}dr 
\le  
\frac{1}{2^{2\e +1}\e(1-\e)\pi^{2}} 
r(-4\tau + 4\e\tau). 
\en 
Therefore, by \kak{6/4-1}, \kak{6/4-6} 
and \kak{6/4-7}, 
one has 
\begin{eqnarray} 
\nonumber 
&{}&  
|\langle \tQ\tgr\, ,\, 
a(f^{\tai}_{\mbox{\tiny UV}})^{2}\tgr\rangle_{\cal H}| \\ 
\nonumber 
&{}&\le  
\frac{\sqrt{6}C_{1}(\alpha,\tau)}{
\e(1-\e)\pi^{2}} 
\Biggl\{ 
\left(\sqrt{2(1-\teppp)} 
+ \sqrt{\frac{2}{\pi}}\alpha^{1/2}r(\tau)
\right)^{2} 
\frac{\alpha}{\e(1-2\e)} 
r(-6\tau) \\ 
&{}& \qquad 
+ 
\frac{\alpha^{2}}{(1-16\e^{2})\pi}
r(-6\tau)
\Biggr\}^{1/2}\|\tgr\|_{\cal H}^{2}. 
\label{6/3-2}
\end{eqnarray}  
Since $\takl \Omega = 0$,
$$|\langle \tQ\tgr\, ,\, 
\takl^{\ast 2}\tgr\rangle_{\cal H}| 
= 0,$$
and 
$$
|\langle \tQ\tgr\, ,\, \takl^\ast \takl 
\tgr\rangle_{\cal H}| = 0.
$$
Hence \kak{ii} follows from \kak{6/3-1} and \kak{6/3-2}. 
\qed

Therefore, by \kak{q1}, \kak{q2} with 
\kak{q2-1} and \kak{q2-2}, \kak{i} 
with \kak{i'}, and \kak{ii} with \kak{ii'}, 
one obtains that 
\begin{eqnarray} 
\nonumber  
&{}& 
\|\tQ\tgr\|_{\cal H}^{2} \\ 
\nonumber 
&{}&\le  
\frac{16\lambda_{1}^{2}C_{1}(\alpha, \tau)^{2}}{3\sqrt{\pi}} 
\sqrt{2(1-\teppp)}\alpha^{1/2}
r(2\tau - 2)\|\tgr\|_{\cal H}^{2}   \\ 
\nonumber 
&{}& 
+ 
\frac{16\lambda_{1}^{2}C_{1}(\alpha, \tau)}{3\pi}
\Bigg[ 
\frac{1}{2}\alpha r(4\tau - 2) 
+ \sqrt{2}\alpha r(3\tau - 2) \\ 
\nonumber 
&{}& 
+  
\frac{\sqrt{6}}{\e(1-\e)} 
\Bigg\{
\left( \sqrt{2\left( 1-\teppp\right)} 
+ \sqrt{\frac{2}{\pi}}
\alpha^{1/2}r(\tau)\right)^{2} 
\frac{\alpha^{3}}{\e(1-2\e)} 
r(2\tau - 4) \\ 
&{}& \qquad\qquad\qquad\qquad 
+ \frac{\alpha^{4}}{(1-16\e^{2})\pi} 
r(2\tau - 4)
\Biggr\}^{1/2} 
\Biggr] 
\|\tgr\|_{\cal H}^{2}.  
\label{cloud}
\end{eqnarray}

Take $\e = 1/5$. 
Then, for $3/4 < \tau \le 1$ and $|\charge| < 
\min\left\{ 
\euv\, ,\, \airi\, ,\, 1\right\}$, 
it is easy to see that 
\eq{10-10-1}
C_{*}(\charge, \tau) 
= 
\frac{\sqrt{6}}{4\pi}\charge^{2(1-\tau)} 
+ \frac{2|\charge|}{\pi}\sqrt{1-\teppp} 
+ \frac{\charge^{2}}{2\pi^{2}} 
\left(\charge^{4\tau} + 3\charge^{2\tau} 
+ 3\right)  
\le C_{\tau}(\charge) \le 1/2
\en 
since 
\eq{10-10-2} 
1 - \teppp \le 1 + \frac{Z^{2}}{32\pi^{2}}.
\en 
Therefore, by \kak{1-overlap}, \kak{cloud}, 
\kak{10-10-1}, and \kak{10-10-2}, 
we have 
\begin{eqnarray*}
&{}& |\langle\gr\, ,\, Q\gr\rangle_{\cal H}| \\ 
&{}&\le 
4\left(\frac{4\pi}{Z}\right)^{2} 
\frac{16}{6\pi}|\charge|^{4\tau-3} 
\sqrt{2(1+Z^{2})}\|\tgr\|_{\cal H}^{2} \\ 
&{}&+ 
4\left(\frac{4\pi}{Z}\right)^{2} 
\frac{8}{3\pi}
\Biggl[ \left(\frac{1}{2} +\sqrt{2}\right)
\frac{\charge^{2}}{4\pi} \\ 
&{}&\qquad\qquad\qquad
+ \frac{25\sqrt{6}}{4}\Biggl\{ 
\left(\sqrt{2(1+Z^{2})} + \frac{1}{\sqrt{2}\pi}\right)^{2} 
\frac{25|\charge|^{4\tau-2}}{3(4\pi)^{3}}  
+ \frac{25|\charge|^{4\tau - 2}}{9\pi(4\pi)^{3}}
\Biggr\}^{1/2}\Biggr]\|\tgr\|_{\cal H}^{2} \\ 
&{}&\le 
8\left(\frac{4\pi}{Z}\right)^{2}
\sqrt{1+Z^{2}} |\charge|^{4\tau-3} 
\|\tgr\|_{\cal H}^{2} \\ 
&{}&+ 
8\left(\frac{4\pi}{Z}\right)^{2} 
\frac{4}{9}
\Biggl[ \frac{1}{6}|\charge|^{2} 
+ \frac{75}{4}\Biggl\{ 
\frac{25}{3}\left(\sqrt{2(1+Z^{2})} + 1\right)^{2} 
\frac{|\charge|}{(4\pi)^{3}}  
+ \frac{25|\charge|}{27(4\pi)^{3}}
\Biggr\}^{1/2}\Biggr]\|\tgr\|_{\cal H}^{2} \\ 
&{}&\le 
8\left(\frac{4\pi}{Z}\right)^{2}
\sqrt{1+Z^{2}} |\charge|^{4\tau-3} 
\|\tgr\|_{\cal H}^{2} \\ 
&{}&+ 
8\left(\frac{4\pi}{Z}\right)^{2} 
\frac{4}{9}
\Biggl[ \frac{1}{6}|\charge|^{2} 
+ \frac{75}{4}\Biggl\{ 
\frac{25}{3}\,\,\frac{1}{12^{3}}\,\, 
2^{3}\left(1+Z^{2}\right)|\charge|  
+ \frac{25}{27}\,\,\frac{1}{12^{4}}|\charge|
\Biggr\}^{1/2}\Biggr]\|\tgr\|_{\cal H}^{2}, 
\end{eqnarray*}
where we used $1 \le \sqrt{2(1+Z^{2})}$ to get 
the last inequality, 
which implies Lemma \ref{5}. 

At the end of this section, we still have to supply 

\begin{lem}
\label{lem:UV-t} 
For arbitrary $\rc > 0$, 
$0 < \kappa, \Lambda < \infty$, 
and for every $\alpha$ with $\alpha < \euv^{2}/4\pi$, 
it holds 
\begin{eqnarray}
\nonumber 
&{}& 
\| (\tG - \tEl + \omega(k))\f  {h}_{\tau,x}({k})\cdot 
\left( p + \sqrt{4\pi\alpha}\,\, r(\tau)\takl
+ \sqrt{4\pi\alpha}\,\, r(\tau)\takl^{*} 
\right)\tgr\|_{\cal H}  \\ 
&{}& 
\nonumber 
\le 
C_{1}(\alpha, \tau)
\Biggl\{ \left(\sqrt{2(1-\teppp)} 
+ \sqrt{2/\pi}\,\,\alpha^{1/2}r(\tau)\right)
\frac{|h_{\tau, 0}(k)|}{\omega(k)} \\ 
&{}& 
\qquad\qquad\qquad 
+ \sqrt{2/\pi}\,\,\alpha^{1/2}r(\tau) 
\frac{|h_{\tau, 0}(k)|}{\sqrt{\omega(k)}} 
\Biggr\} 
\|\tgr\|_{\cal H}
\label{eq:UV1-t}
\end{eqnarray}
in the UV region, $|k| \ge 1$. 
\end{lem} 

\proof 
By using $(\tG - \tEl + 1)^{1/2}\tgr = \tgr$, 
and applying \kak{i1} in the case $\e > 1$, 
it follows from \kak{ineq-p} and 
\kak{maru0} that 
\begin{eqnarray*} 
&{}& 
\| B_{\tau}{h}_{\tau,x}({k})\cdot 
\left( p + \sqrt{4\pi\alpha}\,\, r(\tau)\takl
\right)\tgr \|_{\cal H} \\ 
&\le&  
|{h}_{\tau,0}({k})|\omega(k)^{-1}
\|\left( p + \sqrt{4\pi\alpha}\,\, r(\tau)\takl\right)
(\tho + 1)^{-1/2}\| \\ 
&{}& \qquad \times 
\|(\tho + 1)^{1/2}
(\tG - \tEl + 1)^{-1/2}\|\,\, 
\|(\tG - \tEl + 1)^{1/2}
\tgr \|_{\cal H} \\ 
&\le&   
|{h}_{\tau,0}({k})|\omega(k)^{-1} 
\left\{ \sqrt{2(1 - \teppp)} 
+ \sqrt{2/\pi}\,\,\alpha^{1/2}r(\tau)
\right\} 
C_{1}(\alpha, \tau) 
\|\tgr \|_{\cal H}. 
\end{eqnarray*}  
Moreover, by using \kak{maru0} 
and \kak{i1} in the case $\e > 1$ again,  
\begin{eqnarray*} 
&{}& 
\| B_{\tau}{h}_{\tau,x}({k})\cdot  
\sqrt{4\pi\alpha}\,\, r(\tau)\akl^{*}\tgr \|_{\cal H}  \\ 
&\le&  
\sqrt{4\pi\alpha}\,\, r(\tau)\|B_{\tau}^{1/2}\|\,\, 
\|B_{\tau}^{1/2} 
{h}_{\tau,x}({k})\cdot  
\takl^{*}\tgr \|_{\cal H} \\ 
&\le&  
\sqrt{4\pi\alpha}\,\, r(\tau) 
\omega(k)^{-1/2} 
\|B_{\tau}^{1/2}(\tho + 1)^{1/2}(\tho + 1)^{-1/2}\| \\ 
&{}& \qquad \times  
\|(\hf + 1)^{1/2}(\hf + 1)^{-1/2} 
{h}_{\tau,x}({k})\cdot  
\takl^{*}\tgr \|_{\cal H} \\ 
&\le&   
C_{1}(\alpha, \tau)
\sqrt{4\pi\alpha}\,\, r(\tau) 
\omega({k})^{-1/2} 
\| (\tho + 1)^{-1/2}(\hf + 1)^{1/2}\| \\ 
&{}& \qquad \times  
|{h}_{\tau,0}({k})| \,\, 
\| (\hf + 1)^{-1/2} 
\takl^{*}\| \,\, 
\|\tgr \|_{\cal H} \\ 
&\le&  
C_{1}(\alpha, \tau)
|{h}_{\tau,0}({k})|\omega({k})^{-1/2} 
\sqrt{2/\pi}\,\,\alpha^{1/2}r(\tau) 
\|\tgr \|_{\cal H}. 
\end{eqnarray*} 
Thus we infer \kak{eq:UV1-t}, and the lemma follows.
\qed

\section{Proof of the main theorem} 
\label{sec:open}

We prove Theorem \ref{new-main2}. 
By taking $\varepsilon = 3/4$, so 
$\delta = 1/4$ in Corollaries \ref{cor:1} 
and \ref{cor:2}, we deduce the following 
bound on the number photons in the ground state.

\bl{over} 
There exist positive constants $c_{1}$, $c_{2}$ 
independent of $\charge$, $\kappa$, and  $\Lambda$ 
such that   
\begin{eqnarray*}
&{}& \|\nf^\han \gr\|_{\cal H}^{2} \\  
&{}&\leq 
\Biggl[ 
\biggl( 28C_{D}(\charge) + 39\biggr)^{2} \\ 
&{}& \qquad 
+ 6C_{1}(\charge, 0)^{2}
\biggl( C_{D}(\charge) + 2\biggr)^{2}
\biggl( 9 + 2L(\charge, Z)^{2} 
+ 9\times 10^{2}L(\charge, Z)\biggr)
\Biggr]\,\, \frac{\charge^{2}}{4\pi} \|\gr\|_{\cal H}^{2}
\end{eqnarray*}
for every $\charge$ with $|\charge| < \euv$ and 
arbitrary $\kappa, \Lambda$ with 
$0 < \kappa < \Lambda < \infty$, 
where 
$$L(\charge, Z) = \log\left( 
3 + \frac{400\pi}{\charge^{2}Z}\right)$$ 
and 
$$C_{D}(\charge) =  
\left(1-C_{\mathrm UV}\left(\charge\right)\right)^{-1/2}
\left(
\sqrt{2 + \left(\frac{\charge^{2}Z}{4\pi}\right)^{2}}
+ \frac{\sqrt{2}|\charge|}{\pi}\right).
$$ 
\el

We recall the overlap estimate from Lemma \ref{5}: 
fix $\tau$ in $\left(\left. 3/4\, ,\, 1\right]\right.$. 
There exists a positive constant $\airi$ such that 
for every charge $\charge$ satisfying $|\charge| 
< \min\left\{ \euv\, ,\, \airi\, ,\, 1\right\}$, 
and for arbitrary $\kappa, \Lambda$ with 
$0 < \kappa < \Lambda < \infty$,  
\eq{q'} 
\langle\gr\, ,\, Q \gr\rangle_{\cal H}  
\le 
8\left(\frac{4\pi}{Z}\right)^{2}
F_{\mbox{\tiny IR}}(\charge) 
\|\gr\|_{\cal H}^{2},  
\en
where $F_{\mbox{\tiny IR}}(\charge)$ 
is defined in \kak{new-FIR}. 
We introduce the positive constant 
$\airii$ by 
\eq{AIRII} 
F_{\mbox{\tiny IR}}(\airii) = 
\frac{1}{16}
\left(\frac{Z}{4\pi}\right)^{2}. 
\en
Then, we can prove  Theorem \ref{new-main2} as follows:

We consider normalized $\gr$, i.e., 
$\|\gr\|_{\cal H} = 1$. 
Then, there is a subsequence 
$\left\{ \gr\right\}_{\kappa,\Lambda}$ 
such that 
\eq{ourgs}
\mbox{ w-}\lim_{\kappa\to 0}
\lim_{\Lambda\to\infty}\gr = \psi_{\mathrm g} 
\in {\cal H}.
\en
We establish that this $\psi_{\mathrm g}$ 
is a ground state for $H$ of \kak{9/4-(1)}.
Because of \kak{convgse1}, we only have to show 
that $\psi_{\mathrm g} \ne 0$, see \cite[Lemma 4.9]{ah}, 
which will be obtained by proving that the overlap 
$\langle \psi_{\mathrm g}\, ,\, 
\gsat\otimes\Omega\rangle_{\cal H} \ne 0$ 
for sufficiently small $|\charge|$. 

We have 
\eq{5/17-1} 
|\langle \gsat\o \Omega\, ,\, \gr\rangle_{\cal H}|^{2} 
= \langle \gr\, ,\, P\gr\rangle_{\cal H}, 
\en  
and, by Lemma \ref{over} and \kak{q'},  
\begin{eqnarray} 
\nonumber 
\langle \gr\, ,\, P\gr\rangle_{\cal H} 
&=&  
\langle \gr\, ,\, 1\o P_{\Omega}\gr \rangle_{\cal H} 
- 
\langle \gr\, ,\, Q\gr \rangle_{\cal H} \\  
\nonumber 
&\ge&  
1 -  \langle \gr\, ,\, 1\o \nf\gr \rangle_{\cal H} 
- \langle \gr\, ,\, Q\gr \rangle_{\cal H} \\ 
&\ge& 
1 - c_{0}\alpha^{1/2} 
- 8\left(\frac{4\pi}{Z}\right)^{2}
F_{\mbox{\tiny IR}}(\charge)
=: G_{\mbox{\tiny IR}}(\charge), 
\label{5/17-2}
\end{eqnarray}  
where $c_{0}$ is the constant in the photon number bound 
from Lemma \ref{over}. 
Thus, finally we set 
\eq{eIR}
\eir = \min\left\{ 
\sqrt{\pi}/c_{0}\, ,\, 1\, ,\, 
\airi\, ,\, 
\airii
\right\}. 
\en 
Then we have 
\eq{final}
|\langle \gsat\o \Omega\, ,\, \psi_{\mathrm g}
\rangle_{\cal H}|^{2} 
\ge G_{\mbox{\tiny IR}}(\charge) > 0\qquad  
\mbox{for $0 < |\charge| < \min\left\{\eir\, ,\, \euv\right\}$.} 
\en

\section{Outlook and open problems}
\label{sec:outlook} 

Theorem \ref{new-main2} does not touch the obvious question 
concerning the uniqueness 
of the ground state. 
To prove such a property the natural method is to establish 
that the semigroup $e^{-tH}$, $t > 0$, is positivity improving 
in the Schr\"{o}dinger representation. 
While this property holds formally, 
a complete proof is still under construction. 

The existence of the ground state as such 
provides little information on the binding energy, 
except for the lower bound of Proposition \ref{binding-energy}. 
However on a formal level information is 
available and we discuss it with the hope 
that rigorous bounds will 
be supplied in the future 
(we mention that such bounds are available 
for the quantized Maxwell field \cite{hainzl}, 
which however diverge with $\Lambda\to\infty$). 
Energies are in units of the bare mass $m (= mc^{2})$. 
By definition the (positive) binding energy 
is 
\eq{S-4.1} 
- E_{\mathrm bin} = m(E_{\mathrm g} - E_{0}), 
\en 
where $E_{\mathrm g}$ is the ground state energy of $H$ 
and $E_{0}$ the infimum of the spectrum of $H$ with $Z = 0$. 
We regard $\charge$ as a small parameter. 
It enters in the coupling to the Bose field through 
$\charge A$ and in the strength $\charge^{2}Z$ of 
the Coulomb potential. 
The latter is kept fixed and our strategy is to expand 
$E_{\mathrm g}$ in the former up to order $\charge^{2}$.
Since $\eppp = {\mathcal O}(\charge^{4})$, thereby 
$E_{\mathrm g}$ is determined to order $\charge^{6}$. 
This means $E_{0}$ has to be expanded also to order $\charge^{6}$. 
Taking the difference in \kak{S-4.1} all terms cancel except for 
\eq{10-21-1} 
E^{(6)} = - \frac{\charge^{6}}{6}
\langle\Omega , 
A_{0}\cdot A_{0}\frac{1}{H_{1}}
(P_{\mathrm f}\cdot A_{0} 
+ A_{0}^{*}\cdot P_{\mathrm f})\frac{1}{H_{1}}
(P_{\mathrm f}\cdot A_{0} 
+ A_{0}^{*}\cdot P_{\mathrm f})\frac{1}{H_{1}}
A_{0}^{*}\cdot A_{0}^{*}\Omega
\rangle_{\mathcal F}, 
\en
where $P_{\mathrm f} = \int k a^{*}(k)a(k)d^{3}k$ 
is the total momentum of the bosons, 
$A_{0}$ is $A$ of \kak{9/4-(2)} at $x = 0$, and 
$H_{1} = \hf + \frac{1}{2}P_{\mathrm f}^{2}$. 
The net result is 
\begin{eqnarray} 
\nonumber 
- E_{\mathrm bin} 
&=& 
m\Biggl( E_{\mathrm at} - \charge^{2}(2\pi)^{-3} 
\int_{\BR}(2\omega)^{-1}\beta^{2} 
\langle k\cdot p \psi_{\mathrm at} , \\ 
&{}& 
e^{ik\cdot x}(\pap - \eppp + \omega)^{-1} 
e^{-ik\cdot x} k\cdot p \psi_{\mathrm at}
\rangle_{L^{2}}
d^{3}k  
+ E^{(6)} 
+ {\mathcal O}(\charge^{8})\Biggr).
\label{S-4.9}
\end{eqnarray}
We now use that 
\eq{S-4.11} 
e^{ik\cdot x} (\pap - \eppp + \omega)^{-1} 
e^{-ik\cdot x} 
=  (\pap - \eppp + \omega 
+ \frac{1}{2}k^{2} - k\cdot p)^{-1}.  
\en 
Expanding the resolvent in $k\cdot p$, 
each order picks up an factor $\charge^{2}$ 
through taken the matrix element with $\gsat$. 
Thus 
\begin{eqnarray} 
\nonumber 
- E_{\mathrm bin} 
&=& 
m\Biggl( \eppp - \charge^{2}\frac{1}{3}(2\pi)^{-3} 
\int_{\BR}(2\omega)^{-1}\beta^{2}k^{2}\langle p\gsat , \\ 
&{}& 
(\pap - \eppp + \omega + \frac{1}{2}k^{2})^{-1}
\cdot p\gsat\rangle_{L^{2}}d^{3}k 
+ E^{(6)} + {\mathcal O}(\charge^{8})
\Biggr).
\label{10-21-2}
\end{eqnarray}
Note that the integrand is bounded by 
$\omega^{-2}\beta^{2}k^{2}$ 
which is integrable. 

Physically, energies are calibrated in the effective mass 
$m_{\mathrm eff}$, rather than $m$. 
$m_{\mathrm eff}$ is defined in the following way. 
For $Z = 0$ the Hamiltonian $H$ commutes with $p + P_{\mathrm f} 
= P$. Thus $H$ at fixed total momentum is given by 
\eq{S-4.3} 
H_{P} = \frac{1}{2}
\left( P - P_{\mathrm f}\right)^{2} 
+ \charge\left( P - P_{\mathrm f}\right)\cdot A_{0}  
+ \charge A_{0}^{\ast}\cdot \left( P - P_{\mathrm f}\right) 
+\frac{\charge^{2}}{2}
\left( A_{0}^{\ast 2} + 2A_{0}^{\ast}A_{0} 
+ A_{0}^{2}\right)
\en
as acting on ${\cal F}$.  
Because of infrared divergence $H_{P}$ is expected to have 
a ground state only for $P=0$. 
We set 
$$ 
E_{P} = \inf\sigma (H_{P}). 
$$
By the results in \cite{fro} 
$E_{P}$ is rotation invariant and 
\eq{S-4.4}
E_{P} = E_{0} + \frac{1}{2m_{\mathrm eff}}P^{2} 
+ {\cal O}(P^{4})
\en 
for small $P$. Thus from \kak{S-4.4}
\begin{eqnarray} 
\nonumber 
&{}& 
\frac{m}{m_{\mathrm eff}} 
= 1 - \frac{1}{3}\Delta E_{P}\mid_{P=0} \\  
&{}&  
= 
1 - \frac{2}{3}\langle \psi_{0}\, ,\, 
(P_{\mathrm f} - \charge A_{0})
\cdot (H_{0} - E_{0})^{-1}
(P_{\mathrm f} - \charge A_{0})\psi_{0}\rangle_{\cal F}, 
\label{S-4.7}
\end{eqnarray}  
by second order perturbation theory in $P$ at $P=0$. 
Here $\psi_{0}$ is the ground state of $H_{0}$, 
$H_{0} \psi_{0} = E_{0}\psi_{0}$. 
The inverse operator in \kak{S-4.7} is well defined, 
since $\langle \psi_{0}\, ,\, (P_{\mathrm f} - 
\charge A_{0})\psi_{0}\rangle_{\cal F} 
= (\nabla_{P}E_{P})_{P=0} = 0$. 
If $\omega(k) = (k^{2} + m_{\mathrm b}^{2})^{1/2}$ 
with $m_{\mathrm b} > 0$, 
then $E_{P}$ is an isolated eigenvalue 
\cite{fro} and \kak{S-4.7} follows by standard perturbation 
theory \cite{ka}. 
If $\kappa = 0$ and $\Lambda < \infty$, 
Chen \cite{ch} proves that $E_{P}$ is $C^{2}$ 
close to $P=0$. 
Expanding $m/m_{\mathrm eff}$ from \kak{S-4.7} 
in $\charge$, one obtains 
\eq{S-4.10} 
\frac{m_{\mathrm eff}}{m} 
= 1 + \charge^{2}\frac{2}{3}(2\pi)^{-3} 
\int_{\BR}(2\omega)^{-1}k^{2}\beta^{3}d^{3}k 
+ {\mathcal O}(e^{4}), 
\en
which suggests that the mass renormalization 
in the Nelson model is finite. 
If  $m_{\mathrm b} > 0$, \kak{S-4.10} 
would be the first two terms of a convergent 
power series. 

Writing $- E_{\mathrm bin} = m_{\mathrm eff}(m/m_{\mathrm eff}) 
(E_{\mathrm g} - E_{0})$ and inserting  \kak{10-21-2} 
and \kak{S-4.10} one obtains 
\begin{eqnarray} 
\nonumber 
&{}&  
- E_{\mathrm bin} \\ 
\nonumber 
&{}&= m_{\mathrm eff}
\Biggl( \eppp + \charge^{2}\frac{1}{3}(2\pi)^{-3} 
\int_{\BR}(2\omega)^{-1}\beta^{3} k^{2}\\ 
\nonumber 
&{}&\qquad\qquad   
\langle p\gsat\, ,\, 
(\pap - \eppp + \omega 
+ \frac{1}{2}k^{2})^{-1}
(\pap - \eppp)\cdot p\gsat
\rangle_{L^{2}}d^{3}k + E^{(6)} \\ 
&{}& \qquad
+ {\mathcal O}(\charge^{8})\Biggr). 
\label{S-4.12}
\end{eqnarray}
Thus to order $\charge^{2}$ the large $k$ behavior in 
the matrix element of \kak{S-4.12} is precisely 
canceled by $m/m_{\mathrm eff}$ and the numerical 
correction to $E_{\mathrm bin}$ is reduced considerably. 
For the hydrogen atom the matrix element in \kak{S-4.12} 
is not readily available. 
Approximating the operator ratio by $1$, 
one obtains 
\eq{S-4.13} 
E_{\mathrm bin} \cong 
-E_{\mathrm at}m_{\mathrm eff} 
\left( 1 - \frac{\charge^{2}}{6\pi^{2}}\right) 
+ 
{\mathcal O}(\charge^{8}). 
\en
The one-particle theory predicts 
$- \eppp m_{\mathrm eff}$ 
as binding energy, which is slightly reduced 
through the field fluctuations.

\section{Appendix A} 
\label{sec:Appendix1}

Let $\RI=\{k\in\BR||k|<1\}$ and $\RU=\BR\setminus \RI$. 
For $f \in L^{2}(\BR)$ we split $f = 
f_{\mbox{\tiny IR}} + f_{\mbox{\tiny UV}}$ 
defined by 
${\displaystyle 
f_{\mbox{\tiny {\rm IR}}}
= f\chi_{{}_{|k|<1}}}$ and  
${\displaystyle 
f_{\mbox{\tiny {\rm UV}}}
= f\chi_{{}_{|k|\ge 1}}}$. 

$\hfi$ and $\hfu$ are defined in 
\kak{hfiu}, and $\nfi$ and $\nfu$ are  in 
\kak{nfiu}. 
$N_{\rho}$ is defined in \kak{Nrho'}. 
We note that $N_{0} = \nfu$ and 
$N_{1} = \hfu$ again.
Then, the following lemma 
is a special case of 
\cite[(3.1.21)]{cj} though it has concrete coefficient.

\bl{key-estimate} \quad 
\bi 
\item[](i) 
Let $f, g \in L^{2}(\RU)$. Then, 
for $\e \ge 0$, 
\eq{eq:key-estimate} 
\| (\nfu + 1)^{1/2} 
N_{2\e}^{-1}a^{*}(g)a^{*}(f) 
(\nfu + 1)^{-1/2}\| 
\le 
\sqrt{3}\|f_{\mbox{\tiny UV}}/\omega^{\e}\|_{L^{2}}
\|g_{\mbox{\tiny UV}}/\omega^{\e}\|_{L^{2}},
\en
provided that $f/\omega^{\e}, 
g/\omega^{\e} \in L^{2}(\RU)$. 
\item[](ii) 
For $\e_{i} \ge 0$ and $t_{i} > 0$, $i = 1,2,3$, 
with $t_{1}\e_{1} = t_{2}\e_{2} + t_{3}\e_{3}$, 
$$
N_{\e_{1}}^{t_{1}} \le N_{\e_{2}}^{t_{2}}N_{\e_{3}}^{t_{3}}.
$$
\ei
\el 
\proof 
For (i) we have only to follow the proof of 
\cite[(3.1.21)]{cj} and for (ii) 
that of \cite[(2.32)]{ro}.  
\qed

In the following lemma, (i) is standard. 
In (ii), which is derived from Lemma \ref{key-estimate}, 
we develop a device to decouple 
infrared and ultraviolet problems.

\bl{ine} \quad 
\bi
\item[] (i) For $f \in L^{2}(\BR)$ with 
$\omega^{-1/2}f \in L^{2}(\BR)$, 
$
\|(\hf+1)^{-\han} \add(f) \| 
\leq \|f/\sqrt{\omega}\|_{L^{2}}.$
\item[] (ii) For $f, g \in L^{2}(\BR)$ with 
$f_{\mbox{\tiny {\rm IR}}}$, 
$g_{\mbox{\tiny {\rm IR}}}$, 
$\omega^{-1/2}f_{\mbox{\tiny {\rm IR}}}$, 
$\omega^{-1/2}g_{\mbox{\tiny {\rm IR}}}$, 
$\omega^{-1/4}f_{\mbox{\tiny {\rm UV}}}$, 
$\omega^{-1/4}g_{\mbox{\tiny {\rm UV}}}
 \in L^{2}(\BR)$, 
$$\| (\hf + 1)^{-1/2}
a(f)a(g)(\hf + 1)^{-1/2}\|_{\cal H} 
\le \Xi_{\e}(f,g),$$ \\ 
where   
\begin{eqnarray}
\nonumber
\Xi(f,g) &=& 
\left( 
\|\omega^{-1/2}g_{\mbox{\tiny {\rm IR}}}\|_{L^{2}} 
+ \|g_{\mbox{\tiny {\rm IR}}}\|_{L^{2}}\right)
\|\omega^{-1/2}f_{\mbox{\tiny {\rm UV}}}
\|_{L^{2}}  \\ 
\nonumber 
&{}& 
+ 
\left( \|\omega^{-1/2}f_{\mbox{\tiny {\rm IR}}}\|_{L^{2}} 
+ \|f_{\mbox{\tiny {\rm IR}}}\|_{L^{2}}\right)
\|\omega^{-1/2}g_{\mbox{\tiny {\rm UV}}}\|_{L^{2}} \\ 
\nonumber 
&{}& 
+ 
\|\omega^{-1/2}f_{\mbox{\tiny {\rm IR}}}\|_{L^{2}}
\|\omega^{-1/2}g_{\mbox{\tiny {\rm IR}}}\|_{L^{2}} 
+  
\sqrt{3}\|\omega^{-1/4}f_{\mbox{\tiny {\rm UV}}}\|_{L^{2}}
\|\omega^{-1/4}g_{\mbox{\tiny {\rm UV}}}\|_{L^{2}} \\ 
&{}& 
+ 
\frac{1}{2}
\left( 
\|\omega^{-1/2}g_{\mbox{\tiny {\rm IR}}}\|_{L^{2}}
\|f_{\mbox{\tiny {\rm IR}}}\|_{L^{2}}
+
\|\omega^{-1/2}f_{\mbox{\tiny {\rm IR}}}\|_{L^{2}} 
\|g_{\mbox{\tiny {\rm IR}}}\|_{L^{2}} 
\right).
\label{theta}
\end{eqnarray} 
\ei
\el
\proof 
(i) is a well-known fact. 
We will prove (ii), which uses the division 
of the momentum space 
into IR and UV regions. 
We first the following easy equalities and 
inequalities:  
$$\hf  =\hfi + \hfu \ge H_{{\mathrm f}j}, 
\qquad  
\nf = \nfi + \nfu 
\ge N_{{\mathrm f}j} 
$$ 
for $j = 1, 2$, 
and 
\begin{eqnarray*} 
a(f)a(g) 
&=& 
a(f_{\mbox{\tiny IR}})a(g_{\mbox{\tiny IR}})
+a(f_{\mbox{\tiny IR}})a(g_{\mbox{\tiny UV}})
+a(f_{\mbox{\tiny UV}})a(g_{\mbox{\tiny IR}})
+a(f_{\mbox{\tiny UV}})a(g_{\mbox{\tiny UV}}) \\ 
&=& 
a(f_{\mbox{\tiny IR}})a(g_{\mbox{\tiny IR}})
+a(f_{\mbox{\tiny IR}})a(g_{\mbox{\tiny UV}})
+a(g_{\mbox{\tiny IR}})a(f_{\mbox{\tiny UV}})
+a(f_{\mbox{\tiny UV}})a(g_{\mbox{\tiny UV}}).
\end{eqnarray*} 
We estimate the above four terms separately. 
We have easily   
\begin{eqnarray*} 
&{}&  
\| (\hfi + 1)^{-1/2}a (f_{\mbox{\tiny IR}}) 
a (g_{\mbox{\tiny IR}}) (\hfi + 1)^{-1/2}\|  \\ 
&{}&\le \| a^{\ast} (f_{\mbox{\tiny IR}})(\hfi + 1)^{-1/2}\| 
\|a (g_{\mbox{\tiny IR}}) (\hfi + 1)^{-1/2}\| \\ 
&{}&\le   
\left( \|\omega^{-1/2}f_{\mbox{\tiny IR}}\|_{L^{2}} 
+ \| f_{\mbox{\tiny IR}}\|_{L^{2}} 
\right) 
\|\omega^{-1/2}g_{\mbox{\tiny IR}}\|_{L^{2}}. 
\end{eqnarray*}
Similarly, we get 
\begin{eqnarray*}
&{}& 
\| (\hfi + 1)^{-1/2}a(f_{\mbox{\tiny IR}}) 
a(g_{\mbox{\tiny UV}}) (\hfu + 1)^{-1/2}\|  
\le   
\left( \|\omega^{-1/2}f_{\mbox{\tiny IR}}\|_{L^{2}} 
+ \| f_{\mbox{\tiny IR}}\|_{L^{2}} 
\right) 
\|\omega^{-1/2}g_{\mbox{\tiny UV}}\|_{L^{2}}, \\  
&{}& 
\| (\hfi + 1)^{-1/2}a(g_{\mbox{\tiny IR}}) 
a(f_{\mbox{\tiny UV}}) (\hfu + 1)^{-1/2}\|  
\le   
\left( \|\omega^{-1/2}g_{\mbox{\tiny IR}}\|_{L^{2}} 
+ \| g_{\mbox{\tiny IR}}\|_{L^{2}} 
\right) 
\|\omega^{-1/2}f_{\mbox{\tiny UV}}\|_{L^{2}}.
\end{eqnarray*}    
Thus we have 
\begin{eqnarray*} 
&{}& 
\| (\hf + 1)^{-1/2}a(f)a(g)(\hf + 1)^{-1/2}\| \\ 
&{}&\le 
\left( \|\omega^{-1/2}f_{\mbox{\tiny IR}}\|_{L^{2}} 
+ \| f_{\mbox{\tiny IR}}\|_{L^{2}} 
\right) 
\left( \|\omega^{-1/2}g_{\mbox{\tiny IR}}\|_{L^{2}} 
+ \|\omega^{-1/2}g_{\mbox{\tiny UV}}\|_{L^{2}} 
\right) \\ 
&{}& 
+ 
\left( \|\omega^{-1/2}g_{\mbox{\tiny IR}}\|_{L^{2}} 
+ \| g_{\mbox{\tiny IR}}\|_{L^{2}} 
\right)
\|\omega^{-1/2}f_{\mbox{\tiny UV}}\|_{L^{2}} 
+ 
\| (\hf + 1)^{-1/2}a(f_{\mbox{\tiny UV}})
a(g_{\mbox{\tiny UV}})(\hf + 1)^{-1/2}\|
\end{eqnarray*} 
by using that $H_{{\mathrm f}j} \le \hf$\, 
($ j = 1, 2$). 

From Lemma \ref{key-estimate} (ii) we have    
\begin{eqnarray} 
N_{1/2}^{2} \le 
(\nfu + 1)(\hfu + 1), 
\label{eq:ro} 
\end{eqnarray}
where 
\begin{eqnarray} 
N_{1/2} := 
\int_{|k|\ge 1}\sqrt{\omega(k)} 
a^{*} (k)a(k)d^{3}k. 
\end{eqnarray} 
Moreover, by Lemma \ref{key-estimate} (i), 
we obtain that  
\eq{eq:cj} 
\|(\nfu + 1)^{-1/2}a(f_{\mbox{\tiny UV}})
a(g_{\mbox{\tiny UV}})N_{1/2}^{-1}
(\nfu + 1)^{1/2}\| 
\le 
\sqrt{3}\|\omega^{-1/4}f_{\mbox{\tiny UV}}\|_{L^{2}}
\|\omega^{-1/4}g_{\mbox{\tiny UV}}\|_{L^{2}}. 
\en 
By \kak{eq:ro} and \kak{eq:cj}, we have 
\begin{eqnarray} 
\nonumber 
&{}& 
\| (\nfu + 1)^{-1/2} 
a(f_{\mbox{\tiny UV}})
a(g_{\mbox{\tiny UV}})(\hfu + 1)^{-1/2}\| \\ 
\nonumber 
&{}&= 
\| (\hfu + 1)^{-1/2}
a^{*}(g_{\mbox{\tiny UV}})
a^{*}(f_{\mbox{\tiny UV}})
(\nfu + 1)^{-1/2}\| \\  
\nonumber 
&{}&\le  
\|(\hfu + 1)^{-1/2}N_{1/2}(\nfu + 1)^{-1/2}\|\,\, 
\| (\nfu + 1)^{1/2}N_{1/2}^{-1}
a^{*}(g_{\mbox{\tiny UV}})
a^{*}(f_{\mbox{\tiny UV}})
(\nfu + 1)^{-1/2}\| \\ 
\nonumber 
&{}&\le  
 \|(\nfu + 1)^{-1/2}a(f_{\mbox{\tiny UV}})
a(g_{\mbox{\tiny UV}})N_{1/2}^{-1}
(\nfu + 1)^{1/2}\| \\ 
&{}&\le  
\sqrt{3}
\|\omega^{-1/4}f_{\mbox{\tiny UV}}\|_{L^{2}}
\|\omega^{-1/4}g_{\mbox{\tiny UV}}\|_{L^{2}}, 
\label{eq:am}
\end{eqnarray}  
(see also Lemma 3.3(iii) of \cite{am}), 
where $\nfu \le \hfu$ on $\RU$ is used. 
So, we have  
$$\| (\hf + 1)^{-1/2}a(f_{\mbox{\tiny UV}}) 
a(g_{\mbox{\tiny UV}})(\hf + 1)^{-1/2}\| 
\le  
\sqrt{3}
\|\omega^{-1/4}f_{\mbox{\tiny UV}}\| 
\|\omega^{-1/4}g_{\mbox{\tiny UV}}\|$$
with the inequality 
$\nfu \le \hfu \le \hf$, 
since $1 \le \omega(k)$ on $\RU$.  
Thus, finally, we obtain 
\begin{eqnarray*} 
&{}& 
\| (\hf + 1)^{-1/2}a(f)a(g)(\hf + 1)^{-1/2}\| \\ 
&{}&\le 
\left( \|\omega^{-1/2}f_{\mbox{\tiny IR}}\|_{L^{2}} 
+ \| f_{\mbox{\tiny IR}}\|_{L^{2}} 
\right) 
\left( \|\omega^{-1/2}g_{\mbox{\tiny IR}}\|_{L^{2}} 
+ \|\omega^{-1/2}g_{\mbox{\tiny UV}}\|_{L^{2}} 
\right) \\ 
&{}& 
+ 
\left( \|\omega^{-1/2}g_{\mbox{\tiny IR}}\|_{L^{2}} 
+ \| g_{\mbox{\tiny IR}}\|_{L^{2}} 
\right)\|\omega^{-1/2}f_{\mbox{\tiny UV}}\|_{L^{2}} 
+ 
\sqrt{3}
\|\omega^{-1/4}f_{\mbox{\tiny UV}}\|_{L^{2}}
\|\omega^{-1/4}g_{\mbox{\tiny UV}}\|.
\end{eqnarray*} 
Since $a(f)$ and $a(g)$ commute, 
taking their arithmetic mean results in (ii). 
\qed

The inequality \kak{eq:am} 
is proved here in the same way as in \cite{ne}, 
\cite[(3.1.21)]{cj} or \cite[Corollary 2.7]{am}. 
We also use some well known estimates for massive bosons, 
i.e., $\omega(k) = \sqrt{|k|^2+m_{\mathrm b}^2}$ 
with $m_{\mathrm b} > 0$, 
cf. \cite[Lemma 3.1.3 and (3.1.21)]{cj} and 
\cite[Lemma 3.3(iii)]{am}.

\section{Appendix B} 
\label{sec:Appendix3}

Let $A$ and $B$ be self-adjoint operators 
acting in a Hilbert space ${\mathcal X}$. 

\bl{appendix3}
Suppose that $A$ and $B$ are (strongly) 
commutable in the sense of the definition 
on p.271 in \cite{rs1}. 
If $\varphi \in D(A^{2})\cap D(B^{2})$, 
then $\varphi \in D(AB) \cap D(BA)$.
\el

\proof 
We rewrite $A$ and $B$ with the spectral 
decomposition as 
$$A = \int \xi dE_{A}(\xi), 
\qquad 
A = \int \eta dE_{B}(\eta),
$$
respectively. 
Then, one has 
\begin{eqnarray*} 
&{}& \int \xi^{2}\eta^{2} 
d\| E_{A}\otimes E_{B}(\xi , \eta)\varphi\|_{\mathcal X}^{2}
\equiv 
\int \xi^{2}\eta^{2} 
d\| E_{A}(\xi)E_{B}(\eta)\varphi\|_{\mathcal X}^{2} \\ 
&\le&  
\frac{1}{2} 
\int \xi^{4} 
d\| E_{A}(\xi)E_{B}(\eta)\varphi\|_{\mathcal X}^{2}
+ 
\frac{1}{2}
\int \eta^{4} 
d\| E_{A}(\xi)E_{B}(\eta)\varphi\|_{\mathcal X}^{2} \\ 
&=&  
\frac{1}{2} 
\int \xi^{4} 
d\| E_{B}(\eta)E_{A}(\xi)\varphi\|_{\mathcal X}^{2}
+ 
\frac{1}{2}
\int \eta^{4} 
d\| E_{A}(\xi)E_{B}(\eta)\varphi\|_{\mathcal X}^{2} \\ 
&\le&  
\frac{1}{2}
\int \xi^{4} 
d\| E_{A}(\xi)\varphi\|_{\mathcal X}^{2}
+ 
\frac{1}{2}
\int \eta^{4} 
d\| E_{B}(\eta)\varphi\|_{\mathcal X}^{2}
= 
\frac{1}{2}
\left(\|A^{2}\varphi\|_{\mathcal X}^{2}
+ 
\|B^{2}\varphi\|_{\mathcal X}^{2}
\right) < \infty,
\end{eqnarray*}  
which means that $\varphi \in  D(AB)$. 
In the same way, one has $\varphi \in  D(BA)$. 
\qed 

\hfill\break
\hfill\break
{\large {\bf Acknowledgement}} 
\hfill\break  
M.Hirokawa and F.Hiroshima thank for the hospitality 
at Technische Universit\"{a}t M\"{u}nchen. 
M.Hirokawa is indebted to his colleagues of 
the seminar on constructive field theory and 
renormalization group 
organized by H. Ezawa at Gakushuin University. 
He also thanks Griesemer for explaining him that 
Lemma \ref{SL1} implies Lemma \ref{1stmomentum'}(ii), 
and as a consequence 
Proposition \ref{1stmomentum} (ii), (iii), and  
exponential decay in Proposition \ref{exp-decay}. 
His work in Japan is supported by JSPS, 
Grant-in-Aid for Scientific Research (C) 13640215, 
and his work in Germany by DAAD.  
F. Hiroshima's work is supported by 
the Graduiertenkolleg ``Mathematik in ihrer 
Wechselbeziehung zur Physik'' of the LMU Munich 
and by JSPS, Grant-in-Aid 13740106 for Encouragement 
of Young Scientists.

{\footnotesize


\begin{thebibliography}{99}

\bibitem{am}
Z.~Ammari, 
Asymptotic completeness for a renormalized nonrelativistic 
Hamiltonian in quantum field theory: The Nelson model, 
{\it Math. Phys., Anal. Geom.} 
{\bf 3} (2000), 217--285. 

\bibitem{ar} 
A.~Arai, 
Ground state of the massless Nelson model without infrared cutoff 
in a non-Fock representation,  
{\it Rev.  Math.  Phys. } {\bf 13} (2001), 1075--1094.   

\bibitem{ah}
A.~Arai and M.~Hirokawa,   
On the existence and uniqueness of ground states of 
a generalized spin-boson model,  
{\it J. Funct. Anal.} 
{\bf 151} (1997), 455--503.   


\bibitem{bfs1}
V.~Bach, J.~Fr\"ohlich, and I.~M.~Sigal,   
Quantum electrodynamics of confined non-relativistic particles, 
{\it Adv. Math.} 
{\bf 137} (1998), 299--395.  

\bibitem{bfs2}
V.~Bach, J.~Fr\"ohlich, and I.~M.~Sigal,   
Spectral analysis for systems of atoms and molecules 
coupled to the quantized radiation field, 
{\it Commun. Math. Phys.} 
{\bf 207} (1999), 249--290.  


\bibitem{cj} J. Cannon and A. Jaffe, 
Lorentz covariance of the $\lambda(\phi^4)_2$ quantum field theory, 
Commun. Math. Phys. {\bf 17} (1970), 261--321. 

\bibitem{ch} 
T.~Chen, 
Operator-theoretic infrared renormalization and 
construction dressed $1$-particle states,  
(preprint, 2001, mp\_ arc 01-310).

\bibitem{fro}
J.~Fr\"{o}hlich,   
Existence of dressed electron states in a class 
of persistent models, 
{\it Fortschr. Phys.} {\bf 22} (1974), 159--198.  


\bibitem{gll} 
M.~Griesemer, E.~H.~Lieb, and M.~Loss, 
Ground states in non-relativistic quantum electrodynamics,
{\it Invent. Math.} {\bf 145} (2001), 557--595. 

\bibitem{Gr1} 
E. P. Gross, 
Small oscillation theory of the interaction of 
a particle and scalar field,
{\it Phys. Rev.} {\bf 100} (1955) 1571--1578. 

\bibitem{Gr} 
E. P. Gross,  
Particle-like solutions in field theory,
{\it Ann. Phys. (N.Y.)} {\bf 19} (1962) 219--233. 

\bibitem{grotch}
H. Grotch, 
Lamb shift in nonrelativistic quantum electrodynamics, 
{\it Am. J. Phys.} {\bf 49} (1981), 48--51. 




\bibitem{hainzl}
C.~Hainzl,   
One non-relativistic particle coupled to a photon field, 
arXiv: math-ph/0202001, 2002. 

\bibitem{ka}
T.~Kato, 
{\it Perturbation Theory for Linear Operators},  
Springer-Verlag, 1980.  


\bibitem{llp} 
T.~D. Lee, F.~E.~Low, and D.~Pines, 
The motion of slow electrons in a polar crystal, 
{\it Phys. Rev.} {\bf 90} (1953), 297--302. 

\bibitem{LP1} 
T. D. Lee and D. Pines, 
The motion of slow electrons in polar crystals,
{\it Phys. Rev.} {\bf 88} (1952) 960--961. 

\bibitem{LP2} 
T. D. Lee and D. Pines, 
Interaction of a nonrelativistic particle with 
a scalar field with application to slow electrons 
in polar crystals,
{\it Phys. Rev.} {\bf 92} (1953) 883--889. 

\bibitem{ll} 
E.~H.~Lieb and M.~Loss, 
A bound on binding energies and mass 
renormalization in models of quantum electrodynamics,  
{\it J. Stat. Phys.} {\bf 108} (2002) 1057--1069. 

\bibitem{lms}
J.~ L\H{o}rinczi, R.~A.~Minlos and H.~Spohn,  
The infrared behaviour in Nelson's model 
of a quantum particle coupled to a massless scalar field,  
{\it Ann. Henri Poincar\'{e}} {\bf 3} (2002), 269--295.  

\bibitem{ne}
E.~Nelson, 
Interaction of nonrelativistic particles 
with a quantized scalar field,  
{\it J. Math. Phys.} {\bf 5} (1964), 1190--1197.   
 
\bibitem{rs1}
M.~Reed and B.~Simon, 
{\it Methods of Modern Mathematical Physics I},  
Academic Press, 1980.  

\bibitem{rs2}
M.~Reed and B.~Simon, 
{\it Methods of Modern Mathematical Physics II},  
Academic Press, 1980.  

\bibitem{ro}
L. Rosen, 
The $(\phi^{2n})_{2}$ quantum field theory: Higher order estimates, 
{\it Comm. Pure Appl. Math.} {\bf 24} (1971), 
417--457.  


\bibitem{sp}
H.~Spohn,   
Ground state of quantum particle coupled to a scalar boson field,   
{\it Lett. Math. Phys.} {\bf 44} (1998), 9--16.   

\bibitem{tomo} 
S.~Tomonaga, 
On the effect of the field reactions on the interaction of 
mesotrons and nuclear particles. III,  
{\it Prog. Theoret. Phys.} {\bf 2} (1947), 6--24. 



\end{thebibliography}
\end{document}